\documentclass[12pt]{article}
\usepackage{putex}
\usepackage{graphicx}
\usepackage{latexsym,amsmath,amsfonts,amssymb}
\usepackage{bbm}

\renewenvironment{thebibliography}[1]{%
\begin{oldthebibliography}{#1}%
\setlength\itemsep{-2mm}%
}%
{%
\end{oldthebibliography}%
}

\newcommand{\trans}{\ensuremath{\mathsf T}}

\newcommand{\parfrac}[2]{\frac{\partial #1}{\partial #2}}

\newcommand{\sub}[1]{{\lfloor #1 \rfloor}}

\newcommand{\etc}{\textit{etc}\dots}

\numberwithin{equation}{section}

\newcommand{\Dslash}{D \!\!\!\!\slash\,}
\newcommand{\nn}{\nonumber}
\newcommand{\mat}[1]{\begin{pmatrix} #1 \end{pmatrix}}
\newcommand{\smat}[1]{\big( \begin{smallmatrix} #1 \end{smallmatrix} \big)}
\newcommand{\be}{\begin{equation}} \newcommand{\ee}{\end{equation}}
\newcommand{\bea}{\begin{equation} \begin{aligned}} \newcommand{\eea}{\end{aligned} \end{equation}}
\newcommand{\bmu}{\begin{multline}} \newcommand{\emu}{\end{multline}}

\newcommand{\cD}{\mathcal{D}}

\newcommand{\cG}{\mathcal{G}}

\newcommand{\cL}{\mathcal{L}}
\newcommand{\cM}{\mathcal{M}}
\newcommand{\cN}{\mathcal{N}}
\newcommand{\cO}{\mathcal{O}}

\newcommand{\cQ}{\mathcal{Q}}
\newcommand{\cR}{\mathcal{R}}

\newcommand{\cW}{\mathcal{W}}

\newcommand{\bC}{\mathbb{C}}

\newcommand{\bN}{\mathbb{N}}

\newcommand{\bR}{\mathbb{R}}
\newcommand{\bZ}{\mathbb{Z}}

\newcommand{\fm}{\mathfrak{m}}
\newcommand{\fn}{\mathfrak{n}}
\newcommand{\fu}{\mathfrak{u}}
\newcommand{\unit}{\mathbbm{1}}

\newcommand{\ti}{{\tilde \imath}}

\newcommand{\tM}{{\tilde M}}
\newcommand{\ttau}{{\tilde \tau}}

\newcommand{\bw}{{\overline w}}

\def\su{\mathfrak{su}}

\def\osp{\mathfrak{osp}}

\DeclareMathOperator{\sign}{sign}
\DeclareMathOperator{\Tr}{Tr}
\DeclareMathOperator{\Det}{Det}
\DeclareMathOperator{\rank}{rank}
\DeclareMathOperator{\Res}{Res}
\DeclareMathOperator{\re}{\mathbb{R}e}
\DeclareMathOperator{\im}{\mathbb{I}m}
\DeclareMathOperator{\coker}{coker}
\DeclareMathOperator{\ind}{ind}

\begin{document}


\title{Partition functions of $\cN=(2,2)$ \\ gauge theories on $S^2$ and vortices}

\authors{Francesco Benini$^\diamondsuit$ and Stefano Cremonesi$^\spadesuit$}

\institution{SB}{${}^\diamondsuit$
Simons Center for Geometry and Physics, Stony Brook University, \cr
$\;\:\,$ Stony Brook, NY 11794, USA}
\institution{IC}{${}^\spadesuit$
Theoretical Physics Group, Imperial College London, \cr
$\;\;\,$ Prince Consort Road, London, SW7 2AZ, UK}

\abstract{%
We apply localization techniques to compute the partition function of a two-dimensional $\cN=(2,2)$ R-symmetric theory of vector and chiral multiplets on $S^2$. The path integral reduces to a sum over topological sectors of a matrix integral over the Cartan subalgebra of the gauge group. For gauge theories which would be completely Higgsed in the presence of a Fayet-Iliopoulos term in flat space, the path integral alternatively reduces to the product of a vortex times an antivortex partition functions, weighted by semiclassical factors and summed over isolated points on the Higgs branch. As applications we
evaluate the partition function for some $U(N)$ gauge theories, showing equality of the path integrals for theories conjectured to be dual by Hori and Tong and deriving new expressions for vortex partition functions.}

\preprint{Imperial/TP/12/SC/02}

\maketitle


{\small
\setlength\parskip{-1.2mm}
\tableofcontents
}

\section{Introduction}

Supersymmetric theories are attractive theoretical laboratories allowing the exact computation of many observables irrespective of the strength of interactions. A recent line of development exploits the power of localization techniques, initiated in \cite{Witten:1988ze, Witten:1991zz}, which in many cases make it possible to exactly evaluate the path integral of a supersymmetric quantum field theory placed on a compact manifold, possibly with the insertion of local or non-local operators that respect some supersymmetry. For instance \cite{Pestun:2007rz} computed the path integral of an $\cN=2$ gauge theory on $S^4$; \cite{Kapustin:2009kz, Jafferis:2010un, Hama:2010av} of an $\cN=3$ and $\cN=2$ theory on $S^3$, and \cite{Hosomichi:2012ek} of an $\cN=1$ theory on $S^5$. We stress that in all these cases the gauge theories are not topologically twisted. Other related works are \cite{Kim:2009wb, Pestun:2009nn, Imamura:2011su, Hama:2011ea, Gomis:2011pf, Benini:2011nc, Ito:2011ea, Alday:2012au} where different compact manifolds are considered or various operators are inserted. Many of these papers build on previous work \cite{Moore:1997dj, Moore:1998et} and especially on the $\Omega$-background \cite{Nekrasov:2002qd, Nekrasov:2003rj} which in a sense is another example of this philosophy: even though the manifold on which the field theory is defined is non-compact, the $\Omega$-deformation effectively compactifies it.

We study the Euclidean path integral $Z_{S^2}$ of a two-dimensional $\cN=(2,2)$ theory of vector multiplets and chiral multiplets, placed on the round sphere $S^2$. The $\cN=(2,2)$ supersymmetry algebra on $S^2$ is $\osp^*(2|2) \cong \su(2|1)$, whose bosonic subalgebra has an $\su(2)$ factor of rotations of the sphere and a $\fu(1)$ R-symmetry factor reducing to the vector-like R-symmetry of the theory on $\bR^2$ in the large radius limit.
In particular the R-symmetry is part of the algebra, rather being an outer automorphism of it. When placing the theory on $S^2$ one has the freedom of selecting the R-symmetry, among a family of choices differing by mixing with non-R Abelian symmetries. This choice determines some specific couplings in the Lagrangian. When the theory is actually conformal, the supersymmetry algebra enhances to the superconformal algebra which coincides with the one on flat space. Similar properties have been observed in three, four and five dimensions \cite{Pestun:2007rz, Jafferis:2010un, Hosomichi:2012ek, Festuccia:2011ws}.

The localization technique is based on the observation \cite{Witten:1988ze, Witten:1991zz} that in certain situations the path integral can be exactly equal to its semiclassical approximation. Given a supercharge $\cQ$ that squares to a bosonic symmetry of the theory, the path integral only gets contributions from classical configurations that are fixed points of $\cQ$ and from small quadratic fluctuations around them. The fixed point set is often discrete or of finite dimension, while the contribution from small fluctuations is in terms of a ``one-loop determinant'' which is easy to compute. Therefore in these fortunate situations one is actually able to exactly compute the path integral.
Local or non-local $\cQ$-invariant operators can be included in the path integral as well: in this case one is able to exactly compute their expectation value (VEV).

When localization is at work, the path integral is only sensitive to the cohomology of $\cQ$: $\cQ$-exact operators do not affect the integral. In the case of a two-dimensional $\cN=(2,2)$ theory of vector and chiral multiplets on $S^2$, this implies that the path integral does not depend on the gauge coupling $g$ nor on the parameters of the superpotential $W$ (it does depend on the constraints on R-charges), while it depends on the twisted superpotential $\widetilde W$ -- including Fayet-Iliopoulos (FI) parameters and theta-angles -- and on twisted masses.

We reduce the path integral $Z_{S^2}$ to a matrix integral (\ref{matrix integral}) on the Cartan subalgebra of the gauge group, that we call ``localization on the Coulomb branch'', very similar to the three-dimensional $\cN=2$ case \cite{Kapustin:2009kz, Jafferis:2010un, Hama:2010av} except for a sum over topological sectors which do not appear on 3-spheres.
We perform some simple checks of our expression, in the case of a $U(1)$ gauge theory with chiral multiplets comparing with mirror symmetry, and in case of a pure $U(N)$ gauge theory.

Motivated by the work of \cite{Pasquetti:2011fj} where it was observed in two three-dimensional examples that the path integral on the ellipsoid can be rewritten as a sum of products of a vortex and an antivortex partition function, weighted by semiclassical factors, we wonder whether a similar phenomenon takes place in our setup and look for a different way of performing localization.
A physical way of understanding localization is to add to the Lagrangian a $\cQ$-exact term multiplied by a coupling $t$ that we take to infinity. In that limit the theory becomes semiclassical and the path integral is dominated by the saddle-points of the $\cQ$-exact term; independence of the result on $t$ completes the argument. We find that -- for theories that in flat space are completely Higgsed when a FI term is turned on -- an alternative choice of the deformation Lagrangian and of the path integration contour of auxiliary fields forces the path integral to localize on a finite sum of points on the Higgs branch. We call this ``localization on the Higgs branch''.

For each isolated point on the Higgs branch, the path integral gets contributions from point-like vortices at the north pole and antivortices at the south pole of $S^2$. Close to the poles the supercharge $\cQ$ and the action take the same form as the theory on the so-called $\Omega$-background \cite{Nekrasov:2002qd, Nekrasov:2003rj} studied in two dimensions by Shadchin \cite{Shadchin:2006yz}. Therefore $Z_{S^2}$ can be written as (\ref{Z_S2 Higgs final}) in terms of the vortex partition function $Z_\text{vortex}$. Localization tells us that such expression must agree with the matrix integral discussed before. In fact the expression on the Higgs branch is reminiscent of the $S^4$ case \cite{Pestun:2007rz}, where the instanton partition function $Z_\text{inst}$ of Nekrasov \cite{Nekrasov:2002qd} appears, although in that case $Z_\text{inst}$ is in the integrand of a matrix integral, while in our case $Z_\text{vortex}$ is related to a matrix integral.

We compute $Z_{S^2}$ for a $U(N)$ gauge theory with $N_f$ chiral multiplets in the fundamental representation, $N_a$ antifundamentals and possibly one adjoint, and explicitly check that the result of the matrix integral coincides with the expression in terms of $Z_\text{vortex}$ (when the latter is known). This suggests that the method can be used to compute $Z_\text{vortex}$ in the presence of matter representations for which other means are not available. Already our most general expression has not appeared in the literature before.

Finally, motivated by a duality conjectured by Hori and Tong \cite{Hori:2006dk, Hori:2011pd} between $SU(N)$ gauge theory with $N_f$ fundamentals (and $N_f > N$) and $SU(N_f - N)$ with the same number of fundamentals, we show that $Z_{S^2}$ coincide for the following pairs:
\be
Z_{S^2}^{U(N),\, N_f} \, (\xi, m) \;=\; Z_{S^2}^{U(N_f - N),\, N_f} \, (\xi, -m) \qquad\qquad\text{for $N_f > N$}
\ee
as functions of the FI parameter $\xi$ and the twisted masses $m$, and
\be
Z_{S^2}^{SU(N),\, N_f} \, (b, m) \;=\; Z_{S^2}^{SU(N_f - N),\, N_f} \, (b, -m) \qquad\qquad\text{for $N_f > N$}
\ee
as functions of the twisted mass $b$ associated to the baryon number and those $m$ associated to the flavor symmetry $SU(N_f)$. In the unitary case we extend the equality of partition functions to the case with antifundamentals.

The paper is organized as follows. In section \ref{sec: SUSY S2} we discuss supersymmetry on $S^2$ and the supersymmetric actions for vector and chiral multiplets that we will consider in the paper. In section \ref{sec: localization Coulomb branch} we use localization to reduce the path integral to a matrix integral, summed over topological sectors, while in section \ref{sec: localization Higgs branch} we follow a different route and obtain a sum over Higgs branch points including point-like vortices and antivortices. In section \ref{sec: examples} we present some simple examples, and in section \ref{sec: vortex partition functions} we perform the matrix integral computation for $U(N)$ with $(N_f,N_a)$ flavors and possibly an adjoint chiral multiplet, reproducing the expression in terms of the vortex partition function. In section \ref{sec: duality} we prove the identity of $Z_{S^2}$ for some pairs of gauge theories. In section \ref{sec: dimensional reduction} we compare our two-dimensional matrix integral with the three-dimensional one for the lens space $L(p,1)$ \cite{Benini:2011nc, Alday:2012au} in the $p \to \infty$ limit, and with the three-dimensional integral for the index \cite{Kim:2009wb, Imamura:2011su} in the limit where the radius of $S^1$ shrinks. We conclude in section \ref{sec: discussion} with some comments and open questions. Notations and tedious computations are in appendices.

\paragraph{Note:} During the preparation of this paper, we became aware of a related work \cite{Gomis_talk, Doroud:2012xw} which addresses similar questions.

\section{Supersymmetry on a curved manifold}
\label{sec: SUSY S2}

We start by constructing an $\cN=(2,2)$ supersymmetric Euclidean gauge theory on a curved two-manifold. Our conventions for spinors are summarized in appendix \ref{app: spinor conventions} and follow \cite{Hama:2010av}. We use complex two-dimensional anticommuting Dirac spinors, with multiplication given by
\be
\epsilon\lambda \equiv \epsilon^\alpha C_{\alpha\beta} \lambda^\beta = \lambda \epsilon
\ee
where $C_{\alpha\beta} \equiv -i \varepsilon_{\alpha\beta}$ is the antisymmetric charge conjugation matrix, and $\varepsilon_{12} = - \varepsilon_{21} = 1$ is the antisymmetric tensor. The gamma matrices $\gamma_{a=1,2}$ and the chirality matrix $\gamma_3 \equiv -i\gamma_1\gamma_2$ are Pauli matrices -- where we use $a,b = 1,2$ for flat indices, $i,j=1,2,3$ for flat indices including the chirality matrix, $\mu,\nu$ for curved indices --
such that
\be
\epsilon \gamma_i \lambda = - \lambda \gamma_i \epsilon \;,\qquad \epsilon \gamma_{ij} \lambda = - \lambda \gamma_{ij} \epsilon
\ee
with $\gamma_{ij} \equiv \gamma_{[i} \gamma_{j]}$. They follow from the defining properties of the charge conjugation matrix:
\be
C \gamma_i C = - \gamma_i^\trans \;,\qquad\qquad C^2 = 1 \;.
\ee
Gamma matrices also satisfy $\gamma_3 \gamma^a = i \varepsilon^{ab} \gamma_b$. Later on we will consider commuting spinors as well, and we will need to contract spinors without the matrix $C$ implicit: in that case we will specify either a dagger or a transpose. For instance: $\epsilon \lambda = \epsilon^\trans C \lambda$. The covariant derivative is $D_\mu = \nabla_\mu - i A_\mu$.

The 2d $\cN=(2,2)$ supersymmetry is the dimensional reduction of the 4d $\cN=1$ or the 3d $\cN=2$ supersymmetry. In Euclidean signature the minimal spinor is a complex Weyl spinor, therefore $\cN=(2,2)$ supersymmetry has two Dirac spinor parameters $\epsilon$, $\bar\epsilon$. The algebra is acted upon by a $U(1)_R \times U(1)_A$ R-symmetry outer automorphism, where the action of $U(1)_R$ is vector-like on fermions while $U(1)_A$ is axial. Notice that all flavor symmetries commuting with the supercharges are necessarily vector-like.

The global conformal group%
\footnote{To be more precise, the statements should be referred to the algebras. We follow the conventions of \cite{VanProeyen:1999ni}.}
in two Lorentzian dimensions is $SO(2,2) \cong SL(2,\bR) \times SL(2,\bR)$, and the $\cN=(2,2)$ global superconformal group is $OSp(2|2)^2$ whose bosonic subgroup is indeed $SL(2,\bR)^2 \times U(1)^2$ and it includes the R-symmetries. In Euclidean signature the global conformal group is $SO(3,1) \cong SL(2,\bC)$, and the $\cN=(2,2)$ global superconformal group is the complexification $OSp(2|2,\bC)$ whose bosonic subgroup is $SL(2,\bC) \times U(1)^2$. Let us explicitly construct vector multiplet and chiral multiplet representations of the $OSp(2|2,\bC)$ supergroup.

The gauge multiplet $V$ includes a vector $A_\mu$, two real scalars $\sigma$ and $\eta$ (that could be paired into a complex scalar $\hat\sigma = \sigma + i\eta$), a Dirac fermion $\lambda$ and a real auxiliary scalar $D$, all in the adjoint representation of the gauge group $G$. In Euclidean signature all fields get complexified and we will consider $\bar\lambda$ as a Dirac spinor independent of $\lambda$. We consider the following supersymmetry variations:
\bea
\label{vector SUSY transformations}
\delta A_\mu &= - \frac i2 (\bar\epsilon \gamma_\mu \lambda - \bar\lambda \gamma_\mu \epsilon) \\
\delta \sigma &= \frac12 (\bar\epsilon \lambda - \bar\lambda \epsilon) \\
\delta \eta &= - \frac i2 (\bar\epsilon \gamma_3 \lambda - \bar\lambda \gamma_3 \epsilon) \\
\delta \lambda &= i \gamma_3 \epsilon F_{12} - D\epsilon + i \gamma^\mu \epsilon \, D_\mu \sigma - \gamma_3 \gamma^\mu \epsilon\, D_\mu \eta - \gamma_3 \epsilon\, [\sigma,\eta] + i \gamma^\mu D_\mu \epsilon\, \sigma - \gamma_3 \gamma^\mu D_\mu \epsilon\, \eta \\
\delta \bar\lambda &= i \gamma_3 \bar\epsilon F_{12} + D\bar\epsilon - i \gamma^\mu \bar\epsilon \, D_\mu \sigma - \gamma_3 \gamma^\mu \bar\epsilon\, D_\mu \eta + \gamma_3 \bar\epsilon\, [\sigma,\eta] - i \gamma^\mu D_\mu \bar\epsilon\, \sigma - \gamma_3 \gamma^\mu D_\mu \bar\epsilon\, \eta \\
\delta D &= - \frac i2 \bar\epsilon \gamma^\mu D_\mu \lambda - \frac i2 D_\mu \bar\lambda \gamma^\mu \epsilon + \frac i2 [\bar\epsilon \lambda, \sigma] + \frac i2 [\bar\lambda \epsilon, \sigma] + \frac12 [\bar\epsilon \gamma_3 \lambda, \eta] + \frac12 [\bar\lambda \gamma_3 \epsilon, \eta] \\
&\quad - \frac i2 (D_\mu \bar\epsilon \gamma^\mu \lambda + \bar\lambda \gamma^\mu D_\mu \epsilon)
\eea
where $F_{12} = \frac12 F_{\mu\nu} \varepsilon^{\mu\nu}$.
For constant $\epsilon$, $\bar\epsilon$ they are the dimensional reduction of the 3d $\cN=2$ superalgebra with $\eta$ being the transverse component $A_3$.

Imposing the Killing spinor equation:
\be
D_\mu \epsilon = \gamma_\mu \tilde \epsilon \;,\qquad\qquad D_\mu \bar\epsilon = \gamma_\mu \tilde{\bar\epsilon}
\ee
for some spinors $\tilde\epsilon, \tilde{\bar\epsilon}$,
one can compute the following commutators:
\bea
\label{vector SC algebra mixed}
\,[\delta_\epsilon, \delta_{\bar\epsilon}] A_\mu &= (\cL_\xi^A A)_\mu + D_\mu \Lambda \\
[\delta_\epsilon, \delta_{\bar\epsilon}] \sigma &= \cL^A_\xi \sigma + i [\Lambda,\sigma] + \rho\,\sigma + 2\beta\,\eta \\
[\delta_\epsilon, \delta_{\bar\epsilon}] \eta &= \cL^A_\xi \eta +i [\Lambda,\eta] + \rho\,\eta - 2\beta\,\sigma \\
[\delta_\epsilon, \delta_{\bar\epsilon}] \lambda &= \cL^A_\xi \lambda + i[\Lambda, \lambda] + \frac32 \rho \lambda -i \alpha \lambda + i \beta \gamma_3 \lambda \\
[\delta_\epsilon, \delta_{\bar\epsilon}] \bar\lambda &= \cL^A_\xi \bar\lambda + i[\Lambda, \bar\lambda] + \frac32 \rho \bar\lambda + i \alpha \bar\lambda + i \beta \gamma_3 \bar\lambda \\
[\delta_\epsilon, \delta_{\bar\epsilon}] D &= \cL^A_\xi D + i[\Lambda,D] + 2\rho D \\
&\quad + (\bar\epsilon D_\mu D^\mu \epsilon - D_\mu D^\mu \bar\epsilon \epsilon)\sigma - i (\bar\epsilon \gamma_3 D_\mu D^\mu \epsilon - D_\mu D^\mu \bar\epsilon \gamma_3 \epsilon) \eta
\eea
where we have split $\delta = \delta_\epsilon + \delta_{\bar\epsilon}$ and the parameters are defined as:
\bea
\xi_\mu &= i \bar\epsilon \gamma_\mu \epsilon \qquad\qquad &
\Lambda &= \bar\epsilon \epsilon\, \sigma - i \bar\epsilon \gamma_3 \epsilon\, \eta \\
\alpha &= - \frac 14 (D_\mu\bar\epsilon \gamma^\mu \epsilon - \bar\epsilon \gamma^\mu D_\mu \epsilon) \qquad\qquad &
\rho &= \frac i2 (D_\mu\bar\epsilon \gamma^\mu \epsilon + \bar\epsilon \gamma^\mu D_\mu \epsilon) = \frac12 D_\mu \xi^\mu \\
\beta &= \frac14 ( D_\mu \bar\epsilon \gamma_3 \gamma^\mu \epsilon - \bar\epsilon \gamma_3 \gamma^\mu D_\mu \epsilon ) \;.
\eea
The operator $\cL^A_\xi$ is the gauge-covariant Lie derivative along the vector field $\xi^\mu$, acting on scalars, spinors and connection 1-forms as%
\footnote{The Lie derivative on spinors has been constructed in \cite{Kosmann:1972} and, although not manifestly, it is independent of the metric. $A_\mu$ acts in the correct gauge representation, and $d^A = d - iA$.}
\bea
\cL^A_\xi \sigma &= \xi^\mu (\partial_\mu - iA_\mu) \, \sigma \\
\cL^A_\xi \lambda &= \xi^\mu (\nabla_\mu - i A_\mu)\, \lambda + \frac14 (\nabla_\mu \xi_\nu) \gamma^{\mu\nu} \lambda \\
\cL^A_\xi A &= \cL_\xi A - d^A(\iota_\xi A) = \xi^\rho F_{\rho\mu} \, dx^\mu
\eea
respectively.
The algebra (\ref{vector SC algebra mixed}) implies that the commutator $[\delta_\epsilon, \delta_{\bar\epsilon}]$ is a translation by $\xi_\mu$, a gauge transformation by $\Lambda$, a dilation by $\rho$, a vector-like R-rotation by $\alpha$ and an axial R-rotation by $\beta$. The algebra closes when the second line in the variation of $D$ vanishes, which is achieved by imposing on the Killing spinors the extra conditions:
\be
\label{extra Killing condition}
D_\mu D^\mu \epsilon = h \epsilon \;,\qquad\qquad D_\mu D^\mu \bar\epsilon = h \bar\epsilon
\ee
with the same function $h$. We will see that on $S^2$ this is automatically satisfied by the Killing spinors. The remaining commutators are
\be
\label{vector SC algebra pure}
[\delta_{\epsilon_1}, \delta_{\epsilon_2}] = [\delta_{\bar\epsilon_1}, \delta_{\bar\epsilon_2}] = 0 \;.
\ee

The commutators (\ref{vector SC algebra mixed}) and (\ref{vector SC algebra pure}) describe the superconformal algebra $\osp(2|2,\bC)$. In particular the dimensions of the fields $(A_\mu, \hat\sigma, \lambda, \bar\lambda, D)$ in the vector multiplet are $(1,1,\frac32, \frac32,2)$ respectively, the charges under the vector-like R-symmetry are $(0,0,-1,1,0)$ and those under the axial R-symmetry are $(0, -1, 1,1,0)$.

The chiral multiplet $\Phi$ includes a complex scalar $\phi$, a Dirac spinor $\psi$ and a complex auxiliary scalar $F$, all in some representation $R_\Phi$ of the gauge group $G$. The supersymmetry variations of a chiral multiplet of vector-like R-charge $q$ are:
\bea
\label{matter SUSY transformations}
\delta \phi &= \bar\epsilon \psi \\
\delta \bar\phi &= \bar\psi \epsilon \\
\delta \psi &= i \gamma^\mu \epsilon \, D_\mu \phi + i \epsilon\, \sigma\phi + \gamma_3\epsilon\, \eta\phi + \frac{iq}2 \gamma^\mu D_\mu \epsilon\, \phi + \bar\epsilon F \\
\delta \bar\psi &= i \gamma^\mu \bar\epsilon \, D_\mu \bar\phi + i \bar\epsilon\, \bar\phi\sigma - \gamma_3\bar\epsilon\, \bar\phi\eta + \frac{iq}2 \gamma^\mu D_\mu \bar\epsilon\, \bar\phi + \epsilon \bar F \\
\delta F &= \epsilon \big( i \gamma^\mu D_\mu\psi - i \sigma \psi + \gamma_3 \eta \psi - i \lambda \phi \big) + \frac{iq}2 D_\mu \epsilon \gamma^\mu \psi \\
\delta \bar F &= \bar\epsilon \big( i \gamma^\mu D_\mu \bar\psi - i \bar\psi \sigma - \gamma_3 \bar\psi \eta + i \bar\phi \bar\lambda \big) + \frac{iq}2 D_\mu \bar\epsilon \gamma^\mu \bar\psi \;.
\eea
Indeed imposing the Killing spinor equations one finds the commutators:
\bea
\,[\delta_\epsilon, \delta_{\bar\epsilon}]\phi &= \cL^A_\xi \phi + i\Lambda\phi + \frac q2 \rho\phi + iq \alpha\phi \\
[\delta_\epsilon, \delta_{\bar\epsilon}]\bar\phi &= \cL^A_\xi \bar\phi - i \bar\phi\Lambda + \frac q2 \rho\bar\phi - iq \alpha\bar\phi \\
[\delta_\epsilon, \delta_{\bar\epsilon}] \psi &= \cL^A_\xi \psi + i \Lambda \psi + \frac{q+1}2 \rho \psi + i (q-1) \alpha \psi + i \beta \gamma_3 \psi \\
[\delta_\epsilon, \delta_{\bar\epsilon}] \bar\psi &= \cL^A_\xi \bar\psi - i \bar\psi \Lambda + \frac{q+1}2 \rho \bar\psi - i (q-1) \alpha\bar\psi - i \beta \gamma_3 \bar\psi \\
[\delta_\epsilon, \delta_{\bar\epsilon}] F &= \cL^A_\xi F + i\Lambda F + \frac{q+2}2 \rho F + i (q-2) \alpha F \\
[\delta_\epsilon, \delta_{\bar\epsilon}] \bar F &= \cL^A_\xi \bar F - i \bar F \Lambda + \frac{q+2}2 \rho \bar F - i (q-2) \alpha \bar F \;.
\eea
In particular the dimensions of the fields $(\phi, \bar\phi, \psi, \bar\psi, F, \bar F)$ are $(\frac q2, \frac q2, \frac{q+1}2, \frac{q+1}2, \frac{q+2}2, \frac{q+2}2)$ respectively, the charges under the vector-like R-symmetry are $(q, -q, q-1, 1-q, q-2, 2-q)$ and those under the axial R-symmetry are $(0,0,1,-1,0,0)$.

The remaining commutators are
\be
[\delta_{\epsilon_1}, \delta_{\epsilon_2}] = [\delta_{\bar\epsilon_1}, \delta_{\bar\epsilon_2}] = 0 \;.
\ee
The only exceptions are $[\delta_{\epsilon_1}, \delta_{\epsilon_2}]F = 2q \, \epsilon_{[1} D^\mu D_\mu \epsilon_{2]} \phi$ and $[\delta_{\bar\epsilon_1}, \delta_{\bar\epsilon_2}] \bar F = 2 q \, \bar\epsilon_{[1} D^\mu D_\mu \bar\epsilon_{2]} \bar\phi$: they vanish when the extra condition (\ref{extra Killing condition}) is imposed on the Killing spinors.

It follows from (\ref{matter SUSY transformations}) that if $F$ is the F-term of a neutral chiral multiplet of R-charge $q=2$, then
\be
\delta F = D_\mu ( i\epsilon \gamma^\mu \psi) \;,\qquad \delta \bar F = D_\mu (i \bar\epsilon \gamma^\mu \bar \psi )
\ee
are total derivatives. Such terms are superpotentials, and indeed their superconformal variations vanish up to total derivatives.

\subsection{Supersymmetry on $S^2$}

On the sphere $S^2$ the Euclidean global conformal group is $SO(3,1)$ as on flat space, however it is realized differently than in the latter case (similar observations have been made in four \cite{Pestun:2007rz} and three \cite{Jafferis:2010un} dimensions): the maximal compact subgroup $SO(3)$ acts as rotations of $S^2$.

There are four complex Killing spinors on $S^2$ (that we review in appendix \ref{app: Killing spinors}). A basis can be chosen such that two of them, that we call ``positive'', satisfy
\be\label{positive_Killing}
D_\mu \epsilon = \frac i{2r} \gamma_\mu \epsilon
\ee
where $r$ is the radius of the sphere, while two that we call ``negative'' satisfy $D_\mu \epsilon = - \frac i{2r} \gamma_\mu \epsilon$. None of them is chiral. If we restrict the supersymmetry variations to positive Killing spinors $\epsilon$ and $\bar\epsilon$ (negative spinors would give the same result), one verifies that
\be
\rho = 0 \;,\qquad\qquad
\alpha = \frac i{2r} \bar\epsilon \epsilon \;,\qquad\qquad \beta = 0 \;.
\ee
In particular the algebra does not contain dilations nor axial R-rotations. In fact the superalgebra is $\mathfrak{osp}^*(2|2) \cong \su(2|1)$, embedded as a real form into $\osp(2|2,\bC)$.%
\footnote{The embedding is non-chiral, as non-chiral are the positive Killing spinors.}
Its bosonic subalgebra is $\su(2) \oplus \mathfrak{u}(1)_R$, where the first factor are isometries of $S^2$ and the second is the vector-like R-symmetry. We call such superalgebra the $\cN=(2,2)$ Euclidean supersymmetry on $S^2$. Notice that the R-symmetry is part of the algebra, rather than an outer automorphism of it.

It turns out that a two-dimensional theory -- not necessarily conformal -- that on flat space has $\cN=(2,2)$ supersymmetry with a $U(1)_R$ vector-like R-symmetry, can be placed on $S^2$ preserving the $OSp^*(2|2)$ supersymmetry (a systematic method to do so has been developed in \cite{Festuccia:2011ws}). In doing so one has the freedom to choose the R-charges $q$ in (\ref{matter SUSY transformations}) (in a way which is compatible with superpotential interactions) and this results in different theories on $S^2$. If the theory on flat space is actually superconformal, then the R-charge of a chiral multiplet of dimension $\Delta$ is $q=2\Delta$. For such choice of $q$ in (\ref{matter SUSY transformations}), the supergroup $OSp^*(2|2)$ on $S^2$ enhances to the full superconformal group.

At this point we introduce coordinates $(\theta,\varphi)$ on $S^2$ with metric:
\be
ds^2 = r^2(d\theta^2 + \sin^2\theta\, d\varphi^2) \;.
\ee
We choose a vielbein $e^1 = r\, d\theta$, $e^2 = r \sin\theta\, d\varphi$.

\subsection{Supersymmetric actions on $S^2$}
\label{sec: SUSY actions}

Let us specialize the supersymmetry variations to positive Killing spinors $D_\mu \epsilon = \frac i{2r} \gamma_\mu \epsilon$, $D_\mu \bar\epsilon = \frac i{2r} \gamma_\mu \bar\epsilon$ on $S^2$:
\bea
\delta A_\mu &= - \frac i2(\bar\epsilon \gamma_\mu \lambda - \bar\lambda \gamma_\mu \epsilon) \qquad\qquad\qquad
\delta \sigma = \frac12(\bar\epsilon \lambda - \bar\lambda \epsilon) \qquad\qquad\qquad
\delta \eta = - \frac i2 (\bar\epsilon \gamma_3\lambda - \bar\lambda \gamma_3 \epsilon) \\
\delta \lambda &= i \gamma_3 \epsilon \Big( F_{12} - \frac\eta r + i[\sigma,\eta] \big) - \epsilon \Big( D + \frac\sigma r \Big) + i \Dslash \sigma \epsilon - \gamma_3 \Dslash \eta \epsilon \\
\delta \bar\lambda &= i \gamma_3 \bar\epsilon \Big( F_{12} - \frac\eta r - i[\sigma,\eta] \big) + \bar\epsilon \Big( D + \frac\sigma r \Big) - i \Dslash \sigma \bar\epsilon - \gamma_3 \Dslash \eta \bar\epsilon \\
\delta D &= - \frac i2 \bar\epsilon \gamma^\mu D_\mu \lambda - \frac i2 D_\mu \bar\lambda \gamma^\mu \epsilon + \frac i2 [\bar\epsilon \lambda, \sigma] + \frac i2 [\bar\lambda \epsilon, \sigma] + \frac12 [\bar\epsilon \gamma_3 \lambda, \eta] + \frac12 [\bar\lambda \gamma_3 \epsilon, \eta]
- \frac 1{2r} (\bar\epsilon \lambda - \bar\lambda \epsilon)
\eea
and
\bea
\delta \phi &= \bar\epsilon \psi &
\delta \bar\phi &= \bar\psi \epsilon \\
\delta \psi &= \Big( i \Dslash \phi + i \sigma\phi + \gamma_3 \eta\phi - \frac q{2r} \phi \Big) \epsilon + \bar\epsilon F &
\delta \bar\psi &= \Big( i \Dslash \bar\phi + i \bar\phi\sigma - \gamma_3 \bar\phi\eta - \frac q{2r} \bar\phi \Big) \bar\epsilon + \epsilon \bar F \\
\delta F &= \epsilon \Big( i \Dslash \psi - i \sigma \psi + \gamma_3 \eta \psi + \frac q{2r} \psi - i \lambda \phi \Big) \quad&
\delta \bar F &= \bar\epsilon \Big( i \Dslash \bar\psi - i \bar\psi \sigma - \gamma_3 \bar\psi \eta + \frac q{2r} \bar\psi + i \bar\phi \bar\lambda \Big) \;.
\eea
Supersymmetric Lagrangians on $S^2$ can be constructed by adding suitable $r^{-1}$ and $r^{-2}$ terms to the flat space Lagrangians, either working order by order (as in \cite{Pestun:2007rz, Kapustin:2009kz, Jafferis:2010un, Hama:2010av}) or by applying the method of \cite{Festuccia:2011ws}.

The Yang-Mills (YM) action is $S_{YM} = \int d^2x \, \frac1{g^2} \cL_{YM}$ with
\begin{multline}
\label{YM Lagrangian}
\cL_{YM} = \Tr \bigg\{ \frac12 \Big( F_{12} - \frac \eta r \Big)^2 + \frac12 \Big( D + \frac\sigma r \Big)^2 + \frac12 D_\mu\sigma D^\mu \sigma + \frac12 D_\mu\eta D^\mu \eta - \frac12 [\sigma,\eta]^2 \\
+ \frac i2 \bar\lambda \gamma^\mu D_\mu \lambda + \frac i2 \bar\lambda [\sigma,\lambda] + \frac12 \bar\lambda \gamma_3 [\eta,\lambda] \bigg\}\;.
\end{multline}
Notice that the bosonic part is a sum of squares and therefore positive definite.%
\footnote{The Lorentzian YM Lagrangian has bosonic part:
$$
\cL \,\sim\, - \frac14 (F_{\mu\nu})^2 + \frac12 D^2 - \frac12 (D_\mu\sigma)^2 - \frac12 (D_\mu\eta)^2 + \dots \;.
$$
In the Wick rotation $x^0 = -ix^2$, that implies $d^2x = i d^2x_E$ and $\partial_0 = i \partial_2$, one also redefines some fields: $A_0 = i A_2$, so that $F_{01} = -i F_{12}$ and $D = - i D_E$, so that the $\theta^+ \bar\theta^-$ component of the twisted chiral superfield $\Sigma$ is $D - i F_{01} = -i (D_E - iF_{12})$ and remains complex. This also introduces an unusual $i$ in (\ref{matter Lagrangian}).}
The usual flat space YM Lagrangian is recovered in the $r \to \infty$ limit, and it is invariant under SUSY transformations with $\partial_\mu \epsilon = \partial_\mu \bar\epsilon = 0$.
In fact the action (\ref{YM Lagrangian}) is even exact:
\be
\delta_{\epsilon} \delta_{\bar\epsilon} \int d^2x\, \Tr \Big( \frac12 \bar\lambda \lambda - 2 D\sigma - \frac1r \sigma^2 \Big) = \bar\epsilon\epsilon \int d^2x\, \cL_{YM} \;.
\ee
Since the integrand on the left-hand-side is a neutral scalar invariant under rotations of $S^2$, $\delta_\epsilon \delta_{\bar\epsilon}$ can be commuted and the formula above is valid with $\delta_{\bar\epsilon} \delta_\epsilon$ as well.

Given an Abelian gauge multiplet, we can add a Fayet-Iliopoulos (FI) Lagrangian $\cL_{FI}$. Since $\delta F_{12} = D_\mu (\varepsilon^{\mu\nu} \delta A_\nu)$ and $\delta D = - \frac i2 D_\mu( \bar\epsilon \gamma^\mu \lambda + \bar\lambda \gamma^\mu \epsilon)$ are total derivatives, the FI Lagrangian
\be
\label{FI Lagrangian}
\cL_{FI} = - i \xi D + i \frac\theta{2\pi} F_{12}
\ee
gives a supersymmetric action on $S^2$ without further $\frac1r$ corrections. The reason is that the FI action in two dimensions is superconformal, and the only corrections on $S^2$ for superconformal theories are the conformal couplings of scalars to curvature.

In fact the FI term is a special case of a \emph{twisted superpotential}. This is obtained by rewriting the vector multiplet $V$ in terms of a twisted chiral multiplet $\Sigma$, whose components are $(\sigma + i \eta, \lambda, \bar\lambda, D - i F_{12})$ transforming in the adjoint representation, and considering a holomorphic function $\widetilde W(\Sigma)$. A twisted superpotential interaction can be put supersymmetrically on $S^2$ for any gauge-invariant choice of $\widetilde W$. In the simplest case of a single Abelian vector multiplet, the Lagrangian is:
\be
\label{twisted superpotential Lagrangian}
\cL_{\widetilde W} = i \widetilde W' \, \Big( D - i F_{12} + \frac{\sigma + i\eta}r \Big) - \frac i2 \widetilde W''\, \bar\lambda (1+\gamma_3) \lambda - \frac ir \widetilde W\;,
\ee
where $\widetilde W(\sigma + i\eta)$ is a holomorphic function of $(\sigma + i\eta)$. Its twisted antichiral counterpart is:
\be
\label{anti-twisted superpotential Lagrangian}
\overline{\cL_{\widetilde W}} = i \widetilde W^{*\prime} \, \Big( D + i F_{12} + \frac{\sigma - i\eta}r \Big) - \frac i2 \widetilde W^{*\prime\prime} \, \bar\lambda (1-\gamma_3) \lambda - \frac ir \widetilde W^*\;,
\ee
where $\widetilde W^*(\sigma - i\eta)$ is a holomorphic function of $(\sigma - i\eta)$, which need not be the complex conjugate of $\widetilde W$. On the other hand, actions that would be real in Lorentzian signature are obtained by taking $\widetilde W^*(\bar z)$ as the complex conjugate of $\widetilde W(z)$.

For instance, the FI Lagrangian (\ref{FI Lagrangian}) follows from the twisted superpotential $\widetilde W(z) = \frac12 \big( - \xi + i \frac\theta{2\pi} \big) z$. Another example is $\widetilde W = \frac{L}{4} z^2$, which gives the Lagrangian:
\be
\label{CS Lagrangian}
\cL_{CS} = i L \Tr \Big( F_{12} \eta + D\sigma - \frac12 \bar\lambda \lambda + \frac{\sigma^2}{2r} - \frac{\eta^2}{2r} \Big)
\ee
where the length scale $L$ has been included for dimensional reasons. The flat space limit of \eqref{CS Lagrangian} can be obtained from the 3d Chern-Simons action on $\bR^2\times S^1$, by sending the radius of the circle $R\to 0$ and the CS level $k\to\infty$ while keeping $k R = L$ fixed \cite{Aganagic:2001uw}.
A related Lagrangian is obtained with $\widetilde W = i \frac{\tilde{L}}{4} z^2$:
\be
\cL = i \tilde{L} \Tr \Big( F_{12} \sigma - D\eta - \frac i2 \bar\lambda \gamma_3 \lambda - \frac{\sigma\eta}r \Big)
\ee
which does does not follow from three dimensions.

The kinetic Lagrangian on $S^2$ for a matter chiral multiplet of R-charge $q$ is:
\begin{multline}
\label{matter Lagrangian}
\cL_\text{mat} = D_\mu \bar\phi D^\mu \phi + \bar\phi \sigma^2 \phi + \bar\phi \eta^2 \phi + i \bar\phi D \phi + \bar F F + \frac{iq}r \bar\phi \sigma \phi + \frac{q(2-q)}{4r^2} \bar\phi \phi \\
- i \bar\psi \gamma^\mu D_\mu \psi + i \bar\psi \sigma \psi - \bar\psi \gamma_3 \eta \psi + i \bar\psi \lambda \phi - i \bar\phi \bar\lambda \psi - \frac q{2r} \bar\psi \psi \;.
\end{multline}
Notice that for $q=0$ there are no corrections to the flat space Lagrangian, which is classically superconformal. In fact the matter action is even exact:
\be
\delta_\epsilon \delta_{\bar\epsilon} \int d^2x\, \Big( \bar\psi \psi - 2i\bar\phi \sigma \phi + \frac{q-1}r \bar\phi \phi \Big) = \bar\epsilon \epsilon \int d^2x\, \cL_\text{mat} \;.
\ee
The same formula is valid with $\delta_{\bar\epsilon} \delta_\epsilon$.
The Lagrangian $\cL_\text{mat}$ depends holomorphically on the combination $\sigma + iq/2r$, once one fixes $D + \sigma/r=0$ which is the supersymmetry condition in (\ref{localizing locus vector}). Such on-shell constraint can be imposed only on background fields.
This means that the effect on the path integral  of shifting the R-charges proportionally to some flavor global charge can be obtained by giving an imaginary part to the twisted mass $\sigma^\text{ext}$ associated to that flavor global symmetry (compare with (\ref{R_mixing})).

We can add superpotential interactions:
\be
\cL_W = F^\text{($q=2$, neutral)} \;,\qquad\qquad \overline{\cL_W} = \bar F^\text{($q=2$, neutral)}\;,
\ee
where $F, \bar F$ are the F-terms of a neutral chiral multiplet of charge $q=2$. Since we have $\delta F = D_\mu (i\epsilon \gamma^\mu \psi)$ and $\delta \bar F = D_\mu (i\bar\epsilon \gamma^\mu \bar\psi)$, the action is supersymmetric. In more conventional terms the interaction is defined by the superpotential $W(\Phi_i)$, a holomorphic function of chiral multiplets, and the Lagrangian on $S^2$ is:
\be
\label{superpotential Lagrangian}
\cL_W = \sum_j \parfrac{W}{\phi_j} F_j - \sum_{j,k} \frac12 \parfrac{^2 W}{\phi_j \partial \phi_k} \psi_j \psi_k \;.
\ee
Its supersymmetry variation is a total derivative whenever $W(\Phi_i)$ is neutral and quasi-homogeneous of degree 2, assigning degree $q_i$ to $\Phi_i$.

Finally, if the theory has a global flavor symmetry $G_\text{global}$, we can introduce an external non-dynamical vector multiplet $V^\text{ext}$ coupled in a gauge-invariant way to the conserved current.%
\footnote{At linear order the vector multiplet couples to the current with minimal coupling, but in general gauge invariance might require the addition of seagull terms.}
Additional mass terms, called \emph{twisted masses}, are obtained in flat space by giving VEV to $\sigma^\text{ext}$, $\eta^\text{ext}$ in the external vector multiplet. Supersymmetry on $S^2$ relates $D^\text{ext}$, $F_{12}^\text{ext}$ to $\sigma^\text{ext}$, $\eta^\text{ext}$ as in (\ref{localizing locus vector}) and it requires all four terms to mutually commute. Therefore, up to flavor rotations, twisted masses take values in the complexified Cartan subalgebra of the global symmetry group $G_\text{global}$.
Notice also that $F_{12}^\text{ext}$ should be quantized if we associate it to the background connection of a compact symmetry.

\section[The $S^2$ partition function: localization on the Coulomb branch]{The $S^2$ partition function: \\ localization on the Coulomb branch}
\label{sec: localization Coulomb branch}

We would like to compute the path integral:
\be
Z_{S^2} = \int \cD\varphi\, e^{-S[\varphi]} \;,
\ee
where $\varphi$ are all fields in the Lagrangian, for a theory of vector and chiral multiplets with Lagrangian made of the pieces we discussed: (\ref{YM Lagrangian}) to
(\ref{superpotential Lagrangian}) plus twisted masses. We adopt the localization technique \cite{Witten:1988ze} along the lines of \cite{Pestun:2007rz, Kapustin:2009kz, Jafferis:2010un}.

We choose a (commuting) positive Killing spinor $\epsilon_+$ on $S^2$ (see appendix \ref{app: Killing spinors}) normalized as $\epsilon_+^\dag \epsilon_+ = 1$. It defines a Killing vector $v_\mu$ (in particular $D^\mu v_\mu = 0$), its dual vector $w^\mu = \varepsilon^{\mu\nu} v_\nu$ and a scalar function $s = \frac r2 D_\mu w^\mu$:
\be
v_\mu = \epsilon_+^\dag \gamma_\mu \epsilon_+ \;,\qquad w^\mu = -i \epsilon_+^\dag \gamma_3 \gamma^\mu \epsilon_+ \;,\qquad s = \epsilon_+^\dag \gamma_3 \epsilon_+ \;,
\ee
and we have $\epsilon_+^\trans C \epsilon_+ = 0$.
The Killing vector $v_\mu$ vanishes at two antipodal points that we call north and south pole. In terms of a polar coordinate $\theta$ on $S^2$ such that $\theta = 0$ ($\theta = \pi$) at the north (south) pole, $s=\cos\theta$ and the vector fields in vielbein basis are $v^a = (0,\sin\theta)$ and $w^a = (\sin\theta,0)$.

Then we construct the supercharges $Q,Q^\dag$ corresponding to $\epsilon_+$. To do that, first we extract the supercharges%
\footnote{Recall that $Q^\dag_\alpha$ is not the Hermitian conjugate of $Q_\alpha$.}
$Q_\alpha$, $Q^\dag_\alpha$ from the commuting operator $\delta$:
\be
\delta = \delta_\epsilon + \delta_{\bar\epsilon} = \epsilon^\alpha Q_\alpha + \bar\epsilon^\alpha Q^\dag_\alpha
\ee
where $\epsilon, \bar\epsilon$ are anticommuting. Then we construct
\be
Q \equiv \epsilon_+^\alpha Q_\alpha \;,\qquad\qquad Q^\dag \equiv \epsilon_+^{c\, \alpha} Q^\dag_\alpha
\ee
where now $\epsilon_+$ is commuting and $\epsilon_+^c = C \epsilon_+^*$ is its charge conjugate. In the same way we can transform barred spinors to their charge conjugate: $\bar\lambda = C (\lambda^\dag)^\trans$, recalling that $\lambda^\dag$ is not the Hermitian conjugate of $\lambda$, \etc{}
We will use the two notations interchangeably. Now $Q,Q^\dag$ are anticommuting (their explicit expression is given in appendix \ref{app: supercharges}) and in particular -- up to a gauge transformation $\Lambda$ -- form the $\su(1|1)$ superalgebra
\be
\{Q,Q^\dag\} = M + \frac R2 + i \Lambda \;,\qquad\qquad Q^2 = Q^{\dag\,2} = 0
\ee
where $M$ is the angular momentum that generates translations along $v_\mu$ and $R$ is the R-charge. Finally, we construct the supercharge $\cQ = Q + Q^\dag$ so that:
\be
\label{square of Q}
\cQ^2 = M + \frac R2 + i\Lambda \;,\qquad\qquad \Lambda = -\sigma + is\, \eta \;,
\ee
and the explicit expression for the gauge transformation is computed in appendix \ref{app: supercharges}.

We perform localization with respect to the supercharge $\cQ$. By the usual localization argument, as long as the action is $\cQ$-closed (which is the case if it is supersymmetric) the expectation value of $\cQ$-closed operators is unaffected by a deformation of the path integrand by $\cQ$-exact terms. Moreover the path integral is dominated by the fixed points of $\cQ$ and the small quadratic fluctuations around them (through a one-loop determinant). The second statement follows from the first one, as we now review. We can deform the action by the exact term $S_0 \to S_0 + t\delta S$, where $\delta S = \int (\cL_{YM} + \cL_\psi)$ and
\be
\label{localizing actions}
\cL_{YM} = \Tr \cQ\, \frac{(\overline{\cQ\lambda}) \lambda + \lambda^\dag (\overline{\cQ\lambda^\dag})}4  \;,\qquad\qquad \cL_\psi = \cQ\, \frac{(\overline{\cQ\psi}) \psi + \psi^\dag (\overline{\cQ\psi^\dag}) }2 \;.
\ee
Here $\overline{\phantom{\lambda}}$ is a formal conjugation that acts as $\dag$ on c-numbers and exchanges $\lambda,\phi,\psi,F$ with $\lambda^\dag, \phi^\dag, \psi^\dag, F^\dag$. As the name suggests, $\cL_{YM}$ is the same as in (\ref{YM Lagrangian}) up to total derivatives. The bosonic parts of these actions are
\be
\label{bosonic deformation parts}
\cL_{YM} \big|_\text{bos} = \frac14 \Tr \big( |\cQ\lambda|^2 + |\cQ\lambda^\dag|^2 \big) \;,\qquad\qquad \cL_\psi \big|_\text{bos} = \frac12 \big( |\cQ\psi|^2 + |\cQ\psi^\dag|^2 \big) \;,
\ee
which are formally positive definite and vanish on the fixed points of $\cQ$. In the $t\to\infty$ limit only BPS configurations with $\cL_{YM} = \cL_\psi = 0$ contribute to the path integral: $Z$ equals a sum (or integral) over the BPS solutions, of the Gaussian path integral of the truncated quadratic action around them. The Gaussian path integral gives the on-shell action times a one-loop determinant. Since $\delta S$ vanishes on-shell, only $S_0$ contributes to the on-shell action; on the contrary when $t\to\infty$, $S_0$ is irrelevant with respect to $t\delta S$ in determining the spectrum of fluctuations around a saddle point. The result is independent of $t$, as it should.

The localization argument immediately tells us that the partition function $Z_{S^2}$ will not depend on the YM coupling $g$, since $\cL_{YM}$ in (\ref{YM Lagrangian}) is exact, nor on the superpotential Lagrangian $\cL_W$, since
\be
\cL_W = \int F^\text{($q=2$, neutral)} = \int \cQ \big( \epsilon_+^\trans C \psi \big)
\ee
is exact (it does depend on the constraints that $W$ imposes on the R-charges); $Z_{S^2}$ will depend on twisted masses and the twisted superpotential $\widetilde{W}$.

\

Let us study the BPS configurations. The form (\ref{bosonic deformation parts}) assures that the saddle points of $\delta S$ coincide with its zeros, that is with the BPS configurations.%
\footnote{\label{footnote: reality}
For a fermion $\lambda$, $\cQ\lambda = 0$ coincides with $|\cQ\lambda|^2 \equiv (\overline{\cQ\lambda}) \cQ\lambda = 0$ only if we impose on fields the same reality conditions used to define $\overline{\phantom{\lambda}}$, like $D^\dag = D$, $\sigma^\dag = \sigma$, \etc{} If we relaxed the reality conditions, the BPS equations would have many more solutions.}
For the gauge multiplet we have, up to total derivatives:
\be
\cL_{YM} \big|_\text{bos} = \Tr \Big\{ \frac12 \Big( F_{12} - \frac\eta r\Big)^2 + \frac12 \Big( D + \frac\sigma r \Big)^2 + \frac12 (D_\mu\sigma)^2 + \frac12 (D_\mu\eta)^2 - \frac12 [\sigma,\eta]^2 \Big\} \;.
\ee
The BPS equations $0 = \cQ\lambda = \cQ\lambda^\dag$ are
\be
\label{localizing locus vector}
0 = F_{12} - \frac\eta r = D + \frac\sigma r = D_\mu \sigma = D_\mu \eta = [\sigma,\eta] \;.
\ee
They imply that $\sigma$ and $\eta$ can be simultaneously diagonalized, then $F_{12}$ and $D$ are fixed in terms of them; due to diagonalization, $\sigma$ and $\eta$ are not charged under the background, therefore they are simply constant on $S^2$. The gauge flux
\be
\frac1{2\pi} \int F = \frac1{2\pi} \int F_{12}\, e^1 \wedge e^2 = 2r^2 F_{12}
\ee
is quantized:
\be
F_{12} = \frac{\fm}{2r^2} \;,\qquad\qquad \eta = \frac{\fm}{2r}\;,
\ee
where $\fm$ belongs to a maximal torus of the gauge algebra and is GNO quantized \cite{Goddard:1976qe}, meaning that for any representation $R$ and weight $\rho \in R$, $\rho(\fm) \in \bZ$. The set of BPS configurations is therefore parametrized by a continuous variable $\sigma$ and the discrete fluxes $\fm$ of $F_{12}$. Recall that at this stage $\sigma$ and $F_{12}$ represent dynamical as well as external fields.

For the chiral multiplet we find, up to total derivatives:
\begin{multline}
\cL_\psi \big|_\text{bos} = |D_\mu\phi|^2 + i \frac{1-q}r v^\mu\, \phi^\dag D_\mu \phi + \phi^\dag \Big( \sigma^2 + \eta^2 + \frac{q^2}{4r^2} \Big) \phi + |F|^2 \\
+ s\, \Big( \frac{2-q}r \, \phi^\dag \eta \phi - \phi^\dag F_{12} \phi \Big) + w^\mu \, \phi^\dag D_\mu \eta \phi \;.
\end{multline}
In this case it is easier to study the BPS equations $0 = \cQ\psi = \cQ\psi^\dag$ directly:
\bea
\label{BPS equations matter}
0 &= -\sin\tfrac\theta2\, (2D_+\phi + F) + \cos\tfrac\theta2\, (i\sigma + \eta - \tfrac q{2r})\phi \\
0 &= +\cos\tfrac\theta2\, (2D_-\phi + F) + \sin\tfrac\theta2\, (i\sigma -\eta - \tfrac q{2r})\phi \\
0 &= -\sin\tfrac\theta2\, (2D_-\phi^\dag - F^\dag) + \cos\tfrac\theta2\, \phi^\dag(i\sigma + \eta - \tfrac q{2r}) \\
0 &= +\cos\tfrac\theta2\, (2D_+\phi^\dag - F^\dag) + \sin\tfrac\theta2\, \phi^\dag(i\sigma -\eta - \tfrac q{2r})
\eea
where $D_\pm = (D_1 \mp i D_2)/2$. For generic $q$ (we will study the case $q=0$ in section \ref{sec: localization Higgs branch}) and assuming that the bosonic fields in the vector multiplet are real, the solutions are:
\be
\label{localizing locus matter}
0 = \phi = \bar\phi = F = \bar F \;.
\ee
To be precise, solving the BPS equations (\ref{BPS equations matter}) algebraically one finds another branch with $0 = F = \sigma \phi$ and
\be
2\cos\tfrac\theta2\, D_-\phi = \sin\tfrac\theta2\, \big( \eta + \tfrac q{2r} \big)\phi \;,\qquad\qquad 2 \sin\tfrac\theta2\, D_+\phi = \cos\tfrac\theta2\, \big( \eta - \frac q{2r} \big)\phi \;.
\ee
We combine them with the gauge BPS equations $D_\mu \eta = 0$ and $F_{12} = \eta/r$, whose potential is $A_\varphi = -r\cos\theta\, \eta$. Away from the poles at $\theta=0,\pi$ they imply $\partial_\varphi \phi = -i\frac q2 \phi$ as well as $\partial_\theta \phi = r (\sin\theta)^{-1} (\eta - \frac q{2r}\cos\theta)\phi$. They can be integrated: $\phi = C \big( \tan\frac\theta2 \big)^{r\eta} (\sin\theta)^{-q/2} e^{-iq\varphi/2}$,
however this solution is smooth only for $C = 0$. Even the point-like vortex solutions that we will discuss in section \ref{sec: localization Higgs branch} are not present here due to $D_\mu F_{12} = 0$.

Next we need the one-loop determinant of quadratic fluctuations around the BPS configurations. Instead of using the localizing Lagrangian $\cL_\psi$ in (\ref{localizing actions}), we prefer to use directly $\cL_{YM}$ (\ref{YM Lagrangian}) and $\cL_\text{mat}$ (\ref{matter Lagrangian}) which are $\cQ$-exact: they vanish and are extremized on the BPS configurations. They have the advantage of being invariant under $SO(3)$ rotations of $S^2$, therefore the computation simplifies. We will recompute the one-loop determinants from $\cL_{YM} + \cL_\psi$ using an index theorem in section \ref{sec: index theorem}.

\subsection{The vector multiplet}
\label{sec: one-loop gauge}

We compute the one-loop determinant of quadratic fluctuations of the YM action (\ref{YM Lagrangian}), to which we will soon have to add gauge-fixing terms, around the saddle points:
\be
\label{gauge background Coulomb}
0 = D_\mu\sigma = D_\mu\eta = [\sigma,\eta] \;,\qquad F_{12} = \frac\eta r = \frac{\fm}{2r^2} \;,\qquad D= -\frac\sigma r \;.
\ee
Let us start with the bosonic part of $S = t\int \cL_{YM}$. We separate all fields in a background part and a fluctuation part, as $\varphi = \varphi_0 + \frac{1}{\sqrt{t}} \tilde\varphi$, and expand. Only terms quadratic in the fluctuations appear in the $t\to +\infty$ limit.
We also need to fix the gauge. On a background the gauge-fixing action is:
\be
\cL_\text{g-f} = - \bar c \big( D^\mu D_\mu c - i D^\mu [\tilde A_\mu, c] \big) - \frac1{2\xi} (D^\mu \tilde A_\mu)^2 \;,
\ee
where $c, \bar c$ are anticommuting complex scalar ghost fields in the adjoint representation, $\xi$ is the gauge-fixing parameter%
\footnote{The parameter $\xi$ used in this subsection and in appendix \ref{app: 1-loop determinants} should not be confused with the FI parameter $\xi$ used in the rest of the paper.}
in $R_\xi$-gauge which can be chosen at will, and all covariant derivatives are on the background $A^{(0)}_\mu$. Keeping only terms up to quadratic order in the fluctuations, we get:
\begin{multline}
\nn
\cL = \frac12 \Big( \tilde F_{12} - \frac{\tilde\eta}r \Big)^2 + \frac12 \Big(\tilde D + \frac{\tilde\sigma}r \Big)^2 + \frac12 \big( D_\mu \tilde\sigma - i[\tilde A_\mu, \sigma_0] \big)^2 + \frac12 \big( D_\mu \tilde\eta - i [\tilde A_\mu, \eta_0])^2 \\
- \frac12 \big( [\tilde\sigma,\eta_0] + [\sigma_0, \tilde\eta] \big)^2 - \bar c D^\mu D_\mu c - \frac1{2\xi} (D^\mu \tilde A_\mu)^2 \;.
\end{multline}
We can immediately integrate over $\tilde D$, whose one-loop determinant is 1.

The determinant for the remaining fields is computed in appendix \ref{app: 1-loop determinants}. It turns out to be a product over the roots $\alpha$ of the gauge group $G$. When $\alpha(\fm) \neq 0$ one finds, up to factors that are independent of the background $\sigma, \fm$ that we reabsorb in the normalization:
\be
\label{1-loop gauge}
\frac{\Det \cO_c}{\sqrt{\Det \cO_\text{gauge}}} \Big|_\alpha = \prod_{k = \frac{|\alpha(\fm)|}2}^\infty \frac1{\big( k + ir\alpha(\sigma) \big)^{2k-1} \big( k+1+ir\alpha(\sigma) \big)^{2k+3}} \;,
\ee
where $\cO_c$ and $\cO_\text{gauge}$ are the kinetic operators for ghosts and for $(\tilde A_\mu, \tilde\sigma, \tilde\eta)$ respectively. When $\alpha(\fm) = 0$ there are zero-modes (see appendix \ref{app: 1-loop determinants} for details): after removing the zero eigenvalues one is left with
\be
\label{1-loop gauge zero-modes}
\frac{\Det' \cO_c}{\sqrt{\Det' \cO_\text{gauge}}} \Big|_\alpha = \frac1{|\alpha(\sigma)|} \prod_{k = 0}^\infty \frac1{\big( k + ir\alpha(\sigma) \big)^{2k-1} \big( k+1+ir\alpha(\sigma) \big)^{2k+3}} \;.
\ee
The zero-modes correspond to those of the background scalar $\sigma$, and span the subalgebra $\{\alpha(\fm) = 0\}$ which is left unbroken by the flux $\fm$. To integrate over them, we integrate over a Cartan subalgebra and use a Vandermonde determinant for the unbroken symmetry:
\be
\int \text{zero-modes} =
\frac1{|\cW_\fm|} \int \Big( \prod_{n=1}^{\rank(G)} d\sigma_n \Big) \prod_{\alpha(\fm) = 0} |\alpha(\sigma)|
\ee
where $|\cW_\fm|$ is the order of the Weyl group of the subalgebra $\{\alpha(\fm)=0\}$. Notice that the Vandermonde measure cancels against the extra factor in the one-loop determinant (\ref{1-loop gauge zero-modes}).

Let us now consider the fermions $\lambda, \bar\lambda$. The quadratic expansion of the fermionic part of the YM action (\ref{YM Lagrangian}) around the background is:
$$
\cL = \frac i2 \bar\lambda \gamma^\mu D_\mu \lambda + \frac i2 \bar\lambda[\sigma_0 ,\lambda] + \frac12 \bar\lambda \gamma_3[\eta_0 ,\lambda] \;.
$$
The one-loop determinant is computed in appendix \ref{app: 1-loop determinants}. As before the final result is a product over the roots of $G$:
\be
\label{1-loop gaugino}
\Det \cO_\lambda = \prod_{\alpha \in G} \prod_{k = \frac{|\alpha(\fm)|}2}^\infty \big( k + ir\alpha(\sigma) \big)^{2k} \big( k + 1 + ir\alpha(\sigma) \big)^{2k+2} \;.
\ee

The one-loop determinant for the gauge multiplet is the product of (\ref{1-loop gauge}) and (\ref{1-loop gaugino}), neglecting the possible extra factor that cancels against the Vandermonde determinant. After many cancelations, the final result is:
\be
\label{Z gauge}
Z_\text{gauge} = \prod_{\alpha \in G} \Big( \frac{|\alpha(\fm)|}2 + ir\alpha(\sigma) \Big) = \prod_{\alpha > 0} \Big( \frac{\alpha(\fm)^2}4 + r^2 \alpha(\sigma)^2 \Big) \;.
\ee

\subsection{The chiral multiplet}
\label{sec: one-loop matter}

Now we consider the one-loop determinant of quadratic fluctuations of the matter Lagrangian $\cL_\text{mat}$ (\ref{matter Lagrangian}) around the saddle points: $0 = \phi = \bar\phi = F = \bar F$.
The quadratic expansion of the bosonic part around the background is:
$$
\cL = \bar\phi \Big( - D^\mu D_\mu + \sigma^2 + \eta^2 + iD + \frac{iq\sigma}r + \frac{q(2-q)}{4r^2} \Big) \phi + \bar F F \;.
$$
The auxiliary fields $F, \bar F$ give 1. The scalar $\phi$ in representation $R_\Phi$ is decomposed along the weights $\rho \in R_\Phi$, and in spin spherical harmonics $Y^s_{j,j_3}$ with effective spin $s = - \frac12 \rho(\fm)$. For each mode the eigenvalue is
\be
\cO_\phi = \frac1{r^2} \Big( j + \frac q2 -ir\rho(\sigma) \Big) \Big( j+1-\frac q2 +ir\rho(\sigma) \Big) \;.
\ee
Taking the product over $\rho \in R_\Phi$, $j \geq |s|$ and $|j_3| \leq j$, the determinant is:
\be
\Det \cO_\phi = \prod_{\rho \in R_\Phi} \prod_{j = \frac{|\rho(\fm)|}2}^\infty \Big( j + \frac q2 - ir\rho(\sigma) \Big)^{2j+1} \Big( j+1-\frac q2 + ir\rho(\sigma) \Big)^{2j+1} \;.
\ee

The fermionic part of the matter action (\ref{matter Lagrangian}) expanded at second order around the background is:
$$
\cL = \bar\psi \Big( -i \gamma^\mu D_\mu + i\sigma - \gamma_3\eta - \frac q{2r} \Big) \psi \;.
$$
The one-loop determinant is computed in appendix \ref{app: 1-loop determinants}:
\be
\Det \cO_\psi = \prod_{\rho \in R_\Phi} \prod_{k = \frac{|\rho(\fm)|}2}^\infty (-1)^\sub{\rho(\fm)} \Big( k+ \frac q2 - ir\rho(\sigma)\Big)^{2k} \Big( k+1 - \frac q2 + ir\rho(\sigma) \Big)^{2k+2} \;,
\ee
where we defined the function $\sub{x} = \frac{|x|+x}2$,%
\footnote{This function should \emph{not} be confused with the integer part, sometimes denoted by the same symbol.} which equals $x$ for $x\geq 0$ and vanishes for $x \leq 0$. The one-loop determinant for the chiral multiplet is then given by the ratio
\bea
\label{matter determinant not regularized}
Z_\text{matter} = \frac{\Det \cO_\psi}{\Det \cO_\phi} &= \prod_{\rho \in R_\Phi} (-1)^\sub{\rho(\fm)} \prod_{k=0}^\infty \frac{k + \frac{|\rho(\fm)|}2 + 1 - \frac q2 + ir\rho(\sigma)}{k + \frac{|\rho(\fm)|}2 + \frac q2 - ir\rho(\sigma)} \\
&= \prod_{\rho \in R_\Phi} \prod_{k=0}^\infty \frac{k - \frac{\rho(\fm)}2 + 1 - \frac q2 + ir\rho(\sigma)}{k - \frac{\rho(\fm)}2 + \frac q2 - ir\rho(\sigma)}
\eea
due to many cancelations. For $\rho(\fm) \leq 0$ the last equality is trivial; for $\rho(\fm) \geq 0$ the product in the second line on $0 \leq k \leq \rho(\fm)-1$ equals $(-1)^{\rho(\fm)}$, while the product on $k \geq \rho(\fm)$ equals the product in the first line.

The expression in (\ref{matter determinant not regularized}) is useful to exhibit zeros and poles of the matter one-loop determinant, but does not converge and requires regularization. We choose zeta-function regularization. Consider the Hurwitz zeta function $\zeta(z;q) = \sum_{n=0}^\infty (q+n)^{-z}$, which is absolutely convergent for $\re z >1$ and $\re q >0$. For each $q$, it extends by analytic continuation to a meromorphic function of $z \neq 1$, with a simple pole at $z=1$. In the region of absolute convergence we have $\parfrac{}{z} \zeta(z;q) = - \sum_{n=0}^\infty \frac{\log(q+n)}{(q+n)^z}$, therefore we can define the regularized sum:
\be
\text{`` } \sum_{n=0}^\infty \log(q+n) \text{ ''} \equiv - \parfrac{}{z} \zeta(z;q) \Big|_{z=0} \;.
\ee
We can take the exponential and make use of $\log \Gamma(q) = \parfrac{}{z} \zeta(z;q) \big|_{z=0} + \log\sqrt{2\pi}$
\cite{Lerch:1984} (a short proof is in \cite{Berndt:1985}) to write:
\be
\text{`` } \prod_{n=0}^\infty (q+n) \text{ ''} \equiv \frac{\sqrt{2\pi}}{\Gamma(q)} \;.
\ee
The regularized matter determinant is then:
\be
\label{Z matter}
Z_\text{matter} = \prod_{\rho \in R_\Phi} \frac{\Gamma \big( \frac q2 - ir\rho(\sigma) - \frac{\rho(\fm)}2 \big)}{\Gamma\big( 1 - \frac q2 + ir\rho(\sigma) - \frac{\rho(\fm)}2 \big)} \;.
\ee

\subsection{Matrix integral for $Z_{S^2}$}

The last step is to work out the on-shell value of the classical action on the BPS configurations.
Exact terms do not contribute because they vanish there.

For each Abelian factor we can include a FI term. For instance, for a single $U(1)$ factor:
\be
\cL_{FI} = i \Tr \Big( -\xi D + \frac\theta{2\pi} F_{12} \Big) \;,\qquad S_{FI}^\text{class} = 4\pi i \xi \Tr (r\sigma) + i\theta \Tr \fm \;.
\ee
Since $\Tr \fm \in \bZ$, we see that $\theta \in 2\pi \bZ$ is equivalent to $\theta = 0$.
More generally, we can include a twisted superpotential $\widetilde W(z)$ as in (\ref{twisted superpotential Lagrangian}) and (\ref{anti-twisted superpotential Lagrangian}). Its value on-shell is
\be
\label{on-shell twisted action}
S_{\widetilde W}^\text{class} = -8\pi i r \re \widetilde W \Big(\sigma + i \frac\fm{2r} \Big) \;.
\ee

Let us finally write down the $S^2$ partition function. We have a sum over fluxes $\fm$ and an integral over the gauge group Cartan weights $\sigma$. Since $r$ appears only in front of $\sigma$, we can rescale $r\sigma \to \sigma$ and discard $r$ (the dependence on $r$ is also independent of $\fm$).

If we sum over inequivalent fluxes $\fm$ (for instance for $U(N)$ that would mean over ordered integers), then the symmetry factor is $\frac1{|\cW_\fm|}$ in terms of the order $|\cW_\fm|$ of the Weyl group of the unbroken gauge group. Equivalently, we can sum over all fluxes $\fm$ (for $U(N)$ that means over all integer sequences) and divide by $|\cW|$, the order of the Weyl group of $G$.

If the theory has some non-R global symmetry $G_\text{global}$, let us consider its Cartan subalgebra. We can associate to the global symmetry a background vector multiplet, or equivalently its superfield strength, which is a twisted chiral multiplet.
For each Cartan generator we can turn on the lowest component of the background twisted chiral multiplet $M_a$, which is a complex \emph{twisted mass} $\tilde{m}_a$. Analogously to the lowest component $\frac1r(\sigma + i \frac{\fm}{2})$ of the dynamical superfield strength $\Sigma$, the twisted mass $\tilde{m}_a=\frac1r(\tau_a+ i\frac{n_a}{2})$ has a continuous real part $\tau_a$ and a quantized imaginary part, related to the integer flux $n_a$ for the global symmetry.%
\footnote{$\sigma$ and $\tau_a$ have already been rescaled by $r$ to make them dimensionless.}

Let $f^a[\Phi]$ and $R[\Phi]$ be the non-R-charges and the R-charge of a chiral multiplet $\Phi$. Then the localized partition function is:
\be
\label{matrix integral}
Z_{S^2} = \frac1{|\cW|} \sum_{\fm} \int \Big( \prod_j \frac{d\sigma_j}{2\pi} \Big) \, Z_\text{class}(\sigma, \fm) \, Z_\text{gauge}(\sigma,\fm) \, \prod_\Phi Z_\Phi(\sigma,\fm; \tau,\fn)
\ee
where the one-loop determinants are:
\bea
\label{1-loop determinants}
Z_\text{gauge} &= \prod_{\alpha \in G} \Big( \frac{|\alpha(\fm)|}2 + i\alpha(\sigma) \Big) = \prod_{\alpha > 0} \Big( \frac{\alpha(\fm)^2}4 + \alpha(\sigma)^2 \Big) \\
Z_\Phi &= \prod_{\rho \in R_\Phi} \frac{\Gamma \Big( \dfrac{R[\Phi]}2 - i\rho(\sigma) - i f^a[\Phi]\tau_a - \dfrac{\rho(\fm) + f^a[\Phi]n_a}2 \Big)}{\Gamma \Big( 1 - \dfrac{R[\Phi]}2 + i\rho(\sigma) + i f^a[\Phi]\tau_a - \dfrac{\rho(\fm) + f^a[\Phi]n_a}2 \Big)}
\eea
and the classical piece, for a FI term and a more general twisted superpotential, is:
\be
\label{Z_class}
Z_\text{class} = e^{-4\pi i \xi \Tr \sigma - i \theta \Tr \fm} \exp \big\{ 8\pi i r \re \widetilde W \big( \tfrac\sigma r + i \tfrac\fm{2r} \big) \big\} \;.
\ee
Note that mixing the $R$-symmetry with global Abelian non-R-symmetries amounts to giving an imaginary part to $\tau_a$, like in \cite{Jafferis:2010un}:
\be\label{R_mixing}
\tau_a\to \tau_a + \frac{i}{2}\, c_a \qquad \Longleftrightarrow \qquad R \to R + \sum_a c_a F^a\;.
\ee
The factor of $1/2$ in the mixing, compared to $1$ in the $S^3$ case \cite{Jafferis:2010un}, has to do with the different relation between the superconformal $R$ and the dilation operator. Analyticity in the combination $\tau_a + ic_a/2$ was already observed in the matter Lagrangian $\cL_\text{mat}$ (\ref{matter Lagrangian}).

The integrals in (\ref{matrix integral}) are along $\bR$ when all R-charges satisfy $R[\Phi]>0$, in which case the one-loop determinants $Z_\Phi(\sigma)$ have no poles along the real axis. When some R-charges are zero or negative the integral is defined by analytic continuation, that is the contour has to be deformed to keep the poles on the same side as they were in the former case.

\subsection{One-loop determinants via index theorem}
\label{sec: index theorem}

In section \ref{sec: localization Higgs branch} we will need the one-loop determinants around non-constant configurations, and in such case our previous computation is not valid. We then recompute the determinants -- this time taking as localizing action $\delta S = t \int(\cL_{YM} + \cL_\psi)$ the square of the BPS equations -- using an index theorem. Precisely, we use the Atiyah-Singer index theorem for transversally elliptic operators \cite{Atiyah:1974}. The method has been exploited in \cite{Pestun:2007rz, Gomis:2011pf} to perform localization on $S^4$, and lots of details are contained therein.

Let us repeat the discussion in \cite{Pestun:2007rz, Gomis:2011pf}. The supersymmetry transformations generated by $\cQ$ can be brought to a cohomological form by splitting the fields in four groups $\varphi_{e,o}$, $\hat\varphi_{e,o}$ -- where the subscript refers to even or odd statistics -- and writing
\be
\label{pairing of fields}
\cQ \, \varphi_{e,o} = \hat\varphi_{o,e} \;,\qquad \cQ \, \hat\varphi_{o,e} = \cR\, \varphi_{e,o}
\ee
so that $(\varphi_e, \hat\varphi_o)$ and $(\varphi_o, \hat\varphi_e)$ form multiplets. Here $\cR$ is the action of $U(1)_{M+ \frac R2} \times G$, where $U(1)_{M+\frac R2}$ is a rotation of $S^2$ around the poles combined with an R-symmetry rotation and $G$ is the gauge group.
Therefore $\cQ$ acts as a cohomological equivariant operator because
\be
\cQ^2 \, \varphi_{e,o} = \cR \, \varphi_{e,o}
\ee
to be compared with (\ref{square of Q}), and $\cQ^2$ is nilpotent on $\cR$-invariant field configurations. Since the $\cQ$-exact deformation term, that we can write as $\delta S = \cQ\, P$ for some $\cR$-invariant $P$, is invariant under $\cQ$ and fields are paired as in (\ref{pairing of fields}), many cancelations take place and eventually the one-loop determinant reads \cite{Pestun:2007rz}:
$$
\frac{\det\nolimits_{\coker D_{oe}} \cR_o}{\det\nolimits_{\ker D_{oe}} \cR_e} \;,
$$
where $D_{oe}$ is the differential operator from the space of $\varphi_e$ to the space of $\varphi_o$ obtained by expanding $P$ at quadratic order around a saddle point and picking the term $\varphi_o D_{oe} \varphi_e$. Hence the one-loop determinant equals the product of weights of the group action of $\cR = \cQ^2$. The weights can be extracted from the equivariant index:
\be
\ind D_{oe} = \tr_{\ker D_{oe}} e^\cR - \tr_{\coker D_{oe}} e^\cR \;.
\ee
The index computes the Chern character, while the one-loop determinant is the Euler character. We can read off the weights $w_\alpha(\varepsilon,\hat a)$ from $\ind D_{oe}$, where $\varepsilon, \hat a$ are the equivariant parameters of the $U(1)_{M+ \frac R2} \times G$ action, and apply the formula
\be
\label{character transformation formula}
\sum_\alpha c_\alpha e^{w_\alpha(\varepsilon,\hat a)} \quad\to\quad \prod_\alpha w_\alpha(\varepsilon, \hat a)^{c_\alpha}
\ee
to relate the Chern to the Euler character.

The equivariant index $\ind D_{oe}$ can be computed with the Atiyah-Singer index theorem, that we review. Consider: a manifold $M$, a pair of vector bundles $(E_0,E_1)$ over $M$, and the vector spaces $V_i = \Gamma(E_i)$ of their sections; the action of a Lie group $\cG$, whose maximal torus is $T = U(1)^n$; an elliptic differential operator
\be
D:\, V_0 \to V_1
\ee
commuting with the $\cG$-action. We define the index $\ind D(t)$ as a formal Laurent series in $t = (t_1,\dots,t_n) \in T$ as:
\be
\ind D(t) = \tr_{H^0} t - \tr_{H^1}t
\ee
where $H^0 = \ker D$, $H^1 = \coker D$ and the trace of $t$ is evaluated taking into account the action of $t$ on the subspaces. If $D$ is elliptic and $M$ is compact, then $H^{0,1}$ are finite dimensional vector spaces. In the more general case that $D$ is transversally elliptic, then $H^{0,1}$ can be infinite-dimensional but the eigensubspaces of the $T$-action are finite dimensional (in other terms, one can Laurent-expand in $t$ getting finite coefficients). The Atiyah-Singer index formula allows to compute the index from the fixed points of $\cG$ on $M$. Let $F$ be the set of fixed points of $\cG$ on $M$:
\be
\label{index formula}
\ind D(t) = \sum_{p \,\in\, F} \frac{\tr_{E_0(p)} t - \tr_{E_1(p)} t}{\det_{TM(p)}(1-t)} \;.
\ee

The differential operator $D_{oe}$ is obtained by picking the term $\varphi_o D_{oe} \varphi_e$ in the second order expansion of $P$, when the localizing action is $\delta S = \cQ \, P$. We now consider $\delta S = \int(\cL_{YM} + \cL_\psi)$ that is
\be
P \;\sim\; \sum_\text{fermions $\chi$} (\overline{\cQ \,  \chi}) \, \chi \;,
\ee
therefore $D_{oe}$ is obtained from $\cQ$ by considering its action on $\chi$ and picking the terms with one derivative%
\footnote{The index of $D_\text{oe}$ depends on its symbol.} acting on $\{\varphi_e\}$. Moreover our manifold $M$ is $S^2$, and the only two fixed points of $U(1)_M$ on $S^2$ are the north pole (NP) and south pole (SP). We should then look at the complex of $\cQ$ at those points.

\paragraph{Vector multiplet.} From the action of $\cQ$ on the vector multiplet in (\ref{Q-action gauge}), the splitting of fields is:
\be
\varphi_e = (A_\mu, \eta) \;,\qquad \varphi_o = (\lambda^\dag \epsilon - \epsilon^\dag \lambda) \;,\qquad \hat\varphi_o = (\lambda^\dag \gamma^a \epsilon + \epsilon^\dag \gamma^a \lambda) \;,\qquad \hat\varphi_e = (H)
\ee
where $H = D + \frac1r (\sigma + is\,\eta) - is\, F_{12} + iw^\mu D_\mu \eta$ and $a=1,2,3$. The scalar $\sigma - is\, \eta$, which is the parameter for gauge transformations in $\cR$, is missing and it appears once ghosts fields are considered as well, as explained in \cite{Pestun:2007rz}. Now $\cQ\lambda$ and $\cQ\lambda^\dag$ contain the one-derivative terms $dA$ and $d\eta$, therefore we consider the de Rham complex
\be
\label{de Rham complex}
D_{dR}:\; \Omega^0 \,\xrightarrow{d}\, \Omega^1 \,\xrightarrow{d}\, \Omega^2
\ee
tensored by the gauge bundle in the adjoint representation $\text{adj}(E)$. The equivariant index of the complex is defined as
$$
\ind D_{dR}(t) = \tr_{\ker d_0} t - \tr_{\coker d_0} t - \tr_{\ker d_1} t + \tr_{\coker d_1} t
$$
where $d_p$ acts on $\Omega^p$, and can be folded as
$$
d^* \oplus d:\; \Omega^1 \,\to\, \Omega^0 \oplus \Omega^2
$$
where $d^*$ is the adjoint of $d$, noticing that $\ind D_{dR} = - \ind (d^* \oplus d)$. The index formula (\ref{index formula}) can be applied to the complexification of (\ref{de Rham complex}). We introduce a complex variable $z$ around the north pole, acted upon by $U(1)$ as $z \to tz$ (and $\bar t = t^{-1}$). Now $\Omega^0$ is generated by $1$ and $\tr_{\Omega^0} t = 1$; $\Omega^1$ is generated by $dz,d\bar z$ and $\tr_{\Omega^1} t = t + t^{-1}$; $\Omega^2$ is generated by $dz\wedge d\bar z$ and $\tr_{\Omega^2} t = 1$; $TM$ is generated by $\partial_z, \partial_{\bar z}$ and $\det_{TM}(1-t) = (1-t^{-1})(1-t)$. We get:
\be
\ind_\bC D_{dR}(t) = \frac{2-t-t^{-1}}{(1-t)(1-t^{-1})} = 1 \;.
\ee
For the real complex the index is half of that. We also need to tensor by the adjoint gauge bundle, getting $\ind D_{dR}(t,g) = \frac12 \chi_\text{adj}(g)$, where $g\in G$ and $\chi_\text{adj}$ is the character of the adjoint representation. Writing $t=e^{i\varepsilon}$, $g=e^{i\hat a}$, the $U(1)_\varepsilon \times G_{\hat a}$ equivariant index of the vector multiplet operator is
\be
\label{index for vector}
\ind D^\text{vm} = \frac12 \sum_{\rho \,\in\, \text{adj(G)}} e^{i\rho(\hat a)}
\ee
as a sum over the weights.
Since vanishing weights give a contribution independent of everything, we can discard them and keep a sum over the roots $\alpha$ of $G$.

At the north pole we should expand (\ref{index for vector}) in powers of $t$, extract weights and multiplicities and apply (\ref{character transformation formula}); at the south pole we should instead expand in powers of $t^{-1}$. Their product is the Euler character
\be
Z_\text{gauge} = \prod_{\alpha \,\in\, G} \alpha(\hat a_{NP})^{1/2} \alpha(\hat a_{SP})^{1/2} \;,
\ee
where $\hat a_{NP,SP}$ are the values of the equivariant parameter at the two poles, which are extracted from the cohomology of $\cQ$ (\ref{square of Q}): $\cQ^2 = M + \frac R2 + i\Lambda$, with $i\Lambda = -i\sigma - \cos\theta\, \eta$. We get
\be
\label{equivariant parameters}
\varepsilon = \frac1r \;,\qquad\qquad \hat a = -i\sigma - \cos\theta \, \eta \;.
\ee
Therefore up to an irrelevant phase:
\be
\label{Z gauge index}
Z_\text{gauge} = \prod_{\alpha >0} \big( r\alpha(\eta) + ir\alpha(\sigma) \big)_{NP} \big( r\alpha(\eta) - ir\alpha(\sigma) \big)_{SP} \;.
\ee
In this computation we have neglected ghosts-for-ghosts needed to fix global gauge transformations on $S^2$: we have seen in section \ref{sec: one-loop gauge} that their effect is to introduce a Vandermonde determinant for the unbroken gauge subalgebra in the denominator, which cancels an equal Vandermonde determinant in the numerator when integrating over the zero-modes.

For the constant background (\ref{gauge background Coulomb}), the expression above agrees with (\ref{Z gauge}). However the current expression is more generally valid on non-constant backgrounds as well.

\paragraph{Chiral multiplet.} From the action of $\cQ$ on the chiral multiplet in (\ref{Q-action matter}), the splitting of fields is:
\be
\varphi_e = (\phi) \;,\qquad \varphi_o = (\epsilon^\trans C \psi) \;,\qquad \hat\varphi_o = (-\epsilon^\dag \psi) \;,\qquad \hat\varphi_e = (\tilde H)
\ee
where $\tilde H = F + i \epsilon^\trans C \gamma^\mu\epsilon \, D_\mu\phi + \epsilon^\trans C \gamma_3\epsilon\, \eta\phi$. Let us consider the north pole: $\cQ\psi$ contains $D_-\phi = \bar\partial\phi$ and the relevant complex is the Dolbeault complex
\be
\label{Dolbeault complex}
\bar\partial_q:\; \Omega^{(0,0)} \otimes K^{q/2} \,\to\, \Omega^{(0,1)}\otimes K^{q/2}
\ee
with inverted grading and further tensored by the gauge bundle in the suitable representation $R_\Phi$. Here $K$ is the canonical bundle which is present because $\phi$ has R-charge $q$ and this gives an extra contribution to $\cR$. Notice that for $q=1$, in which case fermions are not charged under $U(1)_R$, (\ref{Dolbeault complex}) is the same as the Dirac complex $D_\text{Dirac}: S_+ \to S_-$ which appears in 4d \cite{Pestun:2007rz}. Now $\Omega^{(0,0)}$ is generated by 1, $K$ by $dz$ and $\Omega^{(0,1)}$ by $d\bar z$, therefore
\be
\ind \bar\partial_q = \frac{t^{q/2}}{1-t} \;.
\ee
The full index for the chiral multiplet is then $\ind D^\text{cm} = - \frac{t^{q/2}}{1-t} \sum_{\rho}e^{i\rho(\hat a)}$. At the north pole we expand in powers of $t$ and extract the Euler character with (\ref{character transformation formula}):
$$
\sum_{k\geq 0} (-t^{\frac q2 + k}) \sum_\rho e^{i\rho(\hat a)} \quad\to\quad \prod_\rho \prod_{k\geq 0} \big[ (\tfrac q2 + k)\varepsilon + \rho(\hat a) \big]^{-1} \;.
$$
At the south pole instead we expand in powers of $t^{-1}$, so that similarly:
$$
\sum_{k\geq 0} (t^{-1})^{1-\frac q2 + k} \sum_\rho e^{i\rho(\hat a)} \quad\to\quad \prod_\rho \prod_{k\geq 0} \big[ (1 - \tfrac q2 + k)(-\varepsilon) + \rho(\hat a) \big] \;.
$$
Multiplying the two, substituting the equivariant parameters $\varepsilon$, $\hat a$ as in (\ref{equivariant parameters}) and regularizing the infinite products as in section \ref{sec: one-loop matter} we get, up to an irrelevant phase:
\be
\label{Z matter index}
Z_\text{matter} = \prod_{\rho\,\in\,R_\Phi} \frac{\Gamma\big( \frac q2 - ir\rho(\sigma) - r\rho(\eta) \big)_{NP}}{\Gamma\big( 1 - \frac q2 +ir\rho(\sigma) - r\rho(\eta) \big)_{SP}} \;.
\ee
For the constant background (\ref{gauge background Coulomb}) this expression agrees with (\ref{Z matter}).

\section{Localization on the Higgs branch}
\label{sec: localization Higgs branch}

It has been observed in \cite{Pasquetti:2011fj} that the path integral on the three-dimensional ellipsoid $S^3_b$ of $\cN=2$ Maxwell-Chern-Simons theory with  $N_f$ chiral multiplets of charge $+1$ and $N_a$ of charge $-1$, can be written as a sum of contributions each of which is the product of a classical action exponential, some one-loop determinants, a partition function for vortices and another for antivortices. This suggests that a similar factorization might take place in our two-dimensional setup. This is in fact the case, as we explicitly show in section \ref{sec: vortex partition functions} for a $U(N)$ gauge theory with $N_f$ fundamentals, $N_a$ antifundamentals and possibly one adjoint. It is natural to ask whether there is an a priori reason for such factorization, and our answer is positive.

In this section we show that when the gauge group admits FI terms that would lead to complete Higgsing in flat space,
it is possible to perform an alternative localization in which the BPS configurations dominating the path integral are a finite number of points on the Higgs branch, supporting point-like vortices at the north pole and antivortices at the south pole. We call this ``localization on the Higgs branch''. This has to be compared to the expression of $Z_{S^2}$ obtained in section \ref{sec: localization Coulomb branch}, which comes from BPS configurations on the Coulomb branch.

To simplify the discussion, in this section we will mainly consider a $U(N)$ gauge group with (anti)fundamentals; the construction can however be generalized to other gauge groups and matter representations (quivers, for instance). Moreover we consider the case that all chiral multiplets have R-charge $q=0$, which is possible unless there is a superpotential. We observed in section \ref{sec: SUSY actions} that the action is holomorphic in $M + i\frac q2$ (where $M$ is a twisted mass in units of $r^{-1}$), and so must be $Z_{S^2}$. Therefore given the answer $Z_{S^2}$ for $q=0$, we can obtain $Z_{S^2}$ for generic R-charges by analytic continuation in the twisted masses. Since a superpotential only affects $Z_{S^2}$ by imposing constraints on the R-charges, we can similarly obtain $Z_{S^2}$ for a theory with superpotential by neglecting the constraints first, computing $Z_{S^2}$ at $q=0$ and then performing the analytic continuation to reinstate the constraints.

At $q=0$ a continuous Higgs branch opens up: besides (\ref{localizing locus matter}), the BPS equations (\ref{BPS equations matter}) with real bosonic vector multiplet fields and deformed by generic twisted masses ($\sigma \to \sigma + M$) admit the solutions
\be
0 = F = D_\mu \phi = \eta \phi = (\sigma + M)\phi \;,
\ee
where it is implied that $\sigma$ and $M$ act on $\phi$ in the correct representation of $G \times G_F$.
If $q>0$ the BPS equations still have Higgs branch solutions but for complex -- not real -- values of $\sigma$, given by $\big( \sigma + \frac{iq}{2r} + M \big)\phi = 0$.
Consequently such solutions are not saddle points of $\cL_\psi$ with $\overline{\phantom{\lambda}}$ defined as after (\ref{localizing actions}), see footnote \ref{footnote: reality}: if we would like to have a Higgs branch at another value of $q$ we should modify $\overline{\phantom{\lambda}}$, that is the path integration contour, accordingly.

We add to the deformation action $\delta S = \int (\cL_{YM} + \cL_\psi)$ a new $\cQ$-exact and $\cQ$-closed term:
\be
\label{L_H}
\cL_H = \cQ \Tr \Big[ \frac{\epsilon_+^\dag \lambda - \lambda^\dag \epsilon_+}{2i} (\phi\phi^\dag - \chi\,\unit) \Big]\;,
\ee
where $\unit$ has rank $N$ as the gauge group, $\phi$ transforms in a possibly reducible representation $R_\Phi$ and includes all chiral multiplets, and $\chi$ is a free parameter. We made a little abuse of notation since $\lambda,\lambda^\dag$ act on $\phi$ in the representation $R_\Phi$ and on $\unit$ in the adjoint.
The parameter $\chi$ is meaningful only if the gauge group has an Abelian factor, otherwise it disappears under the trace.
The bosonic part of $\cL_{YM} + \cL_\psi + \cL_H$ is, up to total derivatives,
\begin{multline}
\label{Higgs bos deformation Lag 1}
\cL_\text{bos} = \Tr \bigg\{ \frac12 \Big(F_{12} - \frac\eta r \Big)^2 + \frac12 \Big( D+\frac\sigma r \Big)^2 + \frac12 (D_\mu\sigma)^2 + \frac12 (D_\mu\eta)^2 - \frac12 [\sigma,\eta]^2 \\
+ \Big[ i\Big(D + \frac\sigma r \Big) + s\, \Big( F_{12} - \frac\eta r \Big) - w^\mu D_\mu \eta \Big] \big( \phi\phi^\dag - \chi\,\unit \big) \bigg\} + \cL_\psi \big|_\text{bos} \;.
\end{multline}
Notice that this action is not positive definite, however we can make it so with a trick.
The path integral over $D$ is Gaussian and can be performed exactly: it generates a term $\frac12 \Tr (\phi\phi^\dag - \chi\unit)^2$ and formally imposes the constraint
\be
D + \frac\sigma r + i(\phi\phi^\dag - \chi\unit) = 0 \;.
\ee
Now at the saddle point $D+\frac\sigma r$ is imaginary, therefore effectively we have changed the path integration contour of the auxiliary field $D$.
The Lagrangian can then be brought to the form
\begin{multline}
\label{Higgs bos deformation Lag}
\cL_\text{bos} = \Tr \bigg\{ \frac12 \Big[ \sin\theta\, \Big( F_{12} - \frac\eta r \Big) + \cos\theta\, D_1\eta \Big]^2 + \frac12 (D_2\eta)^2 + \frac12 (D_\mu\sigma)^2 - \frac12 [\sigma,\eta]^2 \\
+ \frac12 \Big[ \phi\phi^\dag - \chi\unit + \cos\theta\,  \Big( F_{12} - \frac\eta r \Big) - \sin\theta\, D_1\eta \Big]^2 \bigg\} + \cL_\psi \big|_\text{bos} \;,
\end{multline}
which is a sum of squares.

The set of smooth configurations such that $\cL_\text{bos} = 0$, which coincide with its saddle points, consists of two branches. First we have a Higgs branch:
\be
\label{D-term equations}
0 = F_{12} - \frac\eta r = D_\mu \eta = D_\mu\sigma = [\sigma,\eta] = F = D_\mu \phi = \eta \phi = (\sigma + M)\phi = \phi\phi^\dag - \chi\, \unit \;.
\ee
The last one is the usual D-term equation, which spelled out reads: $\phi^\dag T^A_\Phi \phi = \chi\,\delta^{A1}$ for $A=1,\dots,\dim G$, where $T^A_\Phi$ are the gauge generators in representation $R_\Phi$ and $T^1_\Phi$ is the generator of $U(1)$. The set of solutions strongly depends on the gauge group and matter content.
For $U(N)$ with $N_f$ fundamentals $\phi$ and $N_a$ antifundamentals $\tilde \phi$ and for generic twisted masses $M$, up to gauge and flavor rotations the solution is: $\phi_{a\beta} = \sqrt\chi\, \delta_{a\beta}$ ($a$ is a gauge index, $\beta$ a flavor index) for $\chi > 0$ and $N_f \geq N$; $\tilde\phi_{a\gamma} = \sqrt{-\chi}\, \delta_{a\gamma}$ for $\chi<0$ and $N_a \geq N$; empty otherwise. This forces%
\footnote{For simplicity in the text we assume that $0 = \eta^\text{ext} = F_{12}^\text{ext}$. If that is not the case, (\ref{D-term equations}) is modified to $(\eta + \eta^\text{ext})\phi = 0$ and on solutions $F_{12}^\text{ext} = \eta^\text{ext}/r$ will be turned on as well. The following discussion will be modified accordingly.}
$0 = \eta = F_{12}$, and each eigenvalue of $\sigma$ to equal a different eigenvalue of $-M$ (of fundamentals or antifundamentals, depending on $\sign\chi$):
\be
\label{Higgs branch roots}
\sigma_a = -M_{l_a}
\ee
where vacua are parametrized by $N$-combinations $\vec l = (l_1,\dots,l_N) \in C(N,N_f)$ of the first $N_f$ (or $N_a$) integers. Summarizing, for generic twisted masses the set of Higgs branch solutions consists of some number of isolated vacua, in general different for $\chi \gtrless 0$. We will further study these solutions below.

Second we have a Coulomb branch:
\be
0 = D_\mu\sigma = [\sigma,\fm] = \phi \;,\qquad \eta = r\chi\cos\theta + \fm/2r \;,\qquad F_{12} = 2\chi\cos\theta + \fm/2r^2
\ee
for all $\sigma$ and GNO quantized $\fm$. The one-loop determinants have to be computed with the index theorem: the contribution of the gauge multiplet and of the chiral multiplet $\phi$ according to (\ref{Z gauge index})-(\ref{Z matter index}) is%
\footnote{The expression is only valid for $q=0$, which is the situation we are analyzing here.}
$$
\prod\nolimits_{\alpha > 0} \Big( \frac{\alpha(\fm)^2}4 + \alpha(\sigma)^2 \Big) \times \prod\nolimits_{\rho \in R_\Phi} \frac{\Gamma\big(a_\rho - \rho(\chi) \big)}{\Gamma\big( 1 + a_\rho^* + \rho(\chi) \big)}
$$
where $a_\rho = -i\rho(\sigma) - \rho(\fm)/2$, we set $r=1$ and $\alpha(\chi)=0$. For $\chi \to \pm \infty$ we can use the limit:
\be
\frac{\Gamma(a-z)}{\Gamma(1+a^* + z)} \;\xrightarrow{z \to \pm\infty}\; e^{-(2z+1)(\log|z|-1) + \cO(1) + \cO(\im a)} \big(1+ \cO(z^{-1}) \big) \;.
\ee
Let us set $\rho_i(\chi) = q_i \chi$, so that $q_i = \pm1$ are the charges under the $U(1)$ factor. Then the leading contribution is $\exp\big(-2(\sum_i q_i) \chi \log |\chi| \big)$; if $\sum q_i = 0$, the leading contribution is $\exp\big( - (\sum_i 1) \log|\chi| \big)$. In any case it is possible to send $\chi$ to infinity with either one of the two signs so that the Coulomb branch contribution is suppressed and the path integral localizes on the Higgs branch.

We already noticed that the Higgs branches at $\chi \gtrless 0$ are different: suppression of the Coulomb branch only works on (and thus selects) one of the two, depending on the matter content. We will see that this has a nice counterpart in the computation via localization on the Coulomb branch.
In particular we choose $\sign(\chi) = \sign(\sum q_i)$ when the latter is defined. For instance if $q_i$ are all positive, the selected sign of $\chi$ is the one compatible with the existence of Higgs branch solutions.

On top of each of the smooth Higgs branch solutions there are singular point-like vortex solutions at the north pole $\theta = 0$ and antivortex solutions at the south pole $\theta = \pi$:
\bea
\theta &= 0 : \qquad && \{ D + \frac\sigma r = -i(\phi\phi^\dag - \chi\,\unit) = + i F_{12} \,,\, D_-\phi = 0 \} \\
\theta &= \pi : \qquad && \{ D + \frac\sigma r = -i(\phi\phi^\dag - \chi\,\unit) = -i F_{12} \,,\, D_+\phi = 0 \} \;.
\eea
In the vortex and antivortex solutions $\phi$ has a winding and it vanishes at the core; for $\chi >0$ the vortex number $k = \frac1{2\pi} \int \Tr F_{12} e^{12}$ is positive for vortices and negative for antivortices. When there is no continuous Higgs branch but only discrete points -- which is the case for generic twisted masses -- the size of vortices is not a modulus (as instead happens in 4d with instantons) but rather their size is of order $L \sim r/\sqrt{|\chi|}$, assuming that in (\ref{Higgs bos deformation Lag 1}) we adjust the dimensions with powers of $r$. Only in the limit $\chi \to \pm\infty$ we really have point-like vortices.

In fact we can take a deeper perspective on the localizing action $\cL_{YM} + \cL_\psi + \cL_H$: as we change the parameter $\chi$ from zero to infinity we interpolate between localization on the Coulomb branch and localization on the Higgs branch. It is easy to see that for $\chi = 0$ the bosonic action (\ref{Higgs bos deformation Lag 1}) localizes on the Coulomb branch and it is equivalent to $\cL_{YM} + \cL_\psi$ as discussed in section \ref{sec: localization Coulomb branch}: on the one hand the Coulomb branch contribution is no longer suppressed and it precisely coincides with that of section \ref{sec: localization Coulomb branch}; on the other hand the VEV $\phi^\dag\phi$ of vortices vanishes and their size $L \sim r/\sqrt{|\chi|}$ blows up as $\chi \to 0$: vortices simply become the Coulomb branch configurations with constant flux.%
\footnote{One might be worried that some vortices survive in the $\chi \to 0$ limit: that is the case when their size is a zero-mode. However this does not happen for generic twisted masses. See for instance \cite{Tong:2005un}.}

\subsection{The partition function}

Let us now compute the path integral $Z_{S^2}$ by localization on the Higgs branch. We send $\chi \to \pm\infty$ according to the net charge under the $U(1)$ factor, in such a way that anything else than the Higgs branch contribution is suppressed.
We also assume that the gauge group is completely Higgsed.
In this case the classical smooth configurations are such that as many matter components as the rank $N$ of the gauge group get a VEV of order $\chi$, while all other VEVs are zero. To achieve that, we reach first a point (\ref{Higgs branch roots}) on the Coulomb branch where the flavors would be massless before turning on $\chi$: on that vacuum the classical action contribution to $Z_{S^2}$ is
$$
e^{-S_\text{class}} = e^{-4\pi i \xi \sum_{j=1}^N \sigma_j} \;,
$$
in case we do not have a more general twisted superpotential $\widetilde W$.
After turning on $\chi$ the $N$ components get a mass squared of order $\chi$.

The path integral is thus given by a sum over the solutions to the D-term equations (\ref{D-term equations}). In each configuration, the one-loop determinants for the massive W-bosons and the matter fields with vanishing VEV are given by (\ref{Z gauge})-(\ref{Z matter}) (or equivalently (\ref{Z gauge index})-(\ref{Z matter index})) evaluated at $\fm =0$ (or $\eta =0$) and at the values of $\sigma_j$ (\ref{Higgs branch roots}) in that vacuum. The latter combined with $M$ give the effective twisted masses of the matter fields, that we schematically call $m_\text{eff}$. The action $\int \cL_H$ in (\ref{L_H}) linearized around the Higgsed vacuum solutions gives mass of order $\chi$ and mixes the broken Cartan gauge field and the chiral field acquiring VEV. The easiest way to compute their one-loop determinant in the $\chi \to \infty$ limit is to use the Coulomb branch computation for $U(1)$ with one flavor of charge 1 in section \ref{sec:_1chiral}: postponing the explanation to the end of this section, the result is
\be
\label{Z Higgsed}
Z_\text{Higgsed} = e^{- \frac i2\int F_{12}} \;.
\ee
This can be interpreted as an effective shift of the theta-angle by $\pi$. When the chiral getting VEV has higher charge, the formula has to be modified accordingly.

On top of each smooth classical configuration, we have point-like vortices at the north pole and antivortices at the south pole. They differ by the sign of the magnetic flux, which is positive for vortices and negative for antivortices. The complex of $\cQ$ close to the NP $\{\theta=0\}$ is at first order in $\theta$ the same as the complex of the operator $\cQ'$ used in \cite{Shadchin:2006yz} to construct the $\Omega$-background \cite{Nekrasov:2002qd, Nekrasov:2003rj} in $\bR^2$. One has to identify $U(1)_{M + \frac R2}$ here with $U(1)_\varepsilon$ there in terms of the equivariant parameter $\varepsilon = \frac 1r$, as well as $i\Lambda = -i\sigma - \eta$ here with the equivariant parameter $a$ there. Therefore the sum over the singular vortex configurations at the NP is computed by the vortex partition function $Z_\text{vortex}$ in the $\Omega$-background. The one-loop determinants for W-bosons and vanishing matter fields do not depend on the flux at the pole, while $Z_\text{Higgsed}$ above does. The classical weight of each vortex configuration is:
$$
e^{-S} = \exp\Big[ -\int (-i\xi D + i \tfrac\theta{2\pi} F_{12}) \Big] = e^{-2\pi\xi k - i\theta k} = z^k \;,\qquad \text{with}\qquad z = e^{-2\pi \xi - i\theta}
$$
where we used $F_{12} = k/2r^2$, $D = i F_{12}$ at the NP. Taking into account (\ref{Z Higgsed}) as well:
\be
Z_{NP} = Z_\text{vortex} \big( (-1)^N z \,,\; \varepsilon = \tfrac1r \,,\; a = -im_\text{eff} \big) \;.
\ee
At the SP the equivariant rotational parameter is $\varepsilon = -\frac1r$. The flux $F_{12}$ is negative: in fact antivortices are obtained from vortices by charge conjugation%
\footnote{Given a vortex solution characterized by $\int F = 2\pi k > 0$ and $D_-\phi = 0$, we can transform the gauge multiplet as $V \to -V$ and the chiral multiplet as $\Phi \to \Phi^\dag$. Hence $D_\mp \phi \to (D_\pm\phi)^\dag$ and the vortex solution is mapped to an antivortex solution.
}
and this changes sign to the twisted masses. With $F_{12} = -k/2r^2$ and $D = -i F_{12}$ at the SP, we have weight $e^{-S} = \bar z^k$ and therefore:
\be
Z_{SP} = Z_\text{vortex} \big( (-1)^N\bar z \,,\; \varepsilon = -\tfrac1r \,,\; a = im_\text{eff} \big) \;.
\ee
The final expression for the path integral localized on the Higgs branch is:%
\footnote{The formula has been written having $U(N)$ with fundamentals, antifundamentals and adjoints in mind.}
\be
\label{Z_S2 Higgs final}
Z_{S^2} = \sum_{\substack{\text{D-term} \\ \text{solutions}}} \; e^{-4\pi i \xi \sum\limits_{j=1}^N \sigma_j} \; Z_\text{1-loop}' \; Z_\text{vortex} \big( (-1)^N z \,,\, \tfrac 1r \,,\, -im_\text{eff} \big) \; Z_\text{vortex} \big( (-1)^N \bar z \,,\, -\tfrac1r \,,\, im_\text{eff} \big) \;,
\ee
where $Z_\text{1-loop}'$ reminds us that the $N$ non-vanishing chiral multiplets should not be included.

Finally let us justify (\ref{Z Higgsed}). In section \ref{sec:_1chiral} we compute $Z_{S^2}$ for $U(1)$ with one chiral of charge $Q=1$ by localization on the Coulomb branch, obtaining $Z_{S^2} = e^{-z + \bar z}$. On the other hand the vortex partition function is known to be $Z_\text{vortex}(z,\varepsilon) = e^{z/\varepsilon}$ \cite{Shadchin:2006yz, Dimofte:2010tz, Bonelli:2011fq, Fujimori:2012ab}, and the classical contribution is trivial. By comparing the two expressions we deduce (\ref{Z Higgsed}).

The expression (\ref{Z_S2 Higgs final}) is in some sense similar to the four-dimensional result of Pestun \cite{Pestun:2007rz}. However while in that case one has an integral over the Coulomb branch of a partition function of non-perturbative configurations (there instantons), on $S^2$ we find either an integral over the Coulomb branch of perturbative contributions, or a discrete sum of products of perturbative and non-perturbative contributions. We will show in section \ref{sec: vortex partition functions} by direct evaluation that indeed for $U(N)$ with flavors and possibly one adjoint, the two expressions (\ref{matrix integral}) and (\ref{Z_S2 Higgs final}) coincide.

To conclude we notice that (\ref{Z_S2 Higgs final}) could be generalized to the more intricate case that a non-trivial twisted superpotential $\widetilde W$ is present, but we leave that for future work.

\section{Examples}
\label{sec: examples}

In this section we present some properties of our $S^2$ partition function and some examples.

\subsection{Complex masses and decoupling}

Consider a field $\Phi$ in the adjoint representation of the gauge group, with a superpotential mass term $W_\text{mass} = \frac{m}{2} \Tr (\Phi^2)$. $\Phi$ is neutral under Abelian non-R global symmetries and carries R-charge 1, therefore its one-loop determinant is
\be\label{adjoint_cplx_mass}
Z_\Phi(\sigma,\fm) = \prod_{\alpha\in\, G} \frac{ \Gamma( \frac12 -i\alpha(\sigma)- \frac12 \alpha(\fm))}{ \Gamma( \frac12 + i\alpha(\sigma) - \frac12 \alpha(\fm) )} = 1 \;.
\ee
To prove it one splits the products into positive and negative roots, exploits the identity $\Gamma(z+n) = (z)_n \Gamma(z)$  for $n\in \bN$ in terms of the Pochhammer symbol%
\footnote{The Pochhammer symbol is defined as: $(z)_n = \prod_{k=0}^{n-1} (z+k)$.}
$(z)_n$, and is left with $(-1)^{\sum_{\alpha>0} \alpha(\fm)} = (-1)^{2\delta(\fm)} = 1$ where $\delta$ is the Weyl vector.

More generally, a mass term $W_\text{mass} = \frac{m}{2} \Tr (\Phi^2)$ for a chiral field $\Phi$ in gauge representation $R_\Phi$ can be written when $\text{Sym}_2 \, R_\Phi$ contains a singlet, that is there is a symmetric product and therefore $R_\Phi$ is real. The one-loop determinant reduces to
\be
Z_\Phi(\sigma,\fm) = (-1)^{\sum_{\rho>0} \rho(\fm)} \;.
\ee
One can similarly consider two chiral superfields $\Phi_1$, $\Phi_2$ in conjugate representations with superpotential mass term $W_\text{mass} = m \Phi_1 \Phi_2$ and gauge indices contracted. It follows that $R[\Phi_2]=2-R[\Phi_1]$,  $f^a[\Phi_2]=-f^a[\Phi_1]$ and the weights of the gauge representations are opposite. Using the same identity for $\Gamma$ as before, one is left with
\be
Z_{\Phi_1}(\sigma,\fm) Z_{\Phi_2}(\sigma,\fm) = (-1)^{\sum_\rho \rho(\fm)} \;.
\ee
The expression on the right hand side is invariant under the Weyl group, in fact $\sum_\rho \rho = 0$ for semisimple groups while we can have a shift of theta angles by multiples of $\pi$ for Abelian factors. For instance if $\Phi_{1,2}$ are charged under a $U(1)$ gauge group with charges $\pm Q$, we get $Z_{\Phi_1} Z_{\Phi_2} = (-1)^{Q\fm} = e^{\frac i2 Q \int F_{12}}$.

\subsection{Twisted masses and effective twisted superpotential}

Next we show that when integrating out a chiral multiplet with large twisted mass, the expected effective twisted superpotential \cite{D'Adda:1982eh,Hori:2000kt} is generated.

The gamma function $\Gamma(z)$ is a meromorphic function with simple poles at $z \in -\bN$, but without any zero or branch cut. At large $z$ we can use Stirling's formula to approximate $\Gamma(z)$ by an analytic function with a branch cut along $z \in \bR_-$ (which is an approximation to the line of simple poles):
\be
\Gamma(z) = z^{z - \frac12} e^{-z} \sqrt{2\pi}\, \big( 1 + \cO \big( \tfrac 1z \big) \big) \qquad\qquad\text{for}\qquad |\arg z| < \pi \;.
\ee
If we set $w \equiv \rho(\sigma) + f^a[\Phi]\tau_a$, the limit of large twisted mass is achieved in the one-loop determinant (\ref{1-loop determinants}) by tuning $w\to \pm \infty$. We find:%
\footnote{A sum over weights $\sum_{\rho \in R_\Phi}$ is implied in the exponent, and in later formulae. We drop it for notational convenience.}
\be
\label{effective action}
\lim_{w \to \pm\infty} Z_\Phi = \exp\Big\{ -2iw \big( \log|w|-1 \big) + (R[\Phi]-1) \log|w| - \frac{i\pi}2 (\rho(\fm) + f^a n_a) \sign w + \cO\big( \tfrac1w \big) \Big\} \;.
\ee
We would like to interpret the right-hand-side as an on-shell effective action.

For the special case $R[\Phi]=1$, one can verify that the effective action is reproduced by the $f^a[\Phi] \tau_a \to \pm\infty$ limit of the following effective twisted superpotential:
\be
\label{effective twisted superpotential}
\widetilde W_\text{eff}(\Sigma) = - \frac1{4\pi} \Sigma_s \big[ \log(-ir\Sigma_s) - 1 \big] \qquad\qquad\text{with}\qquad \Sigma_s \equiv \rho(\Sigma) + f^a[\Phi] M_a
\ee
where $\Sigma$ is the twisted chiral superfield that describes the $U(1)$ gauge multiplet coupled to $\Phi$, while $M_a$ are external twisted superfields whose VEVs encode the twisted masses and $f^a[\Phi]$ are, as before, the Abelian charges. On-shell we set, as in (\ref{on-shell twisted action}), $\Sigma = (\sigma + i\fm/2)/r$ and $M_a = (\tau_a + i n_a/2)/r$.
In checking the limit one makes use of
$$
z(\log z -1) \;\xrightarrow{y \to \pm\infty}\; \Big[ x \log|y| - \frac\pi2 |y| \Big] + i \Big[ y (\log|y|-1) + \frac\pi2 x \sign y \Big]
$$
with $z = x+iy$. (\ref{effective twisted superpotential}) is the expected effective twisted superpotential generated when a chiral multiplet with effective twisted mass $\Sigma_s$ is integrated out, see for instance \cite{D'Adda:1982eh, Hori:2000kt}.

In the previous particular case, the result is the generation of a purely imaginary Euclidean effective action. This is what one would expect for an action arising from a twisted superpotential in $\bR^2$.
However placing the supersymmetric theory on $S^2$, similarly to $S^3$ \cite{Jafferis:2010un,Festuccia:2011ws}, introduces a complexification due to the R-symmetry.
We encountered a manifestation of such a complexification in \eqref{R_mixing}. The extra term proportional to $R[\Phi]-1$ in \eqref{effective action} is the result of the complexification of the effective real mass (real part of the effective twisted mass)
\be\label{complexification}
w \;\rightarrow\; w_\bC=w + \frac{i}{2}(R(\Phi)-1)\;,
\ee
so that for large $w$
\be
w(\ln|w|-1) \;\rightarrow\; w(\ln|w|-1) + \frac{i}{2}(R(\Phi)-1)\ln|w|\;.
\ee
Note that something analogous occurs for the localized partition function of an $\cN=2$ supersymmetric theory on $S^3$ \cite{Benini:2011mf}: since a CS term at level $k$ for a $U(1)$ with mass parameter $x$ appears as $\exp\left[-i \pi k x^2\right]$, the expected contribution of a chiral multiplet with effective real mass $x$ (neglecting the R-symmetry) at large $x$ is $\exp\left[-\frac i2 \sign(x) x^2\right]$, a level $\sign(x)/2$ CS term for a $U(1)$ with effective mass parameter $x$ induced by the massive fermions.
However $x$ is complexified because of the R-symmetry, and the actual contribution to the partition function is $\exp \big[ \frac i2 \big( x+i(R-1) \big)^2 \big]$, accounting for extra CS terms including the R-symmetry,  whose physical origin has been recently explained in \cite{Closset:2012vg} for theories on $S^3$. A similar phenomenon is at work on $S^2$ for the twisted superpotential.

\subsection[$U(1)$ gauge theory with a charged chiral multiplet and mirror symmetry]{$U(1)$ gauge theory with a charged chiral multiplet \\ and mirror symmetry}
\label{sec:_1chiral}

Consider the $S^2$ partition function of a $U(1)$ gauge theory with a chiral multiplet of gauge charge $Q$ and R-charge $q$:
\be
\label{Z U(1) Q}
Z(\xi,\theta) = \sum_m e^{-i m \theta} \int \frac{d\sigma}{2\pi}\, e^{-4\pi i \xi\sigma} \frac{\Gamma(\frac q2 - i Q \sigma - Qm/2)}{\Gamma(1-\frac q2 + iQ\sigma - Qm/2)} \;.
\ee
Notice that $Z_Q(\xi,\theta) = Z_{-Q}(-\xi,-\theta)$. We compute the integral by closing the contour at infinity: inspection of the asymptotic behavior of the integrand reveals that the contour can be closed in the lower half-plane for $Q>0$, in the upper half for $Q<0$. The numerator has poles at $\sigma = -\frac i{2Q}(2k + q - Qm)$ for $k\geq 0$, however they can cancel against the poles of the denominator $\sigma = \frac i{2Q}(2h + 2 - q -Qm)$ for $h \geq 0$. Therefore the poles of the one-loop determinant are at:
\be
\sigma_k = - \frac i{2Q} \, \big( 2k + |Qm| + q \big) \qquad\qquad\text{for}\qquad k \geq 0 \;.
\ee
Using $\Res_{z_0} f(az+b) = \frac1a \Res_{az_0+b} f(z)$, the residue of the numerator is:
\be
\Res_{\sigma_k} \Gamma \big( \tfrac q2 - iQ\sigma - Qm/2 \big) = \frac{i(-1)^{\tilde k}}{Q \, \tilde k!} \;,\qquad\text{ with }\qquad \tilde k = k + \sub{Qm} \;.
\ee
In the denominator we find $\Gamma\big( 1 - \frac q2 + iQ\sigma_k - Qm/2 \big) = (\tilde k - Qm)!$. We can define
\be
t \equiv \frac{2\pi \xi}Q \;.
\ee
Recall that the integration contour in (\ref{Z U(1) Q}) is along $\bR$ for $q>0$, in which case all poles are included; for $q \leq 0$ the contour has to be deformed so that all poles are still included. Taking into account the direction of the contour, we obtain:
\be
Z = \frac{e^{-tq}}{|Q|} \sum_{m \in \bZ} e^{-im\theta} (-1)^\sub{Qm} \sum_{k=0}^\infty \frac{(-1)^k (e^{-t})^{2k + |Qm|}}{k! (k + |Qm|)!} \;.
\ee
To proceed further (see section \ref{subsec:_U(1)} for an alternative manipulation) we notice that the summation over $k$ produces the Bessel function of the first kind $J_\alpha$:
\be
J_\alpha(x) = \sum_{k=0}^\infty \frac{(-1)^k}{k! \, \Gamma(k+\alpha+1)} \Big( \frac x2 \Big)^{2k+\alpha} \;\stackrel{\text{for }\alpha \in \bZ}{=}\; \frac1{2\pi} \int_{-\pi}^\pi e^{-i\alpha \tau} \, e^{ix\sin\tau} \, d\tau \;,
\ee
which for integer $n$ satisfies $J_{-n}(x) = (-1)^n J_n(x)$. We then obtain:
\be
Z = \frac{e^{-tq}}{|Q|} \sum_{m\in\bZ} e^{im\theta} J_{Qm}(2e^{-t}) = \frac{e^{-qt}}{|Q|} \int_{-\pi}^\pi d\tau \, \sum_{n\in \bZ} \, \delta \big( \theta - Q\tau - 2\pi n \big) \, e^{2i e^{-t} \sin\tau} \;,
\ee
where in the last equality we used Poisson resummation: $\sum_{m\in \bZ} e^{im\alpha} = 2\pi \sum_{n\in \bZ} \delta(\alpha - 2\pi n)$. Evaluation of the integral gives the final expression:
\be
\label{Z U(1) Q final}
Z(\xi,\theta) = \frac{e^{-2\pi\xi q/Q}}{Q^2} \; \sum_{n=0}^{|Q|-1} \; \exp \Big[ 2i \, e^{-2\pi \xi/Q} \, \sin \Big( \frac{\theta -2\pi n}Q \Big) \Big] \;.
\ee
The R-charge $q$ is not physical, as no gauge-invariant operator depend on it: it describes the mixing of the gauge symmetry with the R-symmetry, and it should be fixed to some canonical value, for instance $q=0$.%
\footnote{One might still be bothered by the ambiguity in the dependence of $Z$ on $\xi$. This traces back to the fact that neither $\cL_{YM}$ (\ref{YM Lagrangian}) nor $\cL_{\widetilde W}$ (\ref{twisted superpotential Lagrangian}) are holomorphic in $\sigma + iq/2r$, as noted after (\ref{matter Lagrangian}).}
The $S^3$ partition function of \cite{Jafferis:2010un} has a similar behavior. We kept $q$ in the computation to make the integration contour clear.

We can compare our result with the predictions of mirror symmetry. The mirror description \cite{Hori:2000kt} of this gauge theory is in terms of a twisted chiral multiplet $\Sigma$, the field strength superfield of the original vector multiplet, and another twisted chiral multiplet $Y\sim Y+2\pi i$, coupled by the twisted superpotential
\be
\widetilde W = \frac{1}{4\pi}\left[\Sigma(QY - \tau(\mu)) + i\mu e^{-Y}\right]
\ee
where now $\tau = 2\pi\xi + i\theta$ and $\mu$ is a subtraction scale.
The twisted superpotential is defined modulo $i \Sigma/2$ due to the identification $\theta \sim \theta+2\pi$, therefore integrating out $\Sigma$ by imposing its twisted F-term equation $\frac{\partial\widetilde W}{\partial \Sigma}= in/2$ for $n \in \bZ$, determines $Y$ to be one of the $|Q|$ ``vacua'':
\be
\label{mirror_Q}
Y = \frac{\tau(\mu) - 2\pi i\, n}Q \;,\qquad\qquad \text{for}\qquad n=0,\dots,|Q|-1 \;.
\ee
The on-shell superpotential is thus:
\be
\label{W_n}
\widetilde W_n = \frac{i}{4\pi} \;\mu \; \exp\Big[ \frac{-\tau(\mu) + 2\pi i\, n}Q \Big] \;.
\ee
For the theory on $S^2$ the natural subtraction scale at which parameters are defined is $\mu=1/r$. Plugging this scale in \eqref{W_n} and evaluating the classical action \eqref{on-shell twisted action} we get
\be
\label{S_class_n}
S_\text{class, $n$} = -8\pi i r \re \widetilde{W}_n = -2i \, e^{ - 2\pi\xi/Q } \, \sin \Big( \frac{\theta - 2\pi n}Q \Big)\;.
\ee
To be precise the $|Q|$ ``vacua'' \eqref{mirror_Q} are not physically distinct: they are identified under the ambiguity of $\widetilde W$ which can be implemented by a discrete gauge symmetry. In fact in the original description, a VEV for the chiral multiplet of charge $Q$ breaks the gauge symmetry to $\bZ_{|Q|}$; in the mirror description this manifests itself in \eqref{mirror_Q}.
When computing the partition function we should average over the action of the residual $\bZ_{|Q|}$ gauge symmetry via $\frac{1}{|Q|} \sum_{n=0}^{|Q|-1} e^{-S_\text{class, $n$}}$, which reproduces the sum in \eqref{Z U(1) Q final}. Indeed the $\bZ_{|Q|}$ average is effectively what the $\delta \big( \theta-Q\tau-2\pi n \big)$ in the Poisson resummed expression imposes.

We have considered here a single chiral superfield, and we have found exact agreement with mirror symmetry. To achieve that we have simply evaluated the on-shell twisted superpotential in the mirror, assuming that this gives the exact effective action. In fact this is correct because $\Sigma$ and $Y$ have mass proportional to the gauge coupling $g$ \cite{Hori:2000kt} (while no twisted mass for $Y$ can be introduced) which does not affect $Z_{S^2}$ and can thus be taken to infinity. On the other hand when more than two twisted chiral fields $\Sigma, Y_i$ are present, the masses of the modes tangent to $\sum_i Q_i Y_i = \tau(\mu)$ are proportional to the twisted masses $\tilde m_i$: one should really perform localization in the presence of twisted chiral multiplets.

For instance, consider a $U(1)$ gauge theory with two chirals of charges $\pm 1$ and axial twisted mass $\tilde m$. The partition function is computed in section \ref{subsec:_U(1)} and we anticipate the result (\ref{Z_U(1)_final}):
\be
Z_{U(1)}^{(1,1)} = e^{4\pi i \xi \tilde m} \; \frac{\Gamma(-2i\tilde m)}{\Gamma(1+2i\tilde m)} \; e^{4i\tilde m \log|1+z|} \qquad\quad\text{with:}\qquad z = e^{-2\pi \xi - i\theta} \;.
\ee
The mirror description is in terms of $\Sigma, Y_1, Y_2$ and twisted superpotential \cite{Hori:2000kt}:
\be
\widetilde W = \frac1{4\pi} \Big[ \Sigma ( Y_1 - Y_2 - \tau) + i\mu \big( e^{-Y_1} + e^{-Y_2} \big) + \tilde m (Y_1 + Y_2) \Big] \;.
\ee
There is a single vacuum where the on-shell superpotential is:
\be
\widetilde W_\text{on-shell} = \frac{\tilde m}{2\pi} \Big[ \frac\tau2 - \log\Big( -i \frac{2\tilde m}{\mu (1+z)} \Big) + 1 \Big]
\ee
One can check that $e^{-S} = e^{8\pi i \re \widetilde W_\text{on-shell}}$ is not equal to $Z_{U(1)}^{(1,1)}$ but rather they precisely agree in the limit $\tilde m \to \pm \infty$ (setting $\mu = 1/r$, rescaling $r\tilde m \to \tilde m$ and taking $\tilde m \in \bR$).

\subsection{Pure $U(N)$ gauge theory}

Consider a pure $U(N)$ gauge theory with quadratic twisted superpotential
\be
\widetilde W(z) = \frac{hr}4 \, z^2 \;,
\ee
where we have defined $L = hr$ with respect to the Lagrangian $\cL_{CS}$ (\ref{CS Lagrangian}) so that $h$ is dimensionless. The Lagrangian is the $S^2$ completion of what one obtains from dimensional reduction of Chern-Simons theory, although $h$ is not quantized in two dimensions. According to (\ref{matrix integral}), the path integral is computed by the matrix integral:
\be
Z_{S^2}^{U(N)} = \frac1{N!} \sum_{\vec m \,\in\, \bZ^N} \int \bigg[ \prod_{j=1}^N \frac{d\sigma_j}{2\pi} \bigg] \;  e^{2\pi i h \sum_{k=1}^n \big( \sigma_k^2 - m_k^2/4 \big)} \; \prod_{i<j}^N \Big[ \frac{(m_i - m_j)^2}4 + (\sigma_i - \sigma_j)^2 \Big] \;.
\ee

To perform the computation, one can make the change of variables $\sigma_k = m_k/2 + y_k/2\pi h$, so the exponential becomes $\exp\big[ i\sum_k \big( y_k m_k + y_k^2/2\pi \big)\big]$ and the exponent is linear in $\vec m$. We then apply Poisson resummation $\sum_{m\in\bZ} m^j e^{iym} = 2\pi \sum_{n\in \bZ} \delta(y-2\pi n) \big( i\parfrac{}{y} \big)^j$ and integrate over $\prod dy_k$ obtaining, after a further redefinition $y_k = \sqrt{2\pi h}\, x_k$:
$$
Z = \frac{(2\pi h)^{-\frac{N(N+1)}2}}{N!} \sum_{\vec n \,\in\, \bZ^N} \bigg[ e^{i\sum_k x_k^2} \prod_{i<j}^N \Big( - \frac{(\partial_i - \partial_j)^2}2 + i(\partial_i - \partial_j)(x_i - x_j) + (x_i - x_j)^2 \Big) \bigg]_{x_k = \sqrt{\frac{2\pi}h}\, n_k} \;.
$$
The meaning of this expression is that we should expand the last product as a formal polynomial in $x_i$ and $\partial_i$, and in each monomial order the symbols in such a way that all derivatives come before all variables and before the exponential factor. One can show that all terms with derivatives cancel out, and we are left with:
\be
\label{partition function pure U(N)}
Z_{S^2}^{U(N)} = \frac{h^{-N^2}}{(2\pi)^N N!} \; \sum_{\vec n \,\in\, \bZ^N} \; e^{\frac{2\pi i}h \sum_k n_k^2} \; \prod_{i<j}^N (n_i - n_j)^2 \;.
\ee
We can consider $h\in \bC$, so that the expression is defined in the whole complex plane by analytic continuation.

The theory we are studying is equivalent to pure Yang-Mills. This is because $\frac1{g^2} \int \cL_{YM}$ is exact and can be made arbitrarily small; $\cL_{CS}$ gives mass to $D,\sigma,\lambda,\bar\lambda$ that can be integrated out, and we are left with a quadratic action in $F_{12},\eta$ which is equivalent to pure YM (the YM action is positive definite if we take $h$ negative imaginary).
We would like then to compare (\ref{partition function pure U(N)}) with the partition functions in \cite{Migdal:1975zg, Witten:1991we}. In fact we can rewrite (\ref{partition function pure U(N)}) as a sum over representations of $U(N)$. Let $\alpha = 1$ when $N$ is even, $\alpha = 0$ when $N$ is odd. As in \cite{Aganagic:2004js} we can shift $n_i \to n_i + \frac{N-\alpha+1}2 -i$,  restrict the sum to ordered integers $n_1 \geq \dots \geq n_N$ (because of the last factor in (\ref{partition function pure U(N)}) the permutation group acts freely) and remove $N!$ from the denominator. We thus have a sum over representations of $U(N)$:
\be
Z_{S^2}^{U(N)} = \frac{G(N+1)\, h^{-N^2}}{(2\pi)^N} \; e^{\frac{2\pi i}h \, \frac{N(N^2 +3\alpha -1)}{12}} \sum_{\cR \text{ of } U(N)} e^{\frac{2\pi i}h \, \big[ C_2(\cR) - \alpha C_1(\cR) \big]} \; (\dim \cR)^2 \;.
\ee
Here $G(z)$ is the Barnes G-function such that $G(n+2)$ for $n\in \bN$ is the superfactorial of $n$, while $C_{1,2}(\cR)$ and $\dim \cR$ are the linear and quadratic Casimir and the dimension of the representation $\cR$:
\be
C_1(\cR) = \sum_{i=1}^N n_i \;,\quad\; C_2(\cR) = \sum_{i=1}^N n_i \big( n_i + N -2i + 1 \big) \;,\quad\; \dim \cR = \prod_{i<j}^N \frac{n_i - n_j + j - i}{j-i} \;.
\ee
Unfortunately without extra operators inserted or the genus of the Riemann surface where the theory lives, the dependence of $Z_{S^2}^{U(N)}$ on $h$ does not carry much information because it can be renormalized by an arbitrary function of $h$ \cite{Witten:1991we}.

\section{Vortex partition functions}
\label{sec: vortex partition functions}

In this section we explicitly compute the path integral $Z_{S^2}(\xi,\theta,\tau_i,\tilde\tau_j, \tau_A)$ by localization on the Coulomb branch, for a $U(N)$ gauge theory with $N_f$ chirals in the fundamental representation, $N_a$ chirals in the antifundamental (we will succinctly refer to $(N_f,N_a)$ flavors), possibly one adjoint chiral, twisted masses%
\footnote{The combination $\frac{1}{N_f}\sum_{i=1}^{N_f} \tau_i - \frac{1}{N_a}\sum_{j=1}^{N_a} \tilde\tau_j$ is set to zero since the flavor symmetry is $S[U(N_f)\times U(N_a)]$.}
$\tau_i$, $\tilde \tau_j$, $\tau_A$ respectively, FI term $\xi$ and theta-angle $\theta$. We will see that it reproduces the result of localization on the Higgs branch, in terms of the vortex partition function.
In the following we set to zero the magnetic fluxes for global symmetries, in order to simplify the computations.

\subsection{$U(1)$ with $(N_f,N_a)$ flavors}
\label{subsec:_U(1)}

Consider first a $U(1)$ gauge theory with $N_f$ chirals of charge $+1$ and masses $\tau_j$, and $N_a$ of charge $-1$ and masses $\tilde\tau_f$. We take canonical R-charge $q=0$: generic R-charges are eventually obtained by giving an imaginary part to $\tau_j,\tilde\tau_f$.
The matrix integral (\ref{matrix integral}) is:
\be \label{matrix_integral_U(1)}
Z_{U(1)}^{(N_f,\,N_a)}(\xi,\theta;\tau,\tilde{\tau}) = \sum_{m\in\bZ}e^{-i\theta m} \int \frac{d\sigma}{2\pi} \, e^{-4\pi i \xi \sigma}  \prod_{j=1}^{N_f} \frac{\Gamma(-i\sigma -i\tau_j - \frac m2)}{\Gamma(1 + i\sigma + i\tau_j - \frac m2)}   \prod_{f=1}^{N_a} \frac{\Gamma(i\sigma - i \tilde\tau_f + \frac m2)}{\Gamma(1 -i\sigma + i\tilde\tau_f + \frac m2)}  \;.
\ee
Recall that the integration contour in $\sigma$ is defined along the real line when all R-charges are bigger than zero. The case when some R-charges are zero or negative is obtained by analytic continuation deforming the contour.

We evaluate the integral by closing the integration contour at infinity
according to the asymptotic behavior of the integrand and summing the residues. The computation is similar to the one in section \ref{sec:_1chiral}.
The gamma functions have only poles, no zeros, so we can focus on the numerators.
Assuming $N_f>N_a$, or $N_f=N_a$ and $\xi>0$, we can close the contour in the lower half-plane, therefore only the chirals with positive charge give poles enclosed in the contour. The result for $N_f<N_a$, or $N_f=N_a$ and $\xi<0$, can be obtained thanks to
\be\label{identity_zeta}
Z_{U(1)}^{(N_f,\,N_a)}(\xi,\theta;\tau,\tilde{\tau})=  Z_{U(1)}^{(N_a,\,N_f)}(-\xi,-\theta;\tilde{\tau},\tau)\;,
\ee
which follows from a sign change in the dummy variables of \eqref{matrix_integral_U(1)}, or charge conjugation.
We show in section \ref{sec: analyticity} that a non-trivial identity implies analyticity in $\xi$ for $N_f=N_a$.

There are $N_f$ towers of poles, one for each fundamental $l=1,\dots,N_f$. The $l$-th tower of poles is given by
\be
\sigma = - \tau_l - i k - \tfrac i2 |m| \;,\qquad\qquad k \in \bN \;.
\ee
The residue of the gamma function at a pole is $i(-1)^{k_1}/k_1!$ where we defined $k_1 = k + \sub{m}$, while the value of $\Gamma$ in the denominator is $(k_1 - m)!$. The value of the remaining one-loop determinants at a pole is conveniently written in terms of
\be
\label{effective_masses}
M^l_j \,\equiv\, \tau_j - \tau_l \;,\qquad\qquad \tilde M^l_f \,\equiv\, \tilde\tau_f + \tau_l \;.
\ee
They are the effective masses of the $j$-th fundamental and $f$-th antifundamental at the point of the Coulomb branch $\sigma = -\tau_l$ where the $l$-th fundamental is massless: $M^l_l=0$. We can define $k_2 \equiv k_1 - m$ so that $\sum_{m\in \bZ} \sum_{k \geq 0} = \sum_{k_1 \geq 0} \sum_{k_2 \geq 0}$ and apply the identities
\be
\Gamma(a+n) = (a)_n\, \Gamma(a) \;,\qquad \Gamma(a-n) = \frac1{(a-n)_n} \, \Gamma(a) \qquad\qquad\text{for } n\geq 0
\ee
in terms of the Pochhammer symbol $(a)_n = \prod_{j=0}^{n-1}(a+j)$. Finally we get the expression
\be
\label{Z_U(1)_final}
Z_{U(1)}^{(N_f,\,N_a)} = \sum_{l=1}^{N_f} \; e^{4\pi i \xi \tau_l} \; Z_\text{1-loop}^{(l)} \; Z_\text{v}^{(l)}(z) \, Z_\text{av}^{(l)}(\bar z) \;,
\ee
where the various pieces are:
\bea
\label{Z U(1) pieces}
Z_\text{1-loop}^{(l)} &= \bigg( \prod_{j\, (\neq l)}^{N_f} \frac{ \Gamma(-iM^l_j)}{\Gamma(1+iM^l_j)} \bigg) \bigg( \prod_f^{N_a} \frac{\Gamma(-i \tilde M^l_f )}{\Gamma(1+ i \tilde M^l_f)} \bigg) \\
Z_\text{v}^{(l)}(z) &= \sum_{k=0}^\infty \; \frac{\prod_f^{N_a} (-i \tilde M^l_f)_k}{\prod_{j\,(\neq l)}^{N_f} (1+iM^l_j)_k} \; \frac{ (-1)^{N_f k} \, z^k}{k!} = \,_{N_a}F_{N_f-1} \left( \begin{matrix}  \{-i\tilde M^l_f\}_f^{N_a} \\ \{1+i M^l_j\}_{j\,(\neq l)}^{N_f} \end{matrix} \Bigg| (-1)^{N_f} z  \right)  \\
Z_\text{av}^{(l)}(\bar z) &= \sum_{k=0}^\infty \; \frac{\prod_f^{N_a} (-i \tilde M^l_f)_k}{\prod_{j\, (\neq l)}^{N_f} (1+iM^l_j)_k} \; \frac{ (-1)^{N_a k} \, \bar z^k}{k!} = \,_{N_a}F_{N_f-1} \left( \begin{matrix}  \{-i\tilde M^l_f\}_f^{N_a} \\ \{1+i M^l_j\}_{j\, (\neq l)}^{N_f} \end{matrix} \Bigg| (-1)^{N_a} \bar{z}  \right)
\eea
and $z = e^{-2\pi \xi - i\theta}$. Here $_pF_q$ is the generalized hypergeometric function. For $N_f > N_a$ the two series converge for any $z$; for $N_f = N_a$ they converge for $|z|<1$ (which is the original domain of validity of our computation) and can be extended in the whole $z$-plane analytically. In fact they can be written as:
\bea
Z_\text{v}^{(l)}(z) &= Z_\text{vortex} \big( -z,\, 1,\, \{-iM^l_j\}_{j\,(\neq l)},\, \{-i\tilde M^l_f\}_f \big) \\
Z_\text{av}^{(l)}(\bar z) &= Z_\text{vortex} \big( -\bar z,\, -1,\, \{iM^l_j\}_{j\,(\neq l)},\, \{i\tilde M^l_f\}_f \big)
\eea
in terms of the vortex partition function for $U(1)$ with $(N_f,N_a)$ flavors in $\Omega$-background, that we can write as
\be
\label{vortex partiton function U(1)}
Z_\text{vortex}^{U(1)} (z,\varepsilon,m_j, \tilde m_f) = \sum_{k=0}^\infty \; \frac{z^k}{\varepsilon^{(N_f-N_a)k}} \; \frac{\prod_{f=1}^{N_a} \Big( \dfrac{\tilde m_f}\varepsilon \Big)_k}{k! \, \prod_{j=1}^{N_f-1} \Big( \dfrac{m_j}\varepsilon - k \Big)_k } \;.
\ee
This expression agrees with \cite{Dimofte:2010tz}, up to a redefinition of the masses.
In \cite{Pasquetti:2011fj} a similar structure has been found for the path integral $Z_{S^3_b}$ of 3d $\cN=2$ YM-CS $U(1)$ theory on the three-dimensional ellipsoid.
The expression in (\ref{Z_U(1)_final}) agrees with our general expression (\ref{Z_S2 Higgs final}).

\subsection{$U(N)$ with $(N_f,N_a)$ flavors}
\label{subsec:_U(N)}

Consider now a $U(N)$ gauge theory with $(N_f,N_a)$ flavors. The matrix integral (\ref{matrix integral}) is:
\begin{multline}
\label{Z_U(N)_start}
Z_{U(N)}^{(N_f,\,N_a)} = \frac1{N!}\,\sum_{\vec m \,\in\,\bZ^N} \int \bigg[ \prod_{r=1}^N \frac{d\sigma_r}{2\pi} \, e^{-4\pi i \xi \sigma_r -i\theta m_r} \bigg] \bigg[ \prod_{t<s}^N \Big( \frac{(m_t-m_s)^2}4 +(\sigma_t- \sigma_s)^2 \Big) \bigg] \times  \\
\times \prod_{r=1}^N \prod_{j=1}^{N_f}  \frac{\Gamma\big( -i\sigma_r - i\tau_j - \frac{m_r}2 \big)} {\Gamma\big( 1 + i\sigma_r + i\tau_j - \frac{m_r}2 \big)}  \prod_{f=1}^{N_a} \frac{\Gamma\big( i\sigma_r - i \tilde\tau_f + \frac{m_r}2 \big)}
{\Gamma\big(1 - i\sigma_r + i\tilde\tau_f + \frac{m_r}2 \big)} \;.
\end{multline}
We assume as before that $N_f>N_a$, or $N_f=N_a$ and $\xi>0$, so we can close the integration contours in the lower half-planes and resum residues at the poles arising from fundamentals only.

In this section we compute (\ref{Z_U(N)_start}) in terms of $Z_{U(1)}^{(N_f,N_a)}$ using a trick, while in the next section we will more conventionally sum over all poles.
Note that \eqref{Z_U(N)_start} would be identical to the partition function of $N$ copies of $U(1)$ with $(N_f,N_a)$ flavors -- though all with $\xi_r=\xi$ and $\theta_r=\theta$ -- were not for the one-loop determinant of the non-Abelian gauge fields, which however does not contribute poles to the integrand. We can thus introduce auxiliary distinct FI parameters $\xi^r$ and theta-angles $\theta^r$ for each Cartan $U(1)$ factor, eventually to be set to $\xi$ and $\theta$, and replace the non-Abelian determinant by
\be
\sigma_r \;\leftrightarrow\; \frac i{4\pi}\,\parfrac{}{\xi^r} \;,\qquad\qquad m_r \;\leftrightarrow\; i\parfrac{}{\theta^r}
\ee
acting on the integral. We can then write
\be
Z_{U(N)}^{(N_f,\,N_a)}(\xi,\theta)= \frac{1}{N!}\,\Delta \prod_{r=1}^N Z_{U(1)}^{(N_f,\,N_a)}(\xi_r, \theta_r) \Big|_{\xi_r=\xi,\;\theta_r=\theta} \;,
\ee
in terms of the differential operator $\Delta = \prod_{t<s}^N \big[ -\frac14 \big( \parfrac{}{\theta^t} - \parfrac{}{\theta^s} \big)^2 - \frac1{16\pi^2} \big( \parfrac{}{\xi^t} - \parfrac{}{\xi^s} \big)^2 \big]$ and $Z_{U(1)}$ computed before, with the dependence on twisted masses left implicit. We can further simplify the expression by introducing complex variables $w^r = -\log z^r = 2\pi\xi^r + i \theta^r$. We get:
\be
\label{Z_U(N)_rewritten_hol_antihol}
Z_{U(N)}^{(N_f,\,N_a)}(w,\overline w) = \frac{1}{N!} \, (-1)^{\frac{N(N-1)}2} \,
\cD \overline\cD \, \prod_{r=1}^N Z_{U(1)}^{(N_f,\,N_a)}(w^r, \overline w^r) \Big|_{w^r=w,\, \overline w^r=\overline w} \;,
\ee
where the differential operators are $\cD = \prod_{t<s}^N (\partial_t - \partial_s)$ and $\overline\cD = \prod_{t<s}^N (\overline\partial_t - \overline\partial_s)$,
\be
\label{Z_U(1)_hol_antihol}
Z_{U(1)}^{(N_f,\,N_a)}(w^r, \overline w^r) =  \sum_{l=1}^{N_f} \; e^{ i (w^r+\overline w^r) \tau_l} \; Z^{(l)}_\text{1-loop} \; Z_\text{v}^{(l)}(e^{-w^r}) \; Z_\text{av}^{(l)}(e^{-\overline w^r})
\ee
and the various pieces are as in (\ref{Z U(1) pieces}). In (\ref{Z_U(N)_rewritten_hol_antihol}) effectively we have a sum over $\vec l = (l_1,\dots,l_N)$ where each entry runs from 1 to $N_f$. Each term represents a point on the Coulomb branch where the flavors $(l_1,\dots,l_N)$ have a massless component.
Since both $\cD,\overline\cD$ are totally antisymmetric in $w^r$, $\overline w^r$ respectively, $Z_{U(1)}(w,\bar w)$ factorizes into the product of a holomorphic and an antiholomorphic function of $w$, and eventually we set $w^r=w, \overline w^r = \overline w$, it follows that whenever two $l_r$'s are equal that term vanishes. In particular for $1 \leq N_f < N$ and generic twisted masses the localized partition function $Z_{S^2}$ vanishes.
On the other hand each term is invariant under permutations of $(l_1,\dots,l_N)$ (which is the action of the Weyl group) therefore we can restrict ourselves to sum over $N$-combinations $C(N,N_f)$ of the first $N_f$ integers, and remove $N!$ from the denominator.

To proceed further we split the action of $\cD$ on the exponential in (\ref{Z_U(1)_hol_antihol}) and on $Z_\text{v}^{(l)}$, and similarly for $\overline\cD$. The result can be written as:
\be
\label{Z_U(N)_final}
Z_{U(N)}(w,\overline w) = \sum_{\vec l \,\in\, C(N,N_f)} \; e^{i(w + \overline w) \sum_{r=1}^N \tau_{l_r}} \; Z_\text{1-loop}^{(\vec l)} \; Z_\text{v}^{(\vec l)}(e^{-w}) \; Z_\text{av}^{(\vec l)}(e^{-\overline w})  \;,
\ee
where the one-loop factor now includes the massive W-bosons,
\be
\label{Z U(N) pieces}
Z_\text{1-loop}^{(\vec l)} = \bigg[ \prod_{t<s}^N (M^{l_t}_{l_s})^2 \bigg] \bigg[ \prod_{r=1}^N Z_\text{1-loop}^{(l_r)} \bigg] \;,\qquad\qquad
\begin{aligned}
Z_\text{v}^{(\vec l)}(e^{-w}) &= D^{(\vec l)} \prod_{r=1}^N  Z_\text{v}^{(l_r)}(e^{-w^r}) \Big|_{w^r=w} \\
Z_\text{av}^{(\vec l)}(e^{-\overline w}) &= \overline{D}^{(\vec l)} \prod_{r=1}^N  Z_\text{av}^{(l_r)} (e^{-\overline w^r}) \Big|_{\overline w^r=\overline w}
\end{aligned}
\ee
and the differential operators are:
\be
D^{(\vec l)} = \prod_{t<s}^N  \bigg( 1 - \frac{\partial_t-\partial_s}{iM^{l_t}_{l_s}} \bigg) \;,\qquad\qquad
\overline{D}^{(\vec l)} = \prod_{t<s}^N  \bigg( 1 - \frac{ \overline\partial_t - \overline\partial_s}{iM^{l_t}_{l_s}} \bigg) \;.
\ee
The last expression is only well-defined for $\vec l$ a combination without repetitions, as we are assuming; moreover each term is invariant under permutations of $\vec l$.
Finally we can write the vortex contribution in a more explicit way, acting with the differential to get
\be
Z_\text{v}^{(\vec l)}(e^{-w}) =  \sum_{\vec k \,\in\, \bZ_{\geq 0}^N} \; e^{-w\sum\limits_{r=1}^N k_r} \; \bigg[ \prod_{r=1}^N
\frac{\prod_{f=1}^{N_a} (-i\tilde M^{l_r}_f)_{k_r} }{ \prod_{j=1}^{N_f} (-iM^{l_r}_j - k_r)_{k_r} } \bigg] \,
\bigg[ \prod_{t<s}^N  \bigg( 1 + \frac{k_t-k_s}{iM^{l_t}_{l_s}}  \bigg) \bigg] \;.
\ee
Now that we got rid of differentials, we can go back to $z = e^{-w}$.
Making use of the identity
\be
\label{double Pochhammer identity}
(a-l)_m (-a-m)_l = \Big( 1 + \frac{m-l}a \Big)^{-1} (a+1)_m (-a+1)_l \;,
\ee
the expression above can be recast as:
\be
Z_\text{v}^{(\vec l)}(z) = \sum_{\vec k \,\in\, \bZ_{\geq 0}^N} \; (-1)^{(N-1)|\vec k|} \; z^{|\vec k|} \; \prod_{r\in \vec l} \frac{\prod_{f=1}^{N_a} (-i \tilde M^r_f)_{k_r}}{\big[ \prod_{j\in\vec l}\, (-iM^r_j - k_r)_{k_j} \big] \big[ \prod_{j\not\in \vec l} \, (-iM^r_j - k_r)_{k_r} \big]} \;.
\ee
Let us explain the notation: $r \in \vec l$ is a fundamental flavor index taking one of the $N$ values $(l_1,\dots,l_N)$ picked up by $\vec l$, while $j \not\in \vec l$ runs over the remaining $N_f -N$ values; $\vec k$ is an $N$-tuple whose entries $k_r$ we parameterize by $r\in \vec l$, and $|\vec k| = \sum_r k_r$. Similar manipulations can be done on $Z_\text{av}^{(\vec l)}(e^{-\overline w})$.

Eventually the two (anti)vortex contributions can be written as:
\bea
\label{Z_vortex identification U(N)}
Z_\text{v}^{(\vec l)}(z) &= Z_\text{vortex} \big( (-1)^N z,\, 1,\, \{-i\tau_r\}_{r\in \vec l} \,,\, \{-i\tau_j\}_{j \not\in \vec l} \,,\, \{-i \tilde\tau_f\} \big) \\
Z_\text{av}^{(\vec l)}(\bar z) &= Z_\text{vortex} \big( (-1)^N \bar z,\, -1,\, \{i\tau_r\}_{r\in \vec l} \,,\, \{i\tau_j\}_{j \not\in \vec l} \,,\, \{i \tilde\tau_f\} \big)
\eea
in terms of the vortex partition function for $U(N)$ with $(N_f,N_a)$ flavors:
\begin{multline}
\label{vortex partition function U(N)}
Z^{U(N)}_\text{vortex}\big(z,\, \varepsilon,\, a_r,\, m_j,\, \tilde m_f \big) = \\
= \sum_{\vec k \,\in\, \bZ_{\geq 0}^N} \, \frac{z^{|\vec k|}}{\varepsilon^{(N_f - N_a) |\vec k|}} \; \prod_{r=1}^N \frac{ \prod\limits_{f=1}^{N_a} \Big( \dfrac{\tilde m_f + a_r}\varepsilon \Big)_{k_r}} { k_r! \, \Big[ \prod\limits_{j \, (\neq r)}^N \Big( \dfrac{a_j - a_r}\varepsilon - k_r \Big)_{k_j} \Big] \Big[  \prod\limits_{j=N+1}^{N_f} \Big( \dfrac{m_j-a_r}\varepsilon - k_r \Big)_{k_r} \Big] } \;.
\end{multline}
In this notation we have distinguished between the masses $a_r$ of the $N$ flavors with a component acquiring VEV, and the masses $m_j$ of the $N_f - N$ flavors with vanishing VEV, because they appear differently in the vortex partition function. This asymmetry is inherent to the brane inspired construction of vortex moduli spaces of \cite{Hanany:2003hp}, where the $k$-vortex theory for a 2d $U(N)$ gauge theory with $N_f$ fundamentals is a 0d $U(k)$ gauge theory with one adjoint, $N$ fundamentals and $N_f-N$ antifundamentals.
The expression \eqref{vortex partition function U(N)} agrees with (\ref{vortex partiton function U(1)}) for $N=1$, although the two are written in a slightly different notation.

The vortex partition function (\ref{vortex partition function U(N)}) also agrees with \cite{Shadchin:2006yz, Yoshida:2011au, Bonelli:2011fq, Fujimori:2012ab} for $N_f = N$, apart from a possible minus sign in front of $z$ which can be reabsorbed by a redefinition of the theta angle.
Our final expression (\ref{Z_U(N)_final}) has precisely the form of (\ref{Z_S2 Higgs final}). Finally, the vortex partition function (\ref{vortex partition function U(N)}) can be written as a contour integral, as we spell out in section \ref{sec: duality}.

\subsection{$U(N)$ with flavors and one adjoint: $\cN=(2,2)^*$ SQCD}

Consider a $U(N)$ gauge theory with $(N_f,N_a)$ flavors and one adjoint chiral multiplet. The matrix integral (\ref{matrix integral}) is:
\begin{multline}
Z = \frac1{N!} \sum_{\vec m \,\in\, \bZ^N} \int \bigg[ \prod_{r=1}^N \frac{d\sigma_r}{2\pi} \, e^{-4\pi i\xi \sigma_r - i\theta m_r} \bigg] \bigg[ \prod_{t<s}^N \Big( \frac{(m_t - m_s)^2}4 + (\sigma_t - \sigma_s)^2 \Big) \bigg] \times \\
\times \bigg[ \prod_{r=1}^N \prod_{j=1}^{N_f} \frac{\Gamma\big(-i\sigma_r -i\tau_j - \frac{m_r}2\big)}{\Gamma\big( 1+i\sigma_r +i\tau_j -\frac{m_r}2 \big) } \prod_{f=1}^{N_a} \frac{ \Gamma\big(i\sigma_r -i\tilde\tau_f + \frac{m_r}2 \big)}{\Gamma\big( 1-i\sigma_r +i\tilde\tau_f + \frac{m_r}2\big)} \bigg] \times \\
\times \prod_{r,s=1}^N \frac{\Gamma\big(-i\sigma_r +i\sigma_s -i\tau_A - \frac{m_r - m_s}2\big)}{\Gamma\big(1+ i\sigma_r - i\sigma_s +i\tau_A - \frac{m_r - m_s}2\big) } \;.
\end{multline}
As before $\tau_j, \tilde \tau_f$ are the twisted masses of the flavor symmetry $SU(N_f)\times SU(N_a)$ for fundamentals $Q$ and antifundamentals $\tilde Q$, while $\tau_A$ is the twisted mass of the $U(1)$ flavor symmetry that rotates the adjoint $\Phi$. Moreover we assume $N_f > N_a$, or $N_f = N_a$ and $\xi > 0$, so to close the integration contours in the lower half-planes.

Let us start by listing poles and zeros of the integrand, following section \ref{sec:_1chiral}: fundamentals have poles at $\sigma_r = -\tau_j - i \big( \frac{|m_r|}2 + k \big)$ and zeros at $\sigma_r = -\tau_j + i \big( \frac{|m_r|}2 + k + 1 \big)$, for some $j \in \{1,\dots,N_f\}$ and $k \in \bN$; antifundamentals have poles at $\sigma_r = \tilde\tau_f + i \big( \frac{|m_r|}2 + k \big)$ and zeros at $\sigma_r = \tilde\tau_f - i \big( \frac{|m_r|}2 + k + 1 \big)$; the adjoint has poles at $\sigma_r - \sigma_s = -\tau_A - i \big( \frac{|m_r - m_s|}2 + k \big)$ and zeros at $\sigma_r - \sigma_s = -\tau_A + i \big( \frac{|m_r - m_s|}2 + k + 1 \big)$.
An $N$-pole%
\footnote{By $N$-pole we mean a point in $\bC^N$ where the integrand has a simple pole in each of the $N$ variables $\sigma_r$.}
in the lower half-planes is constructed by picking a number $n \leq N$ of poles from fundamentals to form a set, then adding to the set $N-n$ poles from the adjoint $\Phi_{rs}$: $\sigma_r = \sigma_s - \tau_A + i \big( \frac{|m_r - m_s|}2 + k_{rs} + 1 \big)$ where $\sigma_s$ is already in the set, in such a way to form ``tails''. In fact there is a one-to-one correspondence between poles of the integrand on one hand, and solutions to the D-term equation $\phi\phi^\dag = \chi \unit$ in (\ref{D-term equations}), where we give VEV to the corresponding components, possibly dressed by (anti)vortices on the other hand.

The result of the integral could be computed straightforwardly, however for simplicity we specialize to the case of $\cN=(2,2)^*$ SQCD: a theory with $\cN=(4,4)$ supersymmetry broken to $\cN=(2,2)$ by a twisted mass term $\tau_A$ for the adjoint.%
\footnote{Had we broken to $\cN=(2,2)$ by a superpotential mass term for the adjoint, the resulting partition function would have been the same as for $\cN=(2,2)$ $U(N)$ SQCD with $(N_f,N_f)$ matter and a quartic superpotential, a subcase of what we considered in section \ref{subsec:_U(N)}.}
In particular the matter spectrum consists of hypermultiplets, thus $N_f = N_a$, and we have the superpotential interaction
\be
\label{N=(2,2)* W}
W = \tilde Q \Phi Q
\ee
which reduces the flavor symmetry to $SU(N_f)_\text{diag} \times U(1)_A$ and poses constraints on masses and R-charges:
\be
\label{N=(2,2)* relation}
\tau_j + \tilde\tau_j + \tau_A = i \qquad\forall\, j \qquad\qquad\Rightarrow\qquad\qquad \tilde M^r_f = - M^r_f - \tau_A + i \;.
\ee
These relations imply that poles from the adjoint coincide with zeros from the antifundamentals, so that such poles are not really there. For instance, suppose that $\sigma_s = -\tau_j - i \big( \frac{|m_s|}2 + k_s \big)$ is at a pole from a fundamental $Q_{sj}$ and $\sigma_r = -\tau_j - \tau_A - i \big( \frac{|m_s|}2 + \frac{|m_r - m_s|}2 + k_s + k_{rs} \big)$ at a pole from $\Phi_{rs}$. Because of (\ref{N=(2,2)* relation}) there is always a $\tilde k_r \in \bN$ such that $\sigma_r = \tilde\tau_j - i \big( \frac{|m_r|}2 + \tilde k_r + 1 \big)$, which is a zero from the antifundamental $\tilde Q_{jr}$. The fact that all $N$-poles involving the adjoint disappear is in correspondence with the fact that the F-term equation $\Phi Q = 0$ from $W$ in (\ref{N=(2,2)* W}) forbids those Higgs branch solutions where $\Phi,Q$ get VEV simultaneously.%
\footnote{Concretely in the case of $U(2)$ with $N_f=1$: $Q = \smat{ \sqrt{2\chi} \\ 0}$, $\Phi = \smat{ 0 & 0 \\ \sqrt\chi & 0}$ solves the D-term equation $QQ^\dag + [\Phi,\Phi^\dag] = \chi \unit$ with $\chi>0$, and is in correspondence with a pole for $Q_1$, $\Phi_{21}$. However $\Phi Q = \smat{0 \\ 2\chi}$ so this solution disappears when imposing $\Phi Q =0$.}

Eventually $N$-poles in the lower half-planes are parametrized by $N$-tuples $\vec l = (l_1,\dots,l_N)$ with $l_i =1,\dots,N_f$:
$$
\big\{ \sigma_r = -\tau_{l_r} - ik_r - \tfrac i2 |m_r| \big\}_{r=1,\dots,N} \;.
$$
After carrying out the computation one discovers, as we will show, that $N$-tuples $\vec l$ with repetitions yield vanishing contribution (this is expected from the 1-1 correspondence with Higgs branch solutions, and could also be proven in a way similar to section \ref{subsec:_U(N)}); since residues are invariant under permutations of $\vec l$, we restrict ourselves to sum over $N$-combinations $\vec l \in C(N,N_f)$ and remove $N!$ from the denominator. Then we proceed as in the $U(1)$ case of section \ref{subsec:_U(1)}, and notice that if the adjoint was not there we would reproduce the result of section \ref{subsec:_U(N)}. The residues of the $\Gamma$'s developing a pole are $i(-1)^{k_{1r}}/k_{1r}!$, where $k_{1r} = k_r + \sub{m_r}$, and the value of the corresponding $\Gamma$'s at the denominator is $(k_{1r} - m_r)!$. We define $k_{2r} = k_{1r} - m_r$ so that $\sum_{\vec m \in \bZ^N} \sum_{\vec k \in \bZ_{\geq 0}^N} = \sum_{\vec k_1 \in \bZ_{\geq 0}^N} \sum_{\vec k_2 \in \bZ_{\geq 0}^N}$.

Evaluating all other terms we are left with an expression, that we call $Z_{U(N)}^\text{$N_f$,adj}$, as in (\ref{Z_U(N)_final}) but with the following pieces. The classical term is the same.
The one-loop term is:
$$
Z_\text{1-loop}^{(\vec l)} = \bigg[ \prod_{\substack{t,s\in\vec l \\ t<s}} (M^t_s)^2 \bigg] \bigg[ \prod_{r\in \vec l} \;\prod_{j\,(\neq l_r)}^{N_f} \frac{\Gamma(-iM^r_j)}{\Gamma(1+iM^r_j)} \prod_{f=1}^{N_a} \frac{\Gamma(-i\tilde M^r_f)}{\Gamma(1+i\tilde M^r_f)} \bigg] \bigg[ \prod_{r,s \in \vec l} \frac{\Gamma(-iM^r_s - i\tau_A)}{\Gamma(1+iM^r_s + i\tau_A)} \bigg] \;.
$$
Note that without the last bracket from the adjoint, this would agree with (\ref{Z U(N) pieces}). In the case of $\cN=(2,2)^*$ SQCD instead we have a few simplifications: the W-boson one-loop cancels against fundamentals for $j \in \vec l$ and the adjoint one-loop cancels against antifundamentals for $f \in \vec l$:
\be
Z_\text{1-loop}^{(\vec l)} = \prod_{r\,\in\,\vec l} \prod_{j\,\not\in\,\vec l} \, \frac{\Gamma(-iM^r_j)}{\Gamma(1+iM^r_j)} \prod_{f\,\not\in\,\vec l} \frac{\Gamma(1 + iM^r_f +i\tau_A)}{\Gamma(-iM^r_f - i\tau_A)} \;.
\ee
In particular for $\tau_A = 0$ we get $Z_\text{1-loop}^{(\vec l)} = 1$. The vortex contribution is
$$
Z_\text{v}^{(\vec l)}(z) = \sum_{\vec k \,\in\, \bZ_{\geq 0}^N} z^{|\vec k|} \bigg[ \prod_{r=1}^N \frac{\prod_{f=1}^{N_f} (-i \tilde M^{l_r}_f)_{k_r}}{\prod_{j=1}^{N_f} (-i M^{l_r}_j - k_r)_{k_r}} \prod_{s=1}^N \frac{ (-iM^{l_r}_{l_s} -i\tau_A - k_r)_{k_s}}{ (-iM^{l_r}_{l_s} - i\tau_A - k_r )_{k_r}} \bigg] \bigg[ \prod_{t<s}^N \Big( 1 + \frac{k_t - k_s}{iM^{l_t}_{l_s}} \Big) \bigg]
$$
Looking at this expression multiplied by $\prod_{t<s} M^{l_t}_{l_s}$ from the W-boson one-loop, we see that whenever $\vec l$ has a repetition, $Z^{(\vec l)}_\text{v}(z) = 0$: if some $l_t = l_s$ then $M^{l_t}_{l_s} = 0$, and all terms in the expression are symmetric in $k_t \leftrightarrow k_s$ but the last bracket which is antisymmetric; thus after summing over $\vec k$ we get zero. Coming back to $\vec l \in C(N,N_f)$, we apply the identity (\ref{double Pochhammer identity}) and impose (\ref{N=(2,2)* relation}). Eventually $Z_\text{v}^{(\vec l)}(z)$ and $Z_\text{av}^{(\vec l)}(\bar z)$ can be written in the usual way (\ref{Z_vortex identification U(N)}) in terms of the vortex partition function:
\begin{multline}
\label{vortex partition function adjoint}
Z_\text{vortex}^{(\vec l)}(z, \varepsilon, a, m_A) = \\
\quad = \sum_{\vec k \,\in\, \bZ_{\geq0}^N} \; \frac{z^{|\vec k|}}{(-1)^{(N_f-1) |\vec k|}} \; \prod_{r\,\in\,\vec l}
\frac{ \Big[ \prod\limits_{f\,\in\,\vec l} \Big( \dfrac{a_{fr} + m_A}\varepsilon - k_r \Big)_{k_f} \Big] \Big[ \prod\limits_{f\,\not\in\,\vec l} \Big( \dfrac{a_{fr} + m_A}\varepsilon - k_r \Big)_{k_r} \Big]}
{\Big[ \prod\limits_{j\,\in\,\vec l} \Big( \dfrac{a_{jr}}\varepsilon - k_r \Big)_{k_j} \Big] \Big[ \prod\limits_{j\,\not\in\,\vec l} \Big( \dfrac{a_{jr}}\varepsilon - k_r \Big)_{k_r} \Big]}
\end{multline}
where we defined $a_j = -i\tau_j$, $m_A = -i\tau_A$ and $a_{jr} = a_j - a_r = -iM^r_j$. This expression agrees with \cite{Bonelli:2011fq, Fujimori:2012ab} for $N_f = N$. Notice that here, compared with (\ref{vortex partition function U(N)}), the distinction between masses of flavors getting or not getting VEV is given by $\vec l$.

There are two notable limits to consider in (\ref{vortex partition function adjoint}). First, when $m_A \to \infty$ the vortex partition function reduces to the one (\ref{vortex partition function U(N)}) for $U(N)$ with $N_f$ fundamentals, up to a renormalization of the FI term and a shift of the theta angle. This is expected because the superpotential (\ref{N=(2,2)* W}) imposes $\tilde \tau_j = - \tau_j - \tau_A + i$, therefore in the limit $\tau_A \to \infty$ with $\tau_j$ kept fixed, both the adjoint and the antifundamentals are integrated out. Second, when $m_A = 0$ we simply get:
\be
Z_\text{vortex}^{(\vec l)} = \frac1{\big( 1 + (-1)^{N_f} z \big)^N} \;,
\ee
which does not depend on the twisted masses at all. Indeed for $\tau_A = 0$ the theory has enhanced $\cN=(4,4)$ supersymmetry in the infrared \cite{Witten:1997yu}, which forces the effective twisted superpotential to be a linear function of the twisted masses (see for instance \cite{Billo:1993rd}), in particular $Z_\text{vortex}$ cannot depend on them. We can thus compute $Z_\text{vortex}$ in the limit $\tau_j \to \infty$: the theory flows to $N$ decoupled copies of the $\cN=(4,4)$ $U(1)$ theory with one hyper, with renormalized theta angle $z \to (-1)^{N_f-1}z$ from integrating out the other $N_f-1$ hypers. The partition function of $\cN=(4,4)$ QED is (\ref{vortex partiton function U(1)}) with $\tilde m = \varepsilon$: $Z_\text{vortex} = (1-z)^{-1}$.

\subsection{Analyticity in $\xi$}
\label{sec: analyticity}

In section \ref{subsec:_U(1)} we computed the $S^2$ partition function for a $U(1)$ theory with $(N_f,N_a)$ chirals. While for $N_f\neq N_a$ the result was manifestly analytic in the FI parameter $\xi$
as expected, this analyticity was not manifest for $N_f=N_a$ because the closure of the integration contour depends on the sign of the FI parameter. In this section we show that even in the latter case the partition function is analytic in $\xi$, due to a nontrivial identity satisfied by the relevant generalized hypergeometric functions.

We start with the partition function for $U(1)$ with $(N_f,N_f)$ chirals computed for $\xi>0$:
\be
\label{Z_U(1)_(N_f,N_f)_hol_antihol}
Z_{U(1)}^{(N_f,N_f)}(w, \bw) =  \sum_{l=1}^{N_f} \; e^{ i (w+\bw) \tau_l} \; Z^{(l)}_\text{1-loop} \; Z^{(l)}_\text{v}(e^{-w}) \; Z^{(l)}_\text{av}(e^{-\overline w}) \;,
\ee
where the various pieces are in (\ref{Z U(1) pieces}).
The sum is over the $N_f$ points of the Coulomb branch where a fundamental has vanishing effective twisted mass.
Next we use an identity for generalized hypergeometric functions $\,_{N}F_{N-1}$ that allows to rewrite the function of $x$ in terms of a sum of $N$ hypergeometric functions of $1/x$ (see for instance \cite{Fateev:2007ab}):%
\footnote{We thank Sara Pasquetti for bringing this formula to our attention.}
\begin{multline}
\label{identity_gen_hyper}
\frac{\prod_{k=1}^{N} \Gamma(a_k)}{\prod_{h=1}^{N-1} \Gamma(b_h)} \; \,_{N}F_{N-1} \left( \begin{matrix}  \{a_k\}_{k}^{N} \\ \{b_h\}_{h}^{N-1} \end{matrix} \Bigg| \; x \right) = \\
=\sum_{i=1}^N (-x)^{a_i} \; \frac{\Gamma(a_i)\prod_{j\,(\neq i)}^N \Gamma(a_j-a_i)}{\prod_{l=1}^{N-1} \Gamma(b_l-a_i)} \;
\,_{N}F_{N-1} \left( \begin{matrix} a_i\,,\, \{1+a_i-b_l\}_{l}^{N-1} \\ \{1+a_i-a_j\}_{j\,(\neq 1)}^{N} \end{matrix} \Bigg|\; \frac{1}{x} \right) \;.
\end{multline}
While the Taylor series in $x$ for the LHS converges for $|x|<1$, those for the RHS converge for $|x|>1$, therefore the identity holds for the analytic continuations.

We apply \eqref{identity_gen_hyper} to some of the factors in \eqref{Z_U(1)_(N_f,N_f)_hol_antihol}:
\bea
\label{identity_applied}
& \frac{\prod_{f=1}^{N_f} \Gamma(-i\tM^l_f)}{\prod_{j\,(\neq l)}^{N_f} \Gamma(1+iM^l_j)} \; \,_{N_f}F_{N_f-1} \left( \begin{matrix}  \{-i\tM^l_f\}_f^{N_f} \\ \{1+iM^l_j\}_{j\,(\neq l)}^{N_f} \end{matrix} \Bigg|\, (-1)^{N_f} e^{-\bw} \right) =
\sum_{t=1}^{N_f} e^{-(i\bw+\pi (N_f-1))\tM^l_t} \; \times \\
&\qquad \times \frac{\Gamma(-i\tM^l_t)\prod_{f\,(\neq t)}^{N_f} \; \Gamma(-i \tM^l_f + i \tM^l_t)}{\prod_{j\,(\neq l)}^{N_f} \Gamma(1+i M^l_j + i \tM^l_t)}
\,_{N_f}F_{N_f-1} \left( \begin{matrix} -i\tM^l_t \,,\, \{-iM^l_j - i\tM^l_t\}_{j\,(\neq l)}^{N_f} \\ \{1+i\tM^l_f - i\tM^l_t\}_{f\,(\neq t)}^{N_f} \end{matrix} \Bigg|\, (-1)^{N_f} e^{\bw} \right) = \\
&=\sum_{t=1}^{N_f} e^{-\pi(N_f-1)(\tau_l+\ttau_t)}e^{-i\bw\ttau_t}\, \frac{\pi i}{\sinh(\pi \tM^l_t)}
\frac{\prod_{f\,(\neq t)}^{N_f} \; \Gamma(-i \tilde{\tM}^f_t)}{\prod_j^{N_f} \Gamma(1+i \tM^j_t)} \,_{N_f}F_{N_f-1} \left( \begin{matrix}  \{-i\tM^j_t\}_j^{N_f} \\ \{1+i\tilde{\tM}^f_t\}_{f\,(\neq t)}^{N_f} \end{matrix} \Bigg| \,(-1)^{N_f} e^{\bw} \right)
\eea
where we see the appearance of
\be
\tM^l_f - \tM^l_t = -\ttau_t + \ttau_f \equiv \tilde{\tM}^f_t \;,\qquad\qquad M^l_j + \tM^l_t = \tau_j + \ttau_t = \tM^j_t \;,
\ee
the effective masses of the $f$-th antifundamental and the $j$-th fundamental respectively, in the vacuum in which the $t$-th antifundamental is massless ($\sigma = \ttau_t$), and we used the identity $\Gamma(z)\, \Gamma(1-z) = \pi / \sin(\pi z)$.
The $S^2$ partition function for the $U(1)$ theory with $(N_f,N_f)$ flavors can thus be recast as:
\be
\label{U(1)_final_north_south_mixed}
Z_{U(1)}^{(N_f,N_f)}(w, \bw) = \sum_{l,\,t=1}^{N_f} W^{(l)}(e^{-w}) \; A_{lt} \; \tilde W^{(t)}(e^\bw) \;,\qquad\qquad \text{where} \qquad A_{lt} = \frac{\pi i}{\sinh(\pi\tM^l_t)}
\ee
and
\bea
\label{U(1)_final_north_fund}
W^{(l)}(e^{-w}) &= e^{i[w+i\pi(N_f-1)]\tau_l} \; \frac{\prod_{j\, (\neq l)}^{N_f} \Gamma(-iM^l_j)} {\prod_f^{N_f} \Gamma(1 + i\tM^l_f)} \;
\,_{N_f}F_{N_f-1} \left( \begin{matrix}  \{-i\tilde M^l_f\}_f^{N_f} \\ \{1+i M^l_j\}_{j\,(\neq l)}^{N_f} \end{matrix} \Bigg| \, (-1)^{N_f} e^{-w}  \right)\\
\tilde{W}^{(t)}(e^\bw) &= e^{-i[\bw-i\pi(N_f-1)]\ttau_t} \; \frac{\prod_{f\,(\neq t)}^{N_f} \Gamma(-i\tilde{\tM}^f_t)}{\prod_j^{N_f} \Gamma(1 + i\tM^j_t)} \;
\,_{N_f}F_{N_f-1} \left( \begin{matrix}  \{-i\tilde M^j_t\}_j^{N_f} \\ \{1+i \tilde{\tM}^f_t\}_{f\,(\neq t)}^{N_f} \end{matrix} \Bigg| \, (-1)^{N_f} e^{\bw}  \right)
\eea
In this expression we have two distinct sums over $N_f$ vacua: in the former a fundamental has vanishing effective twisted mass in each vacuum $\sigma=-\tau_l$, and is associated to vortices; in the latter an antifundamental has vanishing mass in each vacuum $\sigma=\ttau_t$, and is associated to antivortices.
Had we applied the identity \eqref{identity_gen_hyper} to $Z^{(l)}_\mathrm{v}$ instead of $Z^{(l)}_\mathrm{av}$, we would have found a similar expression with vortices from antifundamentals and antivortices from fundamentals. Finally, applying the identity both to $Z^{(l)}_\mathrm{v}$ and $Z^{(l)}_\mathrm{av}$ we can get an expression analogous to \eqref{Z_U(1)_(N_f,N_f)_hol_antihol} with a single sum over points of the Coulomb branch were antifundamentals (rather than fundamentals) are massless. That is precisely the expression one finds for the $S^2$ partition function if $\xi<0$ by closing the integration contour in the upper half-plane according to \eqref{identity_zeta}, namely:
\bea
\label{Z_U(1)_(N_f,N_f)_hol_antihol_antifund}
Z_{U(1)}^{(N_f,N_f)}(w, \bw) &=  \sum\nolimits_{t=1}^{N_f} \; e^{ -i (w+\bw) \ttau_t} \; Z^{(t)}_{\text{1-loop}} \; Z^{(t)}_\text{v}(e^w) \; Z^{(t)}_\text{av}(e^\bw) \\
Z^{(t)}_\text{1-loop} &= \bigg( \prod\nolimits_{f\,(\neq t)}^{N_f} \frac{ \Gamma(-i\tilde{\tilde{M}}^f_t)}{\Gamma(1 + i\tilde{\tilde{M}}^f_t)} \bigg) \bigg( \prod\nolimits_j^{N_f} \frac{\Gamma(-i\tilde M^j_t )}{\Gamma(1+ i\tilde M^j_t)} \bigg) \\
Z^{(t)}_\text{v}(e^w) &\equiv  \,_{N_f}F_{N_f-1} \left( \begin{matrix}  \{-i\tilde M^j_t\}_j^{N_f} \\ \{1 + i \tilde{\tilde{M}}^f_t\}_{f\,(\neq t)}^{N_f} \end{matrix} \Bigg|\, (-1)^{N_f} e^w \right) \\
Z^{(t)}_\text{av}(e^\bw) &= \,_{N_f}F_{N_f-1} \left( \begin{matrix}  \{-i\tilde M^j_l\}_j^{N_f} \\ \{1 + i \tilde{\tilde{M}}^f_t\}_{f\,(\neq t)}^{N_f} \end{matrix} \Bigg|\, (-1)^{N_f} e^{\bw} \right)\;.
\eea
Indeed, applying \eqref{identity_gen_hyper} to
$ Z^{(t)}_\mathrm{v}(e^w)$ leads again to \eqref{U(1)_final_north_south_mixed}.
It would be interesting to investigate if a choice of localizing action exists which allows vortices from fundamentals at the NP and antivortices from antifundamentals at the SP, and solitons interpolating between the associated vacua, yielding directly the structure \eqref{U(1)_final_north_south_mixed}.

\section{Dualities}
\label{sec: duality}

In \cite{Hori:2006dk, Hori:2011pd} a duality -- reminiscent of Seiberg duality \cite{Seiberg:1994pq} in four dimensions -- was proposed between the $SU(N)$ gauge theory with $N_f > N$ fundamental flavors and $SU(N_f-N)$ with the same number $N_f$ of fundamentals. By gauging the $U(1)$ baryon symmetry this implies a duality between $U(N)$ and $U(N_f - N)$.%
\footnote{Such duality has recently been considered in \cite{Jockers:2012zr} in order to relate different UV non-Abelian GLSM descriptions of new 2d Calabi-Yau non-linear sigma models. A related duality in $\cN=(2,2)^*$ SQCD has been investigated in \cite{Orlando:2010uu}}
The duality for unitary groups is indeed straightforwardly suggested by the brane construction in \cite{Hanany:1997vm}; it is also related by circle reduction to the 3d $\cN=2$ maximally chiral duality \cite{Benini:2011mf} between $U(N)$ with $N_f$ fundamentals and $U(N_f - N)$ with $N_f$ antifundamentals, up to charge conjugation in the latter.
We show that indeed the $S^2$ partition functions of the two dual theories coincide, as functions of the FI parameter $\xi$ and the twisted masses in the unitary case, and of the twisted masses only (including an extra mass for the baryonic symmetry) in the special unitary case.

\paragraph{Unitary group.} We consider first the case of $U(N)$ gauge theory with $N_f$ fundamental flavors, because we can use the simpler expression of $Z_{S^2}$ in terms of the vortex partition function. To avoid cluttering formulae, we use the notation
\be
-i\tau_j \,\equiv\, a_j
\ee
for the twisted masses. The expression (\ref{Z_U(N)_final}) is a sum over $N$-combinations $\vec l \in C(N,N_f)$. We show that for each combination $\vec l$ the corresponding term in $Z_{U(N)}$ is equal to the term in $Z_{U(N_f-N)}$ corresponding to the dual combination $\vec l^D$, where $\vec l^D \in C(N_f-N,N_f)$ is defined such that $\vec l \cap \vec l^D = \emptyset$. In the duality the map of parameters is:
\be
\label{duality param transformation}
a_j = - a_j^D \;,\qquad\qquad \xi = \xi^D\;,\qquad\qquad \theta = \theta^D + N_f \pi \;,
\ee
where the superscript ${}^D$ labels quantities in the dual theory.
For a fixed vacuum: $\vec l \leftrightarrow \vec l^D$, $a_{r\,\in\,\vec l} = - a^D_{r \,\not\in\, \vec l^D}$ and $a_{s \,\not\in\, \vec l} = - a^D_{s \,\in\, \vec l^D}$.

First, the exponential factor in (\ref{Z_U(N)_final}) is clearly invariant since $\sum_{j=1}^{N_f} a_j = 0$.
Second, the one-loop factor in (\ref{Z U(N) pieces}) takes the following form:
\be
Z_\text{1-loop}^{(\vec l)} = \bigg[ \prod_{\substack{t,s \,\in\, \vec l \\ t\neq s}} (a_t - a_s) \bigg] \bigg[ \prod_{r \,\in\, \vec l} \; \prod_{j\, (\neq l_r)}^{N_f} \frac{\Gamma(a_j - a_r)}{\Gamma( 1 - a_j + a_r)} \bigg] \;.
\ee
The last product over $j$ can be split into $(j \in \vec l, j\neq l_r)$ and $j \not\in \vec l$: the former exactly cancels against the determinant of W-bosons, leaving the latter:
\be
\label{1-loop W fund combined}
Z_\text{1-loop}^{(\vec l)} = \prod_{r\,\in\,\vec l \,,\, j \,\not\in\, \vec l} \frac{\Gamma(a_j - a_r)}{\Gamma(1-a_j + a_r)} \;.
\ee
This is also invariant under the transformation after (\ref{duality param transformation}).

Finally let us consider the vortex partition function in (\ref{vortex partition function U(N)}) for $U(N)$ with $(N_f,0)$ flavors: it is a Taylor series in $z$ in which the $z^k$ term is a sum over $N$-tuples $\vec k$ of positive integers (that we will call partitions) with fixed $|\vec k| \equiv \sum_r k_r = k$. Writing $Z_\text{vortex} = \sum_{k=0}^\infty z^k Z_k$:
\be
\label{Zk partition function}
Z_k = \sum_{\substack{\vec k \,\in\, \bZ_{\geq 0}^N \\ |\vec k| = k}} \; \varepsilon^{-N_f k} \; \prod_{r\,\in\,\vec l} \; \frac1{k_r! \; \Big[ \prod_{\substack{j\,\in\,\vec l \\ j \neq r}} \Big( \dfrac{a_j - a_r}\varepsilon - k_r \Big)_{k_j} \Big] \; \Big[ \prod_{j\,\not\in\,\vec l} \Big( \dfrac{a_j - a_r}\varepsilon - k_r \Big)_{k_r} \Big] }
\ee
Notice that here, as opposed to (\ref{vortex partition function U(N)}), the distinction between masses of flavors getting or not getting VEV is given by $\vec l$. It turns out that $Z_k$ can be written as a contour integral:
\be
\label{contour integral}
Z_k = \frac1{\varepsilon^k k!} \int \bigg[ \prod_{j=1}^k \frac{d\varphi_j}{2\pi i} \bigg] \bigg[ \prod_{i<j}^k \frac{(\varphi_i - \varphi_j)^2}{(\varphi_i - \varphi_j)^2 - \varepsilon^2} \bigg] \prod_{j=1}^k \frac1{\prod_{r \,\in\, \vec l}\: (\varphi_j - a_r) \, \prod_{s \,\not\in\, \vec l}\: (a_s - \varphi_j - \varepsilon)} \;.
\ee
The contours are taken along the real lines $\bR^k$ and closed in the upper half-planes (for $N_f = 1$ there would be an ambiguity because of a pole at infinity).
To begin with, here we have assumed that the twisted masses and $\varepsilon$ are real; we will generalize to complex values below.
The indentations of the integration contours are defined by assigning to
the parameter $\varepsilon$ and the masses $a_{r\,\in\,\vec l}$ and $a_{s\,\not\in\,\vec l}$ small positive imaginary parts, with the imaginary part of $\varepsilon$ larger than that of the masses.

It is easy to see that $k$-poles inside the contours are classified by partitions $\vec k$ of $k$ and each configuration is given by:
\be
\label{poles of contour integral}
\{\varphi_j \} = \left\{ \begin{aligned} & \vdots \\ &a_r,\quad a_r + \varepsilon ,\quad \dots ,\quad a_r + (k_r - 1)\,\varepsilon \\ &\vdots \end{aligned} \right. \qquad\qquad\qquad r\in \vec l \;,
\ee
that is for each $r\in\vec l$ we have a ``tail'' of $k_r$ values starting from $a_r$. For each of such configurations we actually have $k!$ $k$-poles obtained by permutations. The definition of the contours above is valid when $\varepsilon, a_r$ are real: for generic complex parameters the contours are defined to include all and only the poles in (\ref{poles of contour integral}).
One can also show that the residue of a $k$-pole of type $\vec k$ equals the corresponding term in the sum in (\ref{Zk partition function}). For instance, the last product in (\ref{contour integral}) evaluated at a pole gives:
\be
\prod\nolimits_{s \,\not\in\, \vec l} \bigg[ \prod_{j=1}^k (a_s - \varphi_j - \varepsilon) \bigg] = \prod\nolimits_{s \,\not\in\, \vec l} \prod\nolimits_{r\,\in\,\vec l} \; \varepsilon^{k_r} \Big( \frac{a_s - a_r}\varepsilon - k_r \Big)_{k_r} \;,
\ee
and similarly the other pieces.

Given the integral representation of $Z_k$ in (\ref{contour integral}), we can now close the contours in the lower half-planes. The integral is unaffected as long as there are no poles at infinity, which happens for $N_f > 1$. To bring the contour integral back to a form that we can compare with (\ref{contour integral}), we first define ``dual" integration variables $\varphi^D_j = -\varphi_j - \varepsilon$, so that the integration contour of $d\varphi_j^D$ closes in the upper half-plane, and $a_j^D = -a_j$; then we change the direction of the contours to the original orientation; we move the contours back to $\bR^k$ and give a small positive imaginary part to $ a_j^D$ which can be done without crossings. Eventually we reach:
$$
Z_k = \frac1{\varepsilon^k k!} \int \bigg[ \prod_{j=1}^k \frac{d \varphi_j^D}{2\pi i} \bigg] \bigg[ \prod_{i<j}^k \frac{(\varphi_i^D - \varphi^D_j)^2}{(\varphi^D_i - \varphi^D_j)^2 - \varepsilon^2} \bigg] \prod_{j=1}^k \frac1{\prod_{s \not\in \vec l}\, (\varphi^D_j -  a^D_s) \, \prod_{r \in \vec l}\, (a^D_r - \varphi^D_j - \varepsilon)}
$$
with the contours defined as after (\ref{contour integral}). This indeed produces the vortex partition function of $U(N_f - N)$ with $N_f$ flavors, with twisted masses obtained from the map after (\ref{duality param transformation}).

We have thus proved that, at fixed $\vec l$ and $\vec l^D$, the two vortex partition functions of $U(N)$ and $U(N_f-N)$ with $N_f$ flavors coincide under the map after (\ref{duality param transformation}).
Because of the extra sign in (\ref{Z_vortex identification U(N)}), the vortex and antivortex contributions in $Z_{S^2}$ (\ref{Z_U(N)_final}) agree if $z = (-1)^{N_f} z^D$, which amounts to a shift of the theta angle.
Since all factors in (\ref{Z_U(N)_final}) match, the two path integrals $Z_{S^2}$ are equal:
\be
\label{Hori duality U(N)}
Z_{U(N)}^{(N_f,0)} \big(\xi \,,\, \theta \,,\, a_j \big) = Z_{U(N_f-N)}^{(N_f,0)} \big(\xi \,,\, \theta + N_f\pi \,,\, -a_j \big) \;.
\ee

\paragraph{Special unitary group.}
Given the partition function of the $U(N)$ gauge theory with $N_f$ fundamentals, expressed by the matrix integral in (\ref{matrix integral}), we can deduce the partition function of the $SU(N)$ gauge theory with $N_f$ fundamentals by performing an integral over the theta angle and a Fourier transform over the FI parameter:
\be\label{SU_from_U}
Z_{SU(N)}^{(N_f,\,0)} (b \,;\,a_j) = \int_0^{2\pi} \frac{d\theta}{2\pi} \, \int_{-\infty}^{+\infty} 4\pi \,d\xi \; e^{4\pi i\xi b} \; Z_{U(N)}^{(N_f,\,0)} (\xi,\theta \,;\, a_j) \;.
\ee
Eventually $b$ will be the twisted mass for the baryonic symmetry $U(1)_B$ of the $SU(N)$ theory, which gives charge $1$ to baryons.
Indeed the integral over $\theta$ gives a Kronecker delta
\be
\int_0^{2\pi} \frac{d\theta}{2\pi} \; e^{-i\theta\sum_{r=1}^N m_r} = \delta_{0,\,\sum_{r=1}^N m_r}
\ee
restricting the magnetic flux $\fm$ to the Cartan subalgebra of $SU(N)$, whereas the integral over the FI parameter gives a Dirac delta
\be \label{Dirac_delta}
\int_{-\infty}^{+\infty} 4\pi\, d\xi \; e^{4\pi i\xi b} \; e^{-4\pi i\xi\sum_{r=1}^N \sigma_r} = \delta \Big( \sum\nolimits_{r=1}^N \sigma_r - b \Big)
\ee
imposing that the trace of the matrix $\sigma$ is $b$.

We then change basis from the ``natural" Cartan of $U(N)$ to the ``natural" Cartan of $SU(N)$ and $U(1)$ in $U(N)\cong [SU(N)\times U(1)]/\bZ_N$, setting:
\be
\label{change_basis_U(N)_to_SU(N)_U(1)}
\sigma_1 = y + x_1 \;,\qquad
\sigma_j = y - x_{j-1}+x_j \quad \text{for $j=2,\dots,N-1$} \;,\qquad
\sigma_N = y - x_{N-1} \;,
\ee
where $y$ is the center of mass of the fields $\sigma_i$ and $x_1,\dots,x_{N-1}$ are the components with respect to the Cartan of $SU(N)$. Taking into account the Jacobian, the integration measures are related as
$\prod_{r=1}^N d \sigma_r = N dy \,\prod_{i=1}^{N-1} d x_i$.
In terms of the new variables $y$ and $x_i$, the Dirac delta function \eqref{Dirac_delta} becomes $\delta(Ny-b)$. After evaluating the integral $\int dy$ thanks to the delta function, we remain with an integral over the Cartan of $SU(N)$, with the replacement (\ref{change_basis_U(N)_to_SU(N)_U(1)}) and $y=b/N$ inside the integrand. This is precisely the partition function $Z_{SU(N)}$ of an $SU(N)$ gauge theory with $N_f$ flavors and with a $U(1)_B$ baryonic symmetry normalized to give charge $1$ to baryons, to which we associate a twisted mass parameter $b$, expectation value of the corresponding background field $\sigma_B$.

In particular, given the identity of partition functions $Z_{U(N)}^{(N_f,0)} = Z_{U(N_f-N)}^{(N_f,0)}$ proven above, we get the identity:
\be
\label{Hori duality SU(N)}
Z_{SU(N)}^{(N_f,0)} \big(b \,,\, a_j \big) = Z_{SU(N_f-N)}^{(N_f,0)} \big(b \,,\, -a_j \big)
\ee
for the duality in the $SU$ case by applying \eqref{SU_from_U} to both sides of (\ref{Hori duality U(N)}).
Baryons $B$ in the electric theory and $b$ in the magnetic theory are defined as
\be
B^{i_1\dots i_N} = \epsilon^{\alpha_1\dots \alpha_N} \, Q^{i_1}_{\alpha_1}\dots Q^{i_N}_{\alpha_N} \;,\qquad\qquad
b_{j_1\dots j_{\tilde{N}}} = \epsilon^{\beta_1\dots \beta_{\tilde{N}}}\, q_{j_1 \beta_1}\dots b_{j_{\tilde N} \beta_{\tilde N} }
\ee
where $\tilde N = N_f - n$, $\alpha_r$ and $\beta_s$ are electric and magnetic color indices, and $i_k, j_k$ are flavor indices.
The duality map between baryons is
\be
B^{i_1\dots i_N} = \epsilon^{i_1\dots i_N j_1\dots j_{\tilde{N}}}\, b_{j_1\dots j_{\tilde{N}}} \;.
\ee
The quantum numbers of baryonic operators are consistent with the duality map: they have baryon charge 1, canonical R-charge 0 and transform in the same representation of $SU(N_f)$.

\subsection{Duality for $U(N)$ with $(N_f,\,N_a)$ flavors and $N_f>N_a+1$}

Brane constructions \cite{Hanany:1997vm} suggest that a duality could hold between $U$-type gauge theories in the presence of both $N_f$ fundamentals and $N_a$ antifundamentals. If we assume for definiteness that $N_f\geq N_a$, the duality suggested by the brane creation effect \cite{Hanany:1996ie} is between a $U(N)$ gauge theory with $(N_f,N_a)$ flavors and a $U(N_f-N)$ theory with $(N_a,N_f)$ flavors, similarly to the maximally chiral duality of 3d theories studied in \cite{Benini:2011mf}. For the sake of homogeneity with the rest of the section, we relabel the magnetic gauge representations by charge conjugation, so that the magnetic side becomes a $U(N_f-N)$ theory with $(N_f,N_a)$ flavors. We call $Q^i_\alpha$, $\tilde{Q}^\alpha_\ti$ the quark and antiquark superfields of the electric theory and $q_{i\beta}$, $\tilde q^{\beta \ti}$ those of the magnetic theory.
In order for mesonic branches of the putative dual theories to match, the magnetic theory must contain singlets $M^i_\ti$ dual to the electric mesons $\tilde{Q}^\alpha_\ti Q^i_\alpha $, coupled to the dual (anti)quark superfields by the usual superpotential $W_\text{dual}=\tilde{q}^{\beta \ti} M^i_\ti q_{i\beta}$.

Postponing a careful study of the field theories and of a possible duality between them to future work, we show here agreement between the partition functions of a $U(N)$ theory with $(N_f,N_a)$ flavors and $N_f > N$, and a $U(N_f - N)$ theory with $(N_f,N_a)$ flavors, gauge singlets $M^i_\ti$ and superpotential $W_\text{dual} = \Tr \tilde q M q$, up to a FI term for the Abelian global symmetries on whose possible origin we will comment below. We will also require $N_f > N_a+1$. In such cases, even in the absence of twisted masses and bare FI parameter, the classical Coulomb branch of the flat space theory is lifted, up to a set of isolated points, by one-loop logarithmic corrections to the effective FI parameter.
For both theories the flavor symmetry is $SU(N_f) \times SU(N_a) \times U(1)_A$ and the map of parameters is:
\be
\label{map parameters Nf Na}
\xi = \xi^D \;,\qquad \theta = \theta^D + (N_f-N_a)\pi \;,\qquad \tau_j = - \tau_j^D + \frac i2 \;,\qquad \tilde\tau_f = - \tilde\tau_f^D + \frac i2 \;.
\ee
We also impose the constraint $\frac1{N_f} \sum_j \tau_j = \frac1{N_a} \sum_f \tilde\tau_f$ (and similarly in the magnetic theory) because we parametrize the twisted masses as $\tau_j = m_A + \delta\tau_j$, $\tilde \tau_f = m_A + \delta\tilde\tau_f$ and similarly $\tau_j^D = m_A^D + \delta\tau_j^D$, $\tilde\tau_f^D = m_A^D + \delta\tilde\tau_f^D$ in terms of the Cartan of $SU(N_f)\times SU(N_a)\times U(1)_A$, with $0 = \sum_j \delta\tau_j = \sum_f \delta\tilde\tau_f = \sum_j \delta\tau_j^D = \sum_f \delta\tilde\tau_f^D$. The relations among masses are then:
\be
\delta\tau_j = - \delta\tau_j^D \;,\qquad \delta\tilde\tau_f = - \delta\tilde\tau_f^D \;,\qquad m_A = - m_A^D + \frac i2 \;.
\ee
Note that the effective mass of a meson $M_\ti^j=\tilde{Q}_\ti Q^j$ in the electric theory is $\tau_j+\ttau_\ti$, hence the effective mass parameter associated to $W_\text{dual}$ is $i$ which correctly corresponds to R-charge 2 and vanishing flavor charges.

As in the previous section, we compare the contributions to the partition functions from the vacua $\vec l$ and $\vec l^D$.
First, the equality
\be
\exp\Big[ 4\pi i\xi \sum\nolimits_{r\in\vec l} \, \tau_r \Big] = e^{4\pi i \xi^D \big( -N_f m_A^D + \frac i2 N \big)} \; \exp\Big[ 4\pi i \xi^D \sum\nolimits_{r \in \vec l^D} \tau_r^D \Big]
\ee
shows that the classical actions on the two sides agree, up to an extra factor on the magnetic side that could be ascribed to a FI term for the Abelian global symmetries $U(1)_A \times U(1)_R$. If there were a duality between the charge-conjugation-invariant theories with $N_f = N_a$, where such a term could not be generated, one could hope to derive global FI terms in the other cases by integrating out massive matter fields, as done in the 3d case in \cite{Benini:2011mf} starting from Aharony duality \cite{Aharony:1997gp}.

The one-loop determinants of W-bosons and fundamentals combine, as in (\ref{1-loop W fund combined}), into
$$
\prod_{r\in\vec l,\, j \not\in \vec l} \frac{\Gamma(-i\tau_j + i\tau_r)}{\Gamma(1 + i\tau_j - i\tau_r)}
$$
which is manifestly invariant under the duality. On the other hand the equality
$$
\prod_{r\in \vec l} \prod_{\ti=1}^{N_a} \frac{\Gamma(-i\tilde\tau_\ti - i\tau_r)}{\Gamma(1 + i\tilde\tau_\ti + i\tau_r)} = \bigg[ \prod_{r=1}^{N_f}\prod_{\ti=1}^{N_a} \frac{ \Gamma( -i \ttau_\ti -i\tau_r )} {\Gamma( 1 + i\ttau_\ti + i\tau_r )} \bigg] \bigg[ \prod_{r \in \vec l^D} \prod_{\ti=1}^{N_a} \frac{\Gamma( -i \tilde\tau_\ti^D -i \tau_r^D)}{\Gamma( 1 + i\tilde\tau_\ti^D + i \tau_r^D)} \bigg]
$$
shows that the one-loop of antifundamentals in the electric theory (LHS) equals the one-loop of gauge singlets $M^r_\ti$ (first term on the RHS) times the one-loop of antifundamentals (second term) in the magnetic theory.

Finally the partition function $Z_k$ for $k$ vortices in the $U(N)$ theory with $(N_f,N_a)$ flavors is
the term in (\ref{vortex partition function U(N)}) multiplying $z^k$:
$$
Z_k = \sum_{\substack{\vec k \,\in\, \bZ_{\geq 0}^N \\ |\vec k| = k}} \; \varepsilon^{(N_a-N_f) k} \; \prod_{r\,\in\,\vec l} \; \frac{ \prod_{f=1}^{N_a} \Big( \dfrac{\tilde m_f + a_r}\varepsilon \Big)_{k_r}} {k_r! \; \Big[ \prod_{\substack{j\,\in\,\vec l \\ j \neq r}} \Big( \dfrac{a_j - a_r}\varepsilon - k_r \Big)_{k_j} \Big] \; \Big[ \prod_{j\,\not\in\,\vec l} \Big( \dfrac{a_j - a_r}\varepsilon - k_r \Big)_{k_r} \Big] }
$$
where as usual $a_r = -i\tau_r$ and $\tilde m_f = -i\tilde\tau_f$.
Such expression is reproduced by the contour integral:
\be
Z_k = \frac1{\varepsilon^k k!} \int \bigg[ \prod_{j=1}^k \frac{d\varphi_j}{2\pi i} \bigg] \bigg[ \prod_{i<j}^k \frac{ (\varphi_i - \varphi_j)^2}{(\varphi_i - \varphi_j)^2 - \varepsilon^2} \bigg] \; \prod_{j=1}^k \; \frac{ \prod_{f=1}^{N_a} (\tilde m_f + \varphi_j)}
{\prod_{r \,\in\, \vec l}\: (\varphi_j - a_r) \, \prod_{s \,\not\in\, \vec l}\: (a_s - \varphi_j - \varepsilon)} \;.
\ee
The contours are defined in the same way as after (\ref{contour integral}). Repeating the manipulations of the previous section and moving the contours in the lower half-planes, it is straightforward to see that the vortex partition functions in the two dual theories match. For this to work it is important that $N_f > N_a+1$, otherwise there is a pole at infinity that prevents the contours to be harmlessly deformed.

When $N_f = N_a$ or $N_f = N_a+1$ one can check by direct evaluation that the partition functions $Z_k$ do not match under the map (\ref{map parameters Nf Na}), therefore if a duality exists at all, the map has to be modified somehow.

\section{Comparison with 3d partition functions}
\label{sec: dimensional reduction}

Localization of three-dimensional field theories with four supercharges and an R-symmetry on curved manifolds has been a popular subject of research lately \cite{Kapustin:2009kz, Jafferis:2010un, Hama:2010av, Imamura:2011su}, hence it is natural to wonder whether our localized partition function of two-dimensional $\cN=(2,2)$ theories on $S^2$ can be obtained as a limit of a three-dimensional localized partition function on a suitable manifold.%
\footnote{Another limit has been considered in \cite{Spiridonov:2011hf}.}

\subsection{$S^2\times S^1$}

A natural guess is to consider the supersymmetric partition function of a 3d $\cN=2$ theory on $S^2\times S^1$ \cite{Imamura:2011su,Krattenthaler:2011da,Kapustin:2011jm}, which computes a supersymmetric index of the theory, in the limit where the $S^1$ shrinks.
The localized partition function of a 3d $\cN=2$ theory with $U(1)_R$ symmetry placed on $S^2\times S^1$ as in \cite{Imamura:2011su} takes the form%
\footnote{For the sake of simplicity we neglect global non-R symmetries under which the matter fields can be charged. They can be instated in the obvious way.}
\be
\label{matrix_integral_index}
Z_{S^2\times S^1} = \frac1{|\cW|} \sum\nolimits_\fm \int \Big( \prod\nolimits_j \frac{d h_j}{2\pi} \Big) \, e^{-S_\text{class}(h, \fm)} \, Z_\text{gauge}(x,e^{ih},\fm) \, \prod\nolimits_\Phi Z_\Phi(x,e^{ih},\fm)\;,
\ee
where $x$ is the $U(1)_R$ fugacity, $\fm$ the quantized magnetic flux on $S^2$, and $e^{ih}$ the Wilson line on $S^1$ of the maximal torus of the gauge group. Note that the integration variable is the zero-mode of the gauge connection along $S^1$, which becomes $\eta$ in the reduction to two dimensions, while the real scalar $\sigma$ in the 3d vector multiplet is determined by the quantized magnetic flux. Therefore in the 2d limit the roles of $\sigma$ and $\eta$ (but not those of $D$ and $F_{12}$) will be interchanged compared to our setup.
The difference is due to the choice of Killing spinors made in \cite{Imamura:2011su}, which do not reduce to our Killing spinors \eqref{positive_Killing} upon reduction. Consequently the localization locus of \cite{Imamura:2011su} does not reduce to ours.

The contribution to the index of a chiral multiplet $\Phi$ of R-charge $q$, transforming in a representation $R_\Phi$ of the gauge group, can be rewritten as \cite{Krattenthaler:2011da,Kapustin:2011jm}
\be
\label{chiral_3dindex}
Z_{\Phi}(x,e^{ih},\fm)=\prod_{\rho\in R_\Phi} \left( x^{1-q}e^{-i\rho(h)}\right)^{\frac{|\rho(\fm)|}{2}} \frac{(e^{-i\rho(h)}x^{-\rho(\fm)+2-q};x^2)_\infty}{ (e^{i\rho(h)}x^{-\rho(\fm)+q};x^2)_\infty}\;,
\ee
where $(a;q)_n \equiv \prod_{k=0}^{n-1}(1-a q^k)$ is the $q$-Pochhammer symbol and as pointed out in \cite{Dimofte:2011py} we corrected $|\rho(\fm)|\to -\rho(\fm)$ in the $q$-Pochhammer symbols.
For the 2d limit in which the circle shrinks, we set
\be
\label{fugacities_chemical_potentials_2d_limit}
x^2\equiv e^{-\beta}\;,\qquad  h \equiv \beta r\eta
\ee
where $r$ is the radius of $S^2$ and $\beta r$ the circumference of $S^1$,
and send $\beta\to 0$ using%
\footnote{We thank V. Spiridonov and G. Vartanov for correcting a mistake in the original version of this formula.}
\be
\lim_{z\to 1} \frac{(z^s;z)_\infty}{(z^t;z)_\infty} (1-z)^{s-t}= \frac{\Gamma(t)}{\Gamma(s)} \;.
\ee
This implies
\be\label{limit_chiral}
\lim_{\beta\to 0} Z_{\Phi}(e^{-\beta/2},e^{i\beta r\eta},\fm) \,\beta^{\sum_{\rho\in R_\Phi}[2ir\rho(\eta)+1-q]}
= \prod_{\rho\in R_\Phi}\,\frac{ \Gamma( \frac q2 - ir\rho(\eta) - \frac{\rho(\fm)}2) }{\Gamma( 1 - \frac q2 + ir\rho(\eta)-\frac{\rho(\fm)}2)}
\ee
In order to obtain a well-defined $\beta\to 0$ limit and recover our $S^2$ result \eqref{Z matter} up to the replacement $\sigma\leftrightarrow \eta$, the extra power of $\beta$ in the LHS of \eqref{limit_chiral} is needed. While $\beta^{\sum_{\rho\in R_\Phi}(1-q)}$ gives a harmless overall normalization
of the partition function that can be neglected, the linear $\eta$-dependence in the power of $\beta$, which appears if the gauge group has central $U(1)$ factors, requires a more careful treatment which we now explain.

For each central $U(1)$ factor in the gauge group there is an associated topological symmetry under which monopole operators are charged. To simplify the following discussion, let us focus on a single pair of gauge and topological $U(1)$ symmetries.
The sphere index \eqref{matrix_integral_index} can be refined to include a dependence on the background vector multiplet for the topological $U(1)_J$ symmetry, in particular on the magnetic flux $\fm_J$ and the Wilson line $e^{i h_J}$. If one further introduces a mixed Chern-Simon term with level $k_{gJ}$ between the central gauge $U(1)_g$ and the global topological $U(1)_J$  symmetries in the classical action $S_{class}$ \cite{Kapustin:2011jm}, the $S^2\times S^1$ partition function \eqref{matrix_integral_index} contains an additional factor $e^{i k_{gJ} (h \fm_J + h_J \fm)}$.
In order to reproduce the LHS of \eqref{limit_chiral} for all chiral multiplets in the 2d limit $\beta\to 0$, we need to scale $\fm_J\to\infty$ so that $\beta \hat\xi_{3d}\equiv \beta k_{gJ} \fm_J \simeq -2  \log\beta \sum_{\Phi_i} g[\Phi_i] + \hat\xi_{2d}$, where $g[\Phi_i]$ is the $U(1)$ gauge charge of the chiral multiplet $\Phi_i$ and $\hat\xi_{2d}$ is finite. After this renormalization we are left with the 1-loop determinants of chiral multiplets on $S^2$ along with an extra factor of $e^{i \hat\xi_{2d} r \eta}$ which mimics our 2d FI term upon $\sigma\leftrightarrow \eta$. Remark that, despite appearances, $\hat\xi_{3d}$ and $\hat\xi_{2d}$ are \emph{not} FI parameters: in the localization locus of \cite{Imamura:2011su} the auxiliary field $D$ vanishes and supersymmetric CS terms acquire contributions only from gauge vectors; as a consequence one gets a dependence on $\eta$ rather than $\sigma$ on $S^2$.
Similarly one can reproduce a theta-term on $S^2$ if $\eta_J\to\infty$ so that
$k_{gJ} h_J = \beta k_{gJ} r \eta_J \sim \theta $ stays finite in the limit.

The contribution of a vector multiplet is given by a product over roots
\be
\label{vector_3dindex}
Z_{gauge}(x,e^{ih},\fm)=\prod_{\alpha\in G} x^{-\frac{|\alpha(\fm)|}{2}}\left( 1-e^{i\alpha(h)}x^{|\alpha(\fm)|}\right) \;.
\ee
Up to a power of $\beta$ that we can again neglect, it reduces in the 2d limit to
$$
\prod_{\alpha>0} \Big(\frac{\alpha(\fm)^2}{4}+r^2 \alpha(\eta)^2 \Big)\;,
$$
which coincides with \eqref{Z gauge} upon $\sigma\leftrightarrow \eta$. Finally, the integral and sum over topological sectors in \eqref{matrix_integral_index} reduces to the one of \eqref{matrix integral} upon $\sigma\leftrightarrow \eta$.

\subsection{$L(p,1)$}

Another way of obtaining $S^2$ from a 3-manifold is to consider the lens space $L(p,1)=S^3/\bZ_p$, where the quotient is along the Hopf fiber, and send $p\to\infty$.

Consider the vielbein $E^1=\frac R2 d\theta$, $E^2=\frac R2 \sin\theta\, d\varphi$, $ E^3=\frac R{2p}(d\psi+p \cos\theta\, d\varphi)$ for $L(p,1)$ of radius $R$, and the vielbein $e^1=r \,d\theta$, $e^2=r \sin\theta\, d\varphi$ for $S^2$ of radius $r=R/2$.
In these coordinates the positive Killing spinor equation on $L(p,1)$ ($\nabla_M^{L(p,1)} \epsilon = \frac i{2R} \gamma_M^{L(p,1)} \epsilon$) implies that $\epsilon$ is independent of the Hopf fiber coordinate $\psi$ ($\partial_\psi \epsilon = 0$) along with the positive Killing spinor equation on $S^2$ ($\nabla_m^{S^2} \epsilon = \frac i{2r} \gamma_m^{S^2} \epsilon$). The fact that $\epsilon$ is independent of $\psi$ guarantees that we can take the orbifold $S^3/\bZ_p$ preserving some SUSY. The associated Killing vector $v^A = \epsilon^\dag \gamma^A \epsilon$ reads $v = \frac 2R \partial_\varphi$ in coordinates and $v^A = (v^a, s)$ in vielbein basis, where $v^a$ and $s$ are as in (\ref{Killing_vector}). In fact the vector field $v$, of constant norm, represents translations along a different Hopf fiber than the one used in the reduction $S^3/\bZ_p$. In the limit $p\to\infty$, $v^A$ reduces to the vector field $v^a$ (\ref{Killing_vector}) on $S^2$, which has fixed points at the poles.

We would then like to compare our partition function of a 2d $\cN=(2,2)$ theory of vector and chiral multiplets on $S^2$ with the $p\to\infty$ limit of the partition function of a 3d $\cN=2$ theory on $S^3/\bZ_p$ \cite{Benini:2011nc,Alday:2012au}.
The localization on $L(p,1)$ is similar to the one on $S^3$  \cite{Kapustin:2009kz,Jafferis:2010un,Hama:2010av} apart from global aspects. The quotient introduces a sum over topological sectors, because one can turn on a $\bZ_p$-valued Wilson line. This torsion is specified by $\fm=\mathrm{diag}(m_1,\dots,m_N)$, where $m_i$ are integers defined modulo $p$ and $N$ is the rank of the gauge group.
As on $S^3$, all other fields are set to zero except for the real scalar $\sigma$ in the vector multiplet and the auxiliary field $D=-\sigma/R$, with $R$  the radius of $S^3$. Moreover $R \sigma$ is conjugate, by the gauge transformation associated to $\fm$, to a constant matrix $a$ commuting with $\fm$.
The localized partition function on $L(p,1)$ then takes the form
\be
\label{matrix_integral_Lens}
Z_{L(p,1)} = \frac1{|\cW|} \sum\nolimits_\fm \int \Big( \prod\nolimits_j \frac{d a_j}{2\pi} \Big) \, e^{-S_\text{class}(p,a, \fm)} \, Z_\text{gauge}(p,a,\fm) \, \prod\nolimits_\Phi Z_\Phi(p,a,\fm)\;,
\ee
where $a_j$ are the eigenvalues of the matrix $a$.

In the presence of a non-Abelian CS term at level $k$ for the gauge group, the Euclidean classical action is \cite{Alday:2012au}
\be
S_\text{class}(p,a, \fm) = -i \pi \frac{k}{p} \left[ \mathrm{Tr}(a^2) - \mathrm{Tr}(\fm^2) \right] \;.
\ee
If in the $p\to\infty$ limit we scale $k\to\infty$ keeping $k/p$ fixed, and we end up with the contribution of the classical Lagrangian $\cL_{CS}$ \eqref{CS Lagrangian} arising from a quadratic twisted superpotential. The argument applies to mixed CS interactions as well, in particular to a mixed CS interaction between a gauge $U(1)$ and the topological $U(1)_J$, which accounts for a FI term in three dimensions. Taking an analogous scaling limit leads to the contribution of the FI Lagrangian \eqref{FI Lagrangian} on $S^2$. The 2d FI parameter descends from the 3d FI parameter, which is the background real scalar for the vector multiplet of $U(1)_J$: $\xi_{2d}\sim \xi_{3d}/p = a_J/p$; the theta angle descends from the 3d torsion flux for the background $U(1)_J$: $\theta\sim \fm_{J,\,3d}/p$.

The one-loop determinant of the vector multiplet, up to a factor that cancels the Vandermonde determinant, is
\be \label{vector_Lens}
Z_{gauge}(p,a,\fm)=\prod_{\alpha>0} \sinh\Big[ \frac\pi p \big( \alpha(a)+i\alpha(\fm) \big) \Big]
\sinh\Big[ \frac\pi p \big(\alpha(a)-i\alpha(\fm) \big) \Big]\;.
\ee
For $p\to\infty$ it reduces to
\be
Z_{gauge}(a,\fm)=\prod_{\alpha>0} \Big( \frac{2\pi}p \Big)^2 \Big[ \Big( \frac{\alpha(a)}2 \Big)^2 + \Big( \frac{\alpha(\fm)}2 \Big)^2 \Big]
\ee
which agrees, up to a $p$ dependent factor that we renormalize away, with \eqref{Z gauge} under the identification $a=R\sigma=2r\sigma$. Note the factor of 2 relating the radius $R$ of $S^3$ and the radius $r$ of the $S^2$ base of the Hopf fibration. In the limit the discrete Wilson line becomes ordinary magnetic flux on $S^2$.

The one-loop determinant of a chiral multiplet $\Phi$ of R-charge $q$,
transforming in a representation $R_\Phi$ of the gauge group, is
\be \label{chiral_Lens}
Z_{\Phi}(p,a,\fm)=\prod_{\rho\in R_\Phi} \prod_{l=0}^\infty \bigg( (-1)^l \, \frac{l+2-q+ i \rho(a)}{l+q- i \rho(a)} \bigg)^{N_\rho(l)}\;,
\ee
where $N_\rho(l)$ is the number of half-integers $n \in\{ -\frac l2, -\frac l2 + 1, \dots, \frac l2 - 1, \frac l2 \}$ satisfying $2 n = \rho(\fm)$ mod $p$. For $p\to\infty$ this becomes $n = \rho(\fm)/2$, therefore $N_\rho(l)=1$ if $l\geq |\rho(\fm)|$ and $\rho(\fm)-l$ is even, $N_\rho(l)=0$ otherwise. In the case where there are solutions, we can set $\frac{l}{2}=\frac{|\rho(\fm)|}{2}+k$, with $k=0,1,2,\dots$ Then \eqref{chiral_Lens} reduces to
$$
\prod_{\rho\in R_\Phi} \prod_{k=0}^\infty \; (-1)^{|\rho(\fm)| + 2k} \; \frac{k+\frac{|\rho(\fm)|}{2}+1-\frac{q}{2}+i\rho(\frac{a}{2})}{k +\frac{|\rho(\fm)|}{2}+\frac{q}{2}-i\rho(\frac{a}{2})}\;.
$$
The result agrees with our one-loop determinant on $S^2$ before regularization \eqref{matter determinant not regularized} with $a=2r\sigma$, apart from an $\fm$-dependent sign.%
\footnote{The sign mismatch is of similar nature to the one pointed out in \cite{Dimofte:2011py} for the 3d index of \cite{Imamura:2011su}. Since Killing spinors and localization loci on $L(p,1)$ and $S^2$ are directly related by reduction, it is conceivable that this mismatch could be resolved by a more careful analysis of these signs and their regularization.}

The same result can be obtained as a vanishing $S^1$ limit of the $S^2\times S^1$ partition function found in \cite{Benini:2011nc} as the $p\to\infty$ limit of an $L(p,1)\times S^1$ partition function, which differs from the index of \cite{Imamura:2011su} in the localization locus and the resulting classical action.

\section{Discussion}
\label{sec: discussion}

We conclude with some open questions and directions for future work. The question that motivated the present work was to understand from a more fundamental point of view the factorization, observed in \cite{Pasquetti:2011fj} in two examples, of the $S^3_b$ partition function \cite{Hama:2011ea} into products of K-theoretic vortex and antivortex partition functions. Although we have succeeded in explaining a similar equality in two dimensions, the three-dimensional case remains to be addressed. A natural guess would be to consider exactly the same $\cQ$-exact deformation $\cL_H$ (\ref{L_H}) but in three dimensions.

We can gain some intuition by considering how the north and south pole of $S^2$ uplift to $S^3$ or the lens space $L(p,1)$. The preimages of the poles under the projection $S^3\to S^2$ are the Hopf fibers over them, forming two linked great circles in $S^3$. We expect the point-like (anti)vortex configurations at the poles to lift to similar singular configurations on $S^3$ or $L(p,1)$, possibly with KK momentum along the extra circles. The 2d (anti)vortex partition functions would then lift to K-theoretic (anti)vortex partition functions. When the round $S^3$ is squashed into the ellipsoid $S^3_b$ defined by the equation $b^2|z_1|^2 + b^{-2}|z_2|^2=1$ in $\bC^2$, the radii of the two circles become $b$ and $1/b$ respectively (in units of the radius $R$ of $S^3$): they match the sizes of the extra circles in the K-theoretic vortex and antivortex partition functions of \cite{Pasquetti:2011fj}; we expect a similar factorization in terms of K-theoretic (anti)vortex partition functions to occur for $L(p,1)$, and similarly for $S^2 \times S^1$. It would be interesting to make these considerations concrete.

Our result that the matrix integral equals the vortex partition function times the antivortex partition function (weighted by semiclassical factors and summed over isolated points on the Higgs branch) suggests that the matrix integral can be used to determine the vortex partition function for theories where other means are not available. In particular, the computations in the literature of $Z_\text{vortex}$ are based either on an ADHM-like construction of the moduli space \cite{Hanany:2003hp} derived from string theory, which however has applicability limited to some gauge groups and matter representations, or the moduli matrix approach \cite{Eto:2005yh, Fujimori:2012ab}. The matrix integral seems a simple and flexible alternative method.

A useful development of our work would be to include twisted chiral multiplets, possibly with periodic imaginary part, besides chiral multiplets. That would allow a more thorough exploration of mirror symmetry. We hope to report progress in that direction in the near future.
It would also be interesting to include local or non-local operators in our path integral computation, for instance loops or extra vortex operators.

The partition function of an $\cN=2$ gauge theory on $S^4$ has proven very useful to uncover surprising aspects of the dynamics of multiple M5-branes -- whose reduction on a Riemann surface $\Sigma$ can give either an $\cN=2$ \cite{Gaiotto:2009we} or an $\cN=1$ \cite{Benini:2009mz, Bah:2011vv, Bah:2012dg} four-dimensional theory depending on the twist, some aspects being discussed in \cite{Gaiotto:2009hg, Benini:2009gi, Chacaltana:2010ks} -- most notably the AGT correspondence \cite{Alday:2009aq}. A similar role is being played by the $S^3$ and $S^2 \times S^1$ partition functions \cite{Dimofte:2011ju, Dimofte:2011py}, hence one could expect our result on $S^2$ to fit into the pattern. For instance, one can obtain a 2d $\cN=(4,4)$ quiver gauge theory description of M5-branes on $\Sigma \times T^2$ by first compactifying on $\Sigma \times S^1$ (that in general gives a non-Lagrangian theory), then using the 3d mirror symmetry of \cite{Benini:2010uu} and finally compactifying on $S^1$.%
\footnote{Alternatively one could simply consider 4d $\cN=4$ SYM on $\Sigma$ \cite{Bershadsky:1995vm}, which however gives a 2d non-linear sigma model and does not fall within our framework.}
 In this context an adaptation of the anomaly computation of \cite{Alday:2009qq} might be useful to relate the 2d theory to some four-dimensional theory.

\section*{Acknowledgments}

We thank Amihay Hanany and Sara Pasquetti for valuable discussions, Giulio Bonelli, Tudor Dimofte, Lotte Hollands and Alessandro Tanzini for correspondence, and Nikolay Bobev and Cyril Closset for comments on the manuscript.
FB thanks the Newton Institute for Mathematical Sciences in Cambridge (UK), where this work was initiated, for hospitality.
The work of FB is supported in part by the DOE grant DE-FG02-92ER40697. The work of SC was supported in part by the STFC Consolidated Grant ST/J000353/1.

\appendix

\section{Spinor conventions}
\label{app: spinor conventions}

In two Euclidean dimensions the minimal spinor is a complex one-dimensional Weyl spinor, or a real two-dimensional Majorana spinor, however we will use complex two-dimensional Dirac spinors. We choose Dirac matrices in the frame basis as the Pauli matrices: $\gamma_{1,2,3} = \smat{0 & 1 \\ 1 & 0}, \smat{ 0 & -i \\ i & 0}, \smat{ 1 & 0 \\ 0 & -1}$ with $\gamma_3$ being the chirality matrix. We will use $\mu,\nu=1,2$ as curved space indices, $a,b=1,2$ or $i,j=1,2,3$ as flat space indices, and $\alpha,\beta=1,2$ as spinor indices. The matrices satisfy $\gamma_\mu \gamma_\nu = g_{\mu\nu} + i \varepsilon_{\mu\nu} \gamma_3$. They are Hermitian and $\gamma_{1,2,3}^\trans = \{\gamma_1, - \gamma_2, \gamma_3\}$, so that the charge conjugation matrix $C$ that solves $C \gamma_i C^{-1} = - \gamma_i^\trans$ is $C = -i \varepsilon_{\alpha\beta} = \gamma_2$, with $C = C^\dag = C^{-1}$.

A rotationally invariant combination of spinors is $\epsilon^\dag \lambda$. On the other hand, the charge conjugate spinor is
\be
\epsilon^c \equiv C \epsilon^*
\ee
which transforms as $\epsilon$. Therefore another rotationally invariant combination is $\epsilon^\trans C \lambda = \epsilon^{c\,\dag} \lambda$. Notice that $(\epsilon^c)^c = - \epsilon$, therefore the Majorana condition is: $\epsilon^c = \gamma_3 \epsilon$. It is easy to check that $\gamma_3\epsilon = \pm \epsilon$ implies $\gamma_3 \epsilon^c = \mp \epsilon^c$, so that charge conjugation flips the chirality and there are no Majorana-Weyl spinors.

When dealing with anticommuting spinors, we multiply them as
\be
\epsilon \lambda \equiv \epsilon^\alpha C_{\alpha\beta} \lambda^\beta = \lambda \epsilon
\ee
keeping $C$ implicit. When we prefer not to keep $C$ implicit, we specify either $^\dag$ or $^\trans$, so that we can write: $\epsilon \lambda = \epsilon^\trans C \lambda$. On the other hand, barred spinors like $\bar\epsilon$ are simply independent spinors, with no implicit operation on them.
Including gamma matrices in the product, we get for anticommuting spinors:
\be
\epsilon \gamma_i \lambda = - \lambda \gamma_i \epsilon \;,\qquad\qquad \epsilon \gamma_{ij} \lambda = - \lambda \gamma_{ij} \epsilon \;.
\ee
Moreover $\gamma^\mu \gamma_\mu = 2$ and $\gamma^\mu \gamma_\rho \gamma_\mu = 0$.

The Fierz identity for commuting spinors is:
\be
(\epsilon^\dag \lambda_1) \, \lambda_2 = \frac12 \big[ \lambda_1 \, (\epsilon^\dag \lambda_2) + \gamma_3 \lambda_1 \, (\epsilon^\dag \gamma_3 \lambda_2) + \gamma_\rho \lambda_1 \, (\epsilon^\dag \gamma^\rho \lambda_2) \big] \;.
\ee
This is still true if $\epsilon$ is $\bC$-valued and commuting, while $\lambda_{1,2}$ are possibly matrix-valued and anticommuting. The Fierz identity for anticommuting spinors is:
\be
(\bar\epsilon \lambda_1) \, \lambda_2 = - \frac12 \big[ \lambda_1 \, (\bar\epsilon \lambda_2) + \gamma_3 \lambda_1 \, (\bar\epsilon \gamma_3 \lambda_2) + \gamma_\rho \lambda_1 \, (\bar\epsilon \gamma^\rho \lambda_2) \big]
\ee
where all spinors are anticommuting, $\bar\epsilon$ is $\bC$-valued while $\lambda_{1,2}$ can be matrix valued. When $\lambda_1 = \lambda_2$ and matrix-valued, one can derive the two following relations:
\bea
0 &= [\bar\epsilon\lambda, \lambda] - [\bar\epsilon \gamma_3 \lambda, \gamma_3\lambda] - [\bar\epsilon \gamma^\mu \lambda, \gamma_\mu \lambda] \\
0 &= (\bar\lambda\epsilon_2)(\bar\lambda \epsilon_1) - (\bar\lambda \gamma_3 \epsilon_2)(\bar\lambda \gamma_3 \epsilon_1) + (\bar\lambda \gamma_\mu \epsilon_1)(\bar\lambda \gamma^\mu \epsilon_2) \;.
\eea

\section{Killing spinors on $S^2$}
\label{app: Killing spinors}

The metric on $S^2$ of radius $r$ is:
\be
ds^2 = r^2( d\theta^2 + \sin^2\theta\, d\varphi^2) = r^2 \mu^a \mu^a = e^a e^a \;.
\ee
We choose:
\be
\mu^1 = d\theta \;,\qquad\qquad \mu^2 = \sin\theta\, d\varphi \;.
\ee
The equation $de^a + \omega^{ab} \wedge e^b = 0$ determines $\omega^{12} = - \cos\theta\, d\varphi$.

The Killing spinor equation is $D\epsilon = e^a \gamma^a \tilde \epsilon$. It turns out \cite{Fujii:1985bg, Lu:1998nu} that on $S^2$ there is a basis of Killing spinors of definite ``positivity'' that solve $D_\mu \epsilon_\pm = \pm \frac i{2r} \gamma_\mu \epsilon_\pm$. Explicitly:
\be
D\epsilon_\pm \equiv d\epsilon_\pm + \frac14 \omega^{ab} \gamma^{ab} \epsilon_\pm \stackrel{!}{=} \pm \frac i{2r} e^a \gamma^a \epsilon_\pm \;.
\ee
The space of ``positive'' Killing spinors is complex two-dimensional:
\be
\epsilon_+ = C_1 e^{-i\frac\varphi2} \mat{ \sin\theta/2 \\ -i \cos\theta/2} + C_2 e^{i \frac\varphi2} \mat{ \cos\theta/2 \\ i \sin\theta/2} \;.
\ee
Given any $\epsilon_+$, the spinor $\epsilon_- \equiv \gamma_3 \epsilon_+$ is a ``negative'' Killing spinor, therefore the space of Killing spinors on $S^2$ is complex four-dimensional. Notice that Killing spinors of definite positivity do not have definite chirality and viceversa.

Out of a Killing spinor $\epsilon_+$ one can construct a Killing vector $v^\mu$ and a function $s$:
\be
\label{Killing_vector}
v^\mu = \epsilon_+^\dag \gamma^\mu \epsilon_+ \;,\qquad\qquad s = \epsilon_+^\dag \gamma_3 \epsilon_+ \;.
\ee
The Killing vector satisfies $D^{(\mu} v^{\nu)} = 0$, and is related to $s$ by: $\varepsilon_{\mu\nu} D^\mu v^\nu = \frac2r s$, $D_\mu s = - \frac1r \varepsilon_{\mu\nu} v^\nu$. To perform localization we will consider $\epsilon_+ = e^{i\varphi/2} (\cos\theta/2,\, i\sin\theta/2)$. In this case, in vielbein basis, $v^a = (0, \sin\theta)$ and $s = \cos\theta$. Moreover $\epsilon_+^\trans C \epsilon_+ = 0$ where $C = \gamma_2$ is the charge conjugation matrix.

One could also consider the spinors $\epsilon '_\pm = (1+i\gamma_3)\epsilon_\pm$. They satisfy the equation:
\be
D_\mu \epsilon'_\pm = \pm \frac1{2r} \gamma_\mu \gamma_3 \epsilon'_\pm \;.
\ee

\section{Localizing supercharges}
\label{app: supercharges}

The commuting operator $\delta = \delta_\epsilon + \delta_{\bar\epsilon} = \epsilon^\alpha Q_\alpha + \bar\epsilon^\alpha Q^\dag_\alpha$, where $\epsilon, \bar\epsilon$ are anticommuting positive Killing spinors, defines the supercharges $Q_\alpha, Q^\dag_\alpha$. Multiplying them back by commuting Killing spinors, we obtain anticommuting operators. We choose $\epsilon_+$ such that $\epsilon_+^\dag \epsilon_+ = 1$ and $\epsilon_+^\trans C \epsilon_+ = 0$, and form $Q = \epsilon_+^\alpha Q_\alpha$, $Q^\dag = \epsilon_+^{c\,\alpha} Q^\dag_\alpha = - (\epsilon_+^\dag C)^\alpha Q^\dag_\alpha$. For convenience, we also rewrite barred spinors as $\bar\lambda = C(\lambda^\dag)^\trans$, recalling that $\lambda^\dag$ is independent from $\lambda$, and drop the suffix from $\epsilon_+$ recalling that it is here commuting. We get for the gauge multiplet:
\bea
\label{Q-action gauge}
Q A_\mu &= \frac i2 \lambda^\dag \gamma_\mu \epsilon       \qquad\qquad\qquad\qquad
Q^\dag A_\mu = \frac i2 \epsilon^\dag \gamma_\mu \lambda    \qquad\qquad\qquad\qquad
Q \lambda^\dag = 0 \\
Q \sigma &= - \frac12 \lambda^\dag \epsilon       \!\quad\qquad\qquad\qquad\qquad
Q^\dag \sigma = - \frac12 \epsilon^\dag \lambda     \:\qquad\qquad\qquad\qquad
Q^\dag \lambda = 0 \\
Q \eta &= \frac i2 \lambda^\dag \gamma_3 \epsilon    \;\;\;\qquad\qquad\qquad\qquad
Q^\dag \eta = \frac i2 \epsilon^\dag \gamma_3 \lambda \\
Q \lambda &= i\gamma_3\epsilon \Big( F_{12} - \frac \eta r + i [\sigma,\eta]\Big) - \epsilon \Big( D + \frac\sigma r \Big) + i\gamma^\mu \epsilon \, D_\mu \sigma - \gamma_3 \gamma^\mu \epsilon \, D_\mu \eta \\
Q^\dag \lambda^\dag &= - i \epsilon^\dag \gamma_3 \Big( F_{12} - \frac\eta r - i [\sigma,\eta] \Big) + \epsilon^\dag \Big( D + \frac\sigma r \Big) + i \epsilon^\dag \gamma^\mu D_\mu \sigma + \epsilon^\dag \gamma_3 \gamma^\mu D_\mu \eta \\
Q D &= - \frac i2 D_\mu \lambda^\dag \gamma^\mu \epsilon + \frac i2 [\lambda^\dag \epsilon, \sigma] + \frac12 [\lambda^\dag \gamma_3\epsilon, \eta] + \frac1{2r} \lambda^\dag \epsilon \\
Q^\dag D &= \frac i2 \epsilon^\dag \gamma^\mu D_\mu \lambda + \frac i2 [\sigma, \epsilon^\dag \lambda] + \frac12 [\eta, \epsilon^\dag \gamma_3 \lambda] + \frac1{2r} \epsilon^\dag \lambda \\
\eea
and for the chiral multiplet:
\bea
\label{Q-action matter}
Q \phi &= 0 &
Q^\dag \phi &= - \epsilon^\dag \psi \\
Q \phi^\dag &= \psi^\dag \epsilon &
Q^\dag \phi^\dag &= 0 \\
Q \psi &= \Big( i \gamma^\mu D_\mu \phi + i \sigma \phi + \gamma_3 \eta \phi - \frac q{2r} \phi \Big) \epsilon &
Q \psi^\dag &= - \epsilon^\trans C F^\dag \\
Q^\dag \psi^\dag &= \epsilon^\dag \Big( -i \gamma^\mu D_\mu \phi^\dag + i \phi^\dag\sigma + \gamma_3 \phi^\dag \eta - \frac q{2r} \phi^\dag \Big) &
Q^\dag \psi &= C \epsilon^* F \\
Q F &= \epsilon^\trans C \Big( i \gamma^\mu D_\mu \psi - i \sigma \psi + \gamma_3 \eta \psi + \frac q{2r} \psi - i\lambda \phi \Big) &
Q^\dag F &= 0 \\
Q^\dag F^\dag &= \Big( -i D_\mu \psi^\dag \gamma^\mu -i \psi^\dag\sigma + \psi^\dag \gamma_3\eta + \frac q{2r} \psi^\dag + i \phi^\dag\lambda^\dag \Big) C\epsilon^* \qquad &
Q F^\dag &= 0 \;.
\eea
The supercharges $Q,Q^\dag$ satisfy the algebra $\{Q,Q^\dag\} = M + \frac R2$, $Q^2 = Q^{\dag\,2} = 0$, up to a gauge transformation $\Lambda$, where $M$ is the angular momentum that generates rotations along the Killing vector field $v_\mu = \epsilon^\dag \gamma_\mu \epsilon$ and $R$ is R-charge. For instance:
\bea
\{Q, Q^\dag\} A_\mu &= - i v^\nu F_{\nu\mu} + D_\mu \Lambda &
\{Q,Q^\dag\} \phi &= -i v^\mu D_\mu \phi + i \Lambda \phi + \frac q{2r} \phi \\
\{Q, Q^\dag\} \sigma &= - i v^\mu D_\mu \sigma + i [\Lambda,\sigma] &
\{Q,Q^\dag\} \psi &= -i v^\mu D_\mu \psi + \frac s{2r} \gamma_3 \psi + i \Lambda \psi + \frac{q -1}{2r} \psi \\
\{Q, Q^\dag\} \eta &= -i v^\mu D_\mu \sigma + i [\Lambda, \eta] &
\{Q, Q^\dag\} F &= - i v^\mu D_\mu F + i \Lambda F + \frac{q-2}{2r} F \\
\{Q, Q^\dag\} \lambda &= -i v^\mu D_\mu \lambda + \frac s{2r} \gamma_3 \lambda + i [\Lambda,\lambda] \!\!\!\! & - \frac1{2r} \lambda \;. \quad &
\eea
In particular $\{Q,Q^\dag\} = M + \frac R2 + i\Lambda$, with $\Lambda = - \sigma + i \epsilon^\dag \gamma_3 \epsilon\, \eta$. Finally $\cQ = Q + Q^\dag$.

\section{One-loop determinants}
\label{app: 1-loop determinants}

We compute here the one-loop determinant for various kinetic operators appearing in the main text.  Before doing that, let us recall a few facts about spin spherical harmonics on a background.

\subsection{Spin spherical harmonics}

On a sphere $S^2$ of radius $r$, the standard scalar spherical harmonics $Y^0_{j,j_3}$ are the eigenfunctions of the Laplacian
\be
\nabla^2 = \frac1{r^2} \Big( \frac1{\sin\theta} \partial_\theta (\sin\theta\, \partial_\theta) + \frac1{\sin^2\theta} \partial^2_\varphi \Big) \;.
\ee
They are parametrized by $j \in \bN$ and $|j_3| \leq j$, $j_3 \in \bZ$ with eigenvalues $-j(j+1)/r^2$.

Now consider a particle with spin $s_z$ (representations of the spin group $SO(2)$ are one-dimensional) moving in a background magnetic flux $F$ with $\frac1{2\pi} \int F = \fm$, the particle transforming as the weight $\rho$ of some representation. Since $A = \frac\fm2 \omega^{12}$, the covariant derivative is
\be
D_\mu = \partial_\mu + i s_z \omega_\mu^{12} - i \frac{\rho(\fm)}2 \omega^{12}_\mu \;,
\ee
therefore the particle has effective spin $s = s_z - \frac{\rho(\fm)}2$. Let us consider the (anti)holomorphic derivatives
\be
D_\pm = \frac{D_1 \mp i D_2}2 = \frac1{2r} \Big( \partial_\theta \mp \frac i{\sin\theta} \partial_\varphi \Big) \mp \frac s{2r} \, \frac{\cos\theta}{\sin\theta} \;.
\ee
$D_+$ maps a spin $s$ to a spin $s+1$ harmonic, while $D_-$ reduces $s$ by one unit. In fact
\be
D^\mu D_\mu = 2 \{D_+, D_-\} \;,\qquad\qquad [D_+, D_-] = - \frac s{2r^2} \;.
\ee
To be very explicit about the spin of the operators, $D^2_s = 2(D_+^{(s-1)} D_-^{(s)} + D_-^{(s+1)} D_+^{(s)})$.

The eigenfunctions of $D^2_s$ are the spin spherical harmonics $Y^s_{j,j_3}$ with $|s|,|j_3| \leq j$ and $j-s \in \bN$, $j - j_3 \in \bN$, and the eigenvalues are:
\be
r^2 D^2 Y^s_{j,j_3} = - \big[ j(j+1) - s^2 \big] \, Y^s_{j,j_3} \;.
\ee
For integer $s$ they can be constructed from the scalar harmonics: $Y^s_{j,j_3} \propto (D_+)^s Y^0_{j,j_3}$ for $0 \leq s \leq j$ and $Y^s_{j,j_3} \propto (D_-)^s Y^0_{j,j_3}$ for $-j \leq s \leq 0$, while we get zero if $|s|>j$. If $s \in \bN + \frac12$ we can start from $Y^{1/2}_{1/2,1/2} = e^{i\varphi/2} \cos\theta/2$. It will be useful to consider also the operators:
\be
D_+ D_- Y^s_{j,j_3} = - \frac{j(j+1) - s(s-1)}{4r^2} \, Y^s_{j,j_3} \;,\qquad D_- D_+ Y^s_{j,j_3} = - \frac{j(j+1) - s(s+1)}{4r^2} \, Y^s_{j,j_3} \;.
\ee

\subsection{Gauge one-loop determinant}

Consider the action
\be
\begin{split}
\cL = \frac12 \Big( \tilde F_{12} - \frac{\tilde\eta}r \Big)^2 + \frac12 \Big(\tilde D + \frac{\tilde\sigma}r \Big)^2 &+ \frac12 \big( D_\mu \tilde\sigma - i[\tilde A_\mu, \sigma_0] \big)^2 + \frac12 \big( D_\mu \tilde\eta - i [\tilde A_\mu, \eta_0])^2 \\
&- \frac12 \big( [\tilde\sigma,\eta_0] + [\sigma_0, \tilde\eta] \big)^2 - \bar c D^\mu D_\mu c - \frac1{2\xi} (D^\mu \tilde A_\mu)^2\;,
\end{split}
\ee
quadratic in the fluctuations $(\tilde A_\mu, \tilde\sigma, \tilde \eta, \tilde D, \bar c, c)$. First, the term $\frac12 \big( \tilde D + \frac{\tilde\sigma}r \big)^2$ integrated over $\tilde D$ gives determinant 1, and we discard it.
Then let us look at $(\tilde A_\mu, \tilde\sigma, \tilde\eta)$. We write the vectors in holomorphic coordinates, using
\be
g_{\mu\nu} = \mat{ 0 & \frac12 \\ \frac12 & 0 } \;,\qquad g^{\mu\nu} = \mat{ 0 & 2 \\ 2 & 0} \;,\qquad \varepsilon_{\mu\nu} = \mat{ 0 & \frac i2 \\ - \frac i2 & 0} \;,\qquad \varepsilon^{\mu\nu} = \mat{ 0 & -2i \\ 2i & 0}
\ee
and $V_\pm = (V_1 \mp i V_2)/2$. Once we go to holomorphic coordinates, $\tilde A_\pm^\dag = \tilde A_\mp$.
We decompose all modes along the weights $\rho$ of the adjoint representation. The Cartan generators are decoupled from everything else at quadratic order: since their one-loop determinant will not depend on the background $A_\mu^{(0)}, \sigma_0, \eta_0$, we discard them, being left with a decomposition along the roots $\alpha$ of $G$.

Let us choose the gauge $\xi = -1$. Then the Lagrangian can be written in matrix notation as $(\tilde A_+^\dag, \tilde A_-^\dag, \tilde\eta, \tilde\sigma)^\trans \,\cM\, (\tilde A_+, \tilde A_-, \tilde\eta, \tilde\sigma)$ where the matrix $\cM$ is:
\be
\label{matrix gauge operator}
\cM = \mat{ - 4 D_+ D_- + \sigma_0^2 + \eta_0^2 & 0 & -i(\eta_0 + \frac1r)D_+ & -i\sigma_0 D_+ \\
0 & -4 D_- D_+ + \sigma_0^2 + \eta_0^2 & -i(\eta_0 - \frac1r)D_- & -i\sigma_0 D_- \\
-i (\eta_0 + \frac1r)D_- & -i(\eta_0 - \frac1r)D_+ & - \frac12 D^2 + \frac12 \sigma_0^2 + \frac1{2r^2} & - \frac12 \eta_0 \sigma_0 \\
-i\sigma_0 D_- & - i \sigma_0 D_+ & - \frac12 \eta_0 \sigma_0 & - \frac12 D^2 + \frac12 \eta_0^2
}
\ee
This matrix is Hermitian.
To compute the determinant we need a basis of eigenfunctions. We expand $\tilde\eta$, $\tilde\sigma$ on $Y^s_{j,j_3}$, $\tilde A_+$ on $Y^{s+1}_{j,j_3}$ and $\tilde A_-$ on $Y^{s-1}_{j,j_3}$, with $s = - \frac{\alpha(\fm)}2$. We also introduce the notation:
\be
s_\pm = \sqrt{j(j+1) - s(s\pm 1)} \;.
\ee
If we normalize the harmonics as $|Y^s_{j,j_3}|^2 = 1$, then the commutation relations fix the relations among them via $D_\pm$, up to a phase that we choose as follows:
\bea
D_+ Y^s_{j,j_3} &= \frac{s_+}{2r} Y^{s+1}_{j,j_3} \qquad\qquad &
D_- Y^{s+1}_{j,j_3} &= - \frac{s_+}{2r} Y^s_{j,j_3} \\
D_- Y^s_{j,j_3} &= - \frac{s_-}{2r} Y^{s-1}_{j,j_3} \qquad\qquad &
D_+ Y^{s-1}_{j,j_3} &= \frac{s_-}{2r} Y^s_{j,j_3} \;.
\eea
The formulae on the right coincide with those on the left if we shift $s$ by one.

We can now substitute in the matrix, and consider all possible values of $j$.
For $j \geq \frac{|\alpha(\fm)|}2 + 1$ all three harmonics $Y^{s\pm1}_{j,j_3}$, $Y^s_{j,j_3}$ exist: the determinant of (\ref{matrix gauge operator}) is
\be
\label{full determinant gauge matrix}
\det \cM = \frac1{64r^8} \big( j^2 + r^2\alpha(\sigma)^2 \big) \big( (j+1)^2 + r^2 \alpha(\sigma)^2 \big) \Big( j(j+1) - \frac{\alpha(\fm)^2}4 \Big)^2 \;.
\ee
The coefficient of $r$ in front does not depend on the background and can be reabsorbed in the normalization.
These eigenvalues have multiplicity $2j+1$.

For $j = \frac{|\alpha(\fm)|}2 \geq \frac12$ only two harmonics exist. For instance, when $\alpha(\fm) \geq 1$ then $Y^{s+1}_{j,j_3}$ and $Y^s_{j,j_3}$ exist while $D_- Y^s_{j,j_3} = 0$. Removing the second row/column in (\ref{matrix gauge operator}), the determinant is
$$
\frac1{4r^6} \, \frac{\alpha(\fm)^2}4 \Big( \big( \frac{|\alpha(\fm)|}2 + 1 \big)^2 + r^2 \alpha(\sigma)^2 \Big)
$$
with multiplicity $|\alpha(\fm)| + 1$. When $\alpha(\fm) \leq -1$ we reach the same conclusion.
The case $j = \alpha(\fm) = 0$ has to be treated separately. Only the harmonic $Y^{s=0}_{j=0,j_3}$ exists. Removing the first and second row/column in (\ref{matrix gauge operator}), the two eigenvalues are
$$
0\;,\qquad \frac1{2r^2} \Big( 1 + r^2\alpha(\sigma)^2 \Big) \;.
$$
The zero eigenvalue corresponds to the zero-mode of the background $\sigma_0$: we have to remove the eigenvalue from the determinant, and integrate over the zero-mode.

For $j = \frac{|\alpha(\fm)|}2 - 1 \geq 0$ only one harmonic exists. For instance, when $\alpha(\fm) \geq 2$ then $Y^{s+1}_{j,j_3}$ exists while $D_- Y^{s+1}_{j,j_3} = 0$. Removing the second, third and fourth row/column in (\ref{matrix gauge operator}), we are left with
$$
\frac1{r^2} \Big( \frac{\alpha(\fm)^2}4 + r^2 \alpha(\sigma)^2 \Big)
$$
with multiplicity $|\alpha(\fm)|-1$. If $|\alpha(\fm)|=1$ this case does not arise (but we can still formally use the formula if we include the multiplicity); if $\alpha(\fm)=0$ this case does not arise as well, but we have to treat it separately.

Let us notice that, as it should, the determinant only depends on the choice of gauge $\xi$ by an irrelevant multiplicative constant. For generic $\xi$ the derivative contribution in the $2\times2$ upper-left corner of the matrix (\ref{matrix gauge operator}) gets modified to:
$$
\mat{ \tilde A_+^\dag & \tilde A_-^\dag} \mat{ -2(1 - \frac1\xi) D_+ D_- & 2(1 + \frac1\xi) D_+D_+ \\ 2(1 + \frac1\xi) D_- D_- & -2(1 - \frac1\xi) D_- D_+ } \mat{ \tilde A_+ \\ \tilde A_-}
$$
which is still Hermitian. With our choice of phases:
\bea
D_+ D_- Y^{s+1}_{j,j_3} &= - \frac{s_+^2}{4r^2} Y^{s+1}_{j,j_3} \qquad\qquad &
D_+ D_+ Y^{s-1}_{j,j_3} &= \frac{s_+ s_-}{4r^2} Y^{s+1}_{j,j_3} \\
D_- D_- Y^{s+1}_{j,j_3} &= \frac{s_+ s_-}{4r^2} Y^{s-1}_{j,j_3} \qquad\qquad &
D_- D_+ Y^{s-1}_{j,j_3} &= - \frac{s_-^2}{4r^2} Y^{s-1}_{j,j_3} \;,
\eea
the determinant of $\cM$ is the same as in (\ref{full determinant gauge matrix}) but multiplied by $-1/\xi$.

The ghost term is $-\bar c D^2 c \equiv \bar c \cO_c c$: decomposing in a basis of spin spherical harmonics the determinant is:
\be
\Det \cO_c = \prod_{\alpha \in G} \prod_{j \geq \frac{|\alpha(\fm)|}2}^\infty \Big( j(j+1) - \frac{\alpha(\fm)^2}4 \Big)^{2j+1} \;.
\ee

Let us put all pieces together, discarding numerical factors of $r$. When $\alpha(\fm) \neq 0$ we find:
\be
\Det \cO_\text{gauge} \Big|_\alpha = \prod_{k=\frac{|\alpha(\fm)|}2}^\infty \big( k^2 + r^2 \alpha(\sigma)^2 \big)^{2k-1} \big( (k+1)^2 + r^2 \alpha(\sigma)^2 \big)^{2k+3} \Big( k(k+1) - \frac{\alpha(\fm)^2}4 \Big)^{2(2k+1)} \;.
\ee
In the end we have to take the product over all roots $\alpha \in G$. Since the adjoint is a real representation, for each root $\alpha$ we have also $-\alpha$. If we consider the product of a root and its opposite:
$$
\prod_{\{\alpha,-\alpha\}} \big( k^2 + r^2\alpha(\sigma)^2 \big) = \prod_{\{\alpha, - \alpha\}} \big( k + ir\alpha(\sigma) \big)^2 = \prod_{\{\alpha, -\alpha \}} \big( k - ir\alpha(\sigma) \big)^2 \;.
$$
We will use this relation often below. Therefore:
\be
\frac{\Det \cO_c}{\sqrt{\Det \cO_\text{gauge}}} \Big|_\alpha = \prod_{k=\frac{|\alpha(\fm)|}2}^\infty \frac1{ \big( k+ir\alpha(\sigma) \big)^{2k-1} \big( k+1 + ir\alpha(\sigma) \big)^{2k+3}} \;.
\ee
When $\alpha(\fm) = 0$ we find instead:
\be
\Det' \cO_\text{gauge} \Big|_\alpha = r^2 \alpha(\sigma)^2 \prod_{k\geq 0} \big( k^2 + r^2 \alpha(\sigma)^2 \big)^{2k-1} \big( (k+1)^2 + r^2 \alpha(\sigma)^2 \big)^{2k+3} \prod_{k\geq 1} \big( k(k+1) \big)^{2(2k+1)}
\ee
where we wrote $\Det'$ because there is a zero eigenvalue, corresponding to a zero-mode of the background $\sigma$ along the root $\alpha$, that we removed. Also $\cO_c$ has a zero eigenvalue. The determinants give:
\be
\frac{\Det' \cO_c}{\sqrt{\Det' \cO_\text{gauge}}} \Big|_\alpha = \frac1{|\alpha(\sigma)|} \prod_{k=\frac{|\alpha(\fm)|}2}^\infty \frac1{ \big( k+ir\alpha(\sigma) \big)^{2k-1} \big( k+1 + ir\alpha(\sigma) \big)^{2k+3}}
\ee

\subsection{Gaugino one-loop determinant}

The fermionic part of the YM action (\ref{YM Lagrangian}) expanded at quadratic order around the background is:
$$
\cL = \frac i2 \bar\lambda \gamma^\mu D_\mu \lambda + \frac i2 \bar\lambda[\sigma_0 ,\lambda] + \frac12 \bar\lambda \gamma_3[\eta_0 ,\lambda] \;.
$$
As before, modes along the Cartan generators give a background-independent constant that we discard. We are left with a decomposition along the roots $\alpha$ of $G$.

The fermions $\lambda, \bar\lambda$ have two components of spin $\pm \frac12$, so we decompose them into spin spherical harmonics.
For $j \geq \frac12 + \frac{|\alpha(\fm)|}2$ the effective spins are $s = \pm \frac12 - \frac{\alpha(\fm)}2$. The matrix acting on the subspace $(Y^{\frac12 - \frac{\alpha(\fm)}2}_{j,j_3}, Y^{-\frac12 - \frac{\alpha(\fm)}2}_{j,j_3})$ is
$$
\cM = \frac1{2r} \mat{ ir \alpha(\sigma) + \frac{\alpha(\fm)}2 & 2irD_+ \\ 2irD_- & ir\alpha(\sigma) - \frac{\alpha(\fm)}2} \;.
$$
However the product of spinors is $\bar\lambda \lambda = \bar\lambda^\alpha \varepsilon_{\alpha\beta} \lambda^\beta = - \bar\lambda^2 \lambda^1 + \bar\lambda^1 \lambda^2$. Therefore we should really compute the determinant of $\cO_\lambda \equiv\smat{-1 & 0 \\ 0 & 1} \cM$:
\be
\cO_\lambda = \frac1{2r} \mat{ -ir \alpha(\sigma) - \frac{\alpha(\fm)}2 & -2irD_+ \\ 2irD_- & ir\alpha(\sigma) - \frac{\alpha(\fm)}2} \;.
\ee
The determinant is:
\be
\det \cO_\lambda = \frac1{4r^2} \Big(  j(j+1) + r^2\alpha(\sigma)^2 + \frac14 \Big) = \frac1{4r^2} \big( j + \tfrac12 - ir\alpha(\sigma) \big) \big( j+\tfrac12 + ir\alpha(\sigma) \big)
\ee
with multiplicity $2j+1$.

For $j = \frac{|\alpha(\fm)|}2 - \frac12 \geq 0$ only one harmonic exists. For instance, when $\alpha(\fm) \geq 1$ then $\lambda =  \big( Y^{\frac12 - \frac{\alpha(\fm)}2}_{j,j_3}, 0 \big)$ and the eigenvalue is
$$
- \frac1{2r}  \Big( \frac{|\alpha(\fm)|}2 + ir\alpha(\sigma) \Big)
$$
with multiplicity $|\alpha(\fm)|$. When $\alpha(\fm) \leq -1$ then $\lambda = \big( 0, Y^{-\frac12 - \frac{\alpha(\fm)}2}_{j,j_3} \big)$ and the eigenvalue is the same as above but with opposite sign. For $\alpha(\fm) = 0$ this case does not exist, but we can still formally use the formula as long as we include the multiplicity.

The one-loop determinant, discarding numerical factors of $r$, is then:
\bea
\Det \cO_\lambda &= \prod_{\alpha \in G} (-1)^\sub{\alpha(\fm)} \big( \tfrac{|\alpha(\fm)|}2 + ir\alpha(\sigma) \big)^{|\alpha(\fm)|} \prod_{j=\frac{|\alpha(\fm)|}2 + \frac12}^\infty \big( j + \tfrac12 + ir\alpha(\sigma) \big)^{2j+1} \big( j + \frac12 - ir\alpha(\sigma) \big)^{2j+1} \\
&= \prod_{\alpha \in G} \prod_{k = \frac{|\alpha(\fm)|}2}^\infty \big( k + ir\alpha(\sigma) \big)^{2k} \big( k + 1 + ir\alpha(\sigma) \big)^{2k+2} \;,\label{last}
\eea
In the last equality \eqref{last} we changed sign to $\alpha(\sigma)$ since we take the product over $\alpha$ and the adjoint is a real representation. We  used the function $\sub{x} = \frac{|x|+x}2$ and
\be
\prod_{\alpha \in G} (-1)^\sub{\alpha(\fm)}= (-1)^{\sum_{\alpha>0} \alpha(\fm)} =(-1)^{2\delta(\fm)}= 1
\ee
where $\delta$ is the Weyl vector.

\subsection{Fermion one-loop determinant}

The fermionic part of the matter action (\ref{matter Lagrangian}) expanded at second order around the background is:
$$
\cL = \bar\psi \Big( -i \gamma^\mu D_\mu + i\sigma - \gamma_3\eta - \frac q{2r} \Big) \psi \;.
$$
There are two components in $\psi$ of spin $\pm \frac12$, therefore the effective spins are $s = \pm \frac12 - \frac{\rho(\fm)}2$. As in the previous section, we compute the determinant of the operator multiplied by $\smat{1 & 0 \\ 0 & -1}$.

For $j \geq \frac12 + \frac{|\rho(\fm)|}2$, both harmonics $Y^{\frac12 - \frac{\rho(\fm)}2}_{j,j_3}$ and $Y^{-\frac12 - \frac{\rho(\fm)}2}_{j,j_3}$ exist, then the operator is:
\be
\label{fermionic matrix}
\cO_\psi = \frac1r \mat{ ir\rho(\sigma) - \frac q2 - \frac{\rho(\fm)}2 & - 2ir D_+ \\ 2ir D_- & -ir\rho(\sigma) + \frac q2 - \frac{\rho(\fm)}2} \;.
\ee
The determinant of the matrix is:
\be
\det \cO_\psi = \frac1{r^2} \Big( j + \frac{q+1}2 - ir\rho(\sigma) \Big) \Big( j+1 - \frac{q+1}2 + ir\rho(\sigma) \Big)
\ee
and each mode has degeneracy $2j+1$.

For $j = \frac{|\rho(\fm)|}2 - \frac12 \geq 0$, one of the two harmonics with $s = \pm \frac12 - \frac{\rho(\fm)}2$ does not exist because $|s|$ exceeds $j$. For instance, when $\rho(\fm) \geq 1$ then $Y^{\frac12 - \frac{\rho(\fm)}2}_{j,j_3}$ exists while $D_- Y^{\frac12 - \frac{\rho(\fm)}2}_{j,j_3} = 0$. Then we take $\psi = (Y^{\frac12 - \frac{\rho(\fm)}2}_{j,j_3}, 0)$, only the upper-left corner in (\ref{fermionic matrix}) is non-trivial and the eigenvalue is
\be
\label{extra eigenvalue}
- \frac1r \Big( \frac{|\rho(\fm)|}2 + \frac q2 - ir\rho(\sigma) \Big)
\ee
with degeneracy $|\rho(\fm)|$. When $\rho(\fm) \leq -1$ only $Y^{-\frac12 - \frac{\rho(\fm)}2}_{j,j_3}$ exists. Taking $\psi = (0, Y^{-\frac12 - \frac{\rho(\fm)}2}_{j,j_3})$ we pick the lower-right corner in (\ref{fermionic matrix}) getting (\ref{extra eigenvalue}) with an extra minus sign, and same degeneracy $|\rho(\fm)|$. For $\rho(\fm) = 0$ this case does not exist, but we can still formally use the formula as long as we include the multiplicity.

Multiplying all eigenvalues, up to multiplicative factors of $r$ that we disregard, we get the one-loop determinant:
\be
\Det \cO_\psi = \prod_{\rho \in R_\Phi} (-1)^\sub{\rho(\fm)} \prod_{k = \frac{|\rho(\fm)|}2}^\infty \Big( k+ \frac q2 - ir\rho(\sigma)\Big)^{2k} \Big( k+1 - \frac q2 + ir\rho(\sigma) \Big)^{2k+2} \;,
\ee
where we used the function $\sub{x} = \frac{|x|+x}2$.

{
\bibliographystyle{utphys}
\bibliography{S2_localization}
}

\end{document}